\documentclass{article}
\ifdefined\pdfinfoomitdate\pdfinfoomitdate=1\fi
\ifdefined\pdftrailerid\pdftrailerid{}\fi
\ifdefined\pdfsuppressptexinfo\pdfsuppressptexinfo=-1\fi
\usepackage[T1]{fontenc}
\usepackage{microtype}
\usepackage{graphicx}
\usepackage{booktabs}
\usepackage{array}
\usepackage{amsmath,amssymb}
\usepackage{xcolor}
\usepackage{xr-hyper}   
\usepackage{hyperref}
\usepackage{tikz}
\usetikzlibrary{arrows.meta,positioning,calc,decorations.pathreplacing,fit,backgrounds}
\usepackage{pgfplots}
\pgfplotsset{compat=1.17}

\ifdefined\ANONYMOUS
  \usepackage{mlsys2025}
\else
  \usepackage[accepted]{mlsys2025}
  \makeatletter
  \renewcommand{\mlsys@appearing}{Preprint.}
  \makeatother
\fi
\usepackage[capitalise]{cleveref}

\newcommand{\BlkEarlyEarly}{0.08}
\newcommand{\BlkEarlyLate}{0.00}
\newcommand{\BlkEarlyMid}{0.30}
\newcommand{\BlkLateLate}{0.23}
\newcommand{\BlkMidLate}{0.22}
\newcommand{\BlkMidMid}{0.88}
\newcommand{\BlkMidShare}{51}
\newcommand{\BorrowAggrDfourall}{0.66}

\newcommand{\BorrowAggrRtwoh}{0.33}

\newcommand{\CapRatio}{1.94}
\newcommand{\CohHi}{1.39}

\newcommand{\CohLo}{1.13}
\newcommand{\ConfAggrBfsixteen}{0.94}
\newcommand{\ConfAggrDfourall}{0.99}

\newcommand{\ConfAggrFeightall}{0.89}
\newcommand{\ConfAggrRfourhgonetwoeight}{0.79}
\newcommand{\ConfAggrRtwoh}{1.55}
\newcommand{\ConfArms}{24}
\newcommand{\ConfDevRfourhgonetwoeight}{0.21}
\newcommand{\ConfFinalShare}{72.8}
\newcommand{\ConfHillDfourall}{2.45}
\newcommand{\ConfHillDfourhall}{1.55}
\newcommand{\ConfHillFeightall}{2.06}
\newcommand{\ConfHillRfourhgonetwoeight}{1.27}
\newcommand{\ConfHillRtwoh}{0.94}

\newcommand{\ConfMedianInBand}{24}

\newcommand{\ConfRows}{14{,}474}
\newcommand{\ConfSeqs}{136}

\newcommand{\ConfTopShareRfourhgonetwoeight}{6}

\newcommand{\ConfVShare}{94.6}
\newcommand{\ConvCells}{32}

\newcommand{\ConvKrylov}{72}

\newcommand{\ConvLateNeg}{8/12}
\newcommand{\ConvMidNeg}{10/20}
\newcommand{\ConvOf}{32}

\newcommand{\CosKLoopOne}{0.83}

\newcommand{\CosVLoopOne}{0.61}

\newcommand{\CrossoverKTok}{12.8}
\newcommand{\DasIdentity}{0.966}
\newcommand{\DasMatureV}{0.438}

\newcommand{\DedupDecHalfSixteen}{0.913}

\newcommand{\DedupSeqHalfSixteen}{0.978}

\newcommand{\DiagCpuOverGh}{1.002}
\newcommand{\DiagEpsQuarter}{0.95}
\newcommand{\DiagEpsSixteenth}{1.93}
\newcommand{\DiagEpsSixtyFourth}{7.28}
\newcommand{\DiagFdLarge}{0.99}
\newcommand{\DiagFdSmall}{11.1}
\newcommand{\DiagJvpOverKl}{11.5}
\newcommand{\DiagKDfourhall}{88.8}
\newcommand{\DiagMaxTangent}{152}
\newcommand{\DiagRowShare}{92.6}
\newcommand{\DiagRows}{95}

\newcommand{\DiagTangentHi}{7322}
\newcommand{\DiagTangentLo}{828}
\newcommand{\DiagTangentRmsLarge}{1.45}
\newcommand{\DiagTangentRmsSmall}{4.07}
\newcommand{\DiagTangentSeqs}{3}
\newcommand{\DiagTopP}{0.9998}
\newcommand{\DirBias}{2.37}

\newcommand{\DirCosInteight}{$-0.0002$}
\newcommand{\DirCosIntfour}{$0.0026$}

\newcommand{\DirHi}{2.44}
\newcommand{\DirLo}{2.21}
\newcommand{\DirShrink}{2.21}
\newcommand{\DirShrinkBatchHi}{5.64}
\newcommand{\DirShrinkBatchLo}{0.87}

\newcommand{\DirSr}{2.44}
\newcommand{\DirSrBatchHi}{7.28}
\newcommand{\DirSrBatchLo}{0.56}

\newcommand{\DrawsNew}{32}
\newcommand{\DrawsOld}{8}
\newcommand{\EOneArms}{21}

\newcommand{\EOneDraws}{8}

\newcommand{\EOneRows}{6{,}658}

\newcommand{\EOneSeqs}{64}
\newcommand{\ETwoBatches}{17}

\newcommand{\ETwoSeqs}{32}
\newcommand{\EarlyPairs}{0.26}
\newcommand{\EarlyPairsNew}{0.19}
\newcommand{\EightBelowOne}{5}

\newcommand{\EightClose}{2\times10^{-6}}

\newcommand{\EightEvenNine}{2.21}
\newcommand{\EightEvenShare}{0.13}
\newcommand{\EightIntLateCov}{0.35}

\newcommand{\EightIntMidCov}{0.53}
\newcommand{\EightIntMidEnrich}{3.5}
\newcommand{\EightIntMidMeas}{0.55}
\newcommand{\EightIntOddShare}{0.56}
\newcommand{\EightLateCov}{0.63}
\newcommand{\EightLateEnrich}{0.75}
\newcommand{\EightLateGeom}{0.80}
\newcommand{\EightLateMed}{0.59}
\newcommand{\EightLateN}{15}
\newcommand{\EightMidCov}{0.29}
\newcommand{\EightMidEnrich}{2.2}
\newcommand{\EightMidGeom}{0.14}
\newcommand{\EightMidMeas}{0.31}
\newcommand{\EightOdd}{0.88}
\newcommand{\EightOddExp}{$-0.18$}
\newcommand{\EightOddNine}{$-2.42$}
\newcommand{\EightOddPtwo}{0.11}
\newcommand{\EightOddShare}{0.87}
\newcommand{\EightOddSub}{0.64}
\newcommand{\EightOddSubN}{10}
\newcommand{\EightReOdd}{$-0.33$}
\newcommand{\EighteenAccExact}{68.1}
\newcommand{\EighteenAccKfour}{67.5}
\newcommand{\EighteenAccKl}{65.0}
\newcommand{\EighteenCriteria}{2}
\newcommand{\EighteenP}{0.013}
\newcommand{\EighteenPassed}{2}
\newcommand{\EighteenX}{88}
\newcommand{\EighteenY}{60}
\newcommand{\EighthCriteria}{3}

\newcommand{\ElevenAddKl}{0.85}
\newcommand{\ElevenCalChains}{64}
\newcommand{\ElevenCalMin}{33}
\newcommand{\ElevenCellMedRatio}{1.5}
\newcommand{\ElevenCellRankFisher}{95}

\newcommand{\ElevenCellRatio}{972}
\newcommand{\ElevenCellTopN}{10}
\newcommand{\ElevenCellTopShare}{99.9}
\newcommand{\ElevenCriteria}{17}

\newcommand{\ElevenFisherThirtyoneAgg}{13.6}

\newcommand{\ElevenFortyeightAggExact}{0.3032}
\newcommand{\ElevenForwards}{768}
\newcommand{\ElevenInFisherAgg}{9.0}
\newcommand{\ElevenKTwoCells}{2}
\newcommand{\ElevenKeyShareHi}{143}
\newcommand{\ElevenKeyShareLo}{89}

\newcommand{\ElevenKlThirtyoneAgg}{0.69}

\newcommand{\ElevenKlThirtyoneWorse}{6}
\newcommand{\ElevenKlTwentyfiveAgg}{0.84}

\newcommand{\ElevenLinMax}{1.7}
\newcommand{\ElevenMath}{176}
\newcommand{\ElevenNoVTwoAgg}{14.5}

\newcommand{\ElevenOutlierShare}{44}

\newcommand{\ElevenPassed}{15}
\newcommand{\ElevenPredFisherAgg}{10.8}
\newcommand{\ElevenVTwoCells}{26}
\newcommand{\ElevenValShareHi}{0.6}
\newcommand{\ExactBatchHi}{8.0}
\newcommand{\ExactBatchHiInteight}{3.3}
\newcommand{\ExactBatchLo}{2.2}
\newcommand{\ExactBatchLoInteight}{0.98}
\newcommand{\ExactCohInteight}{1.74}
\newcommand{\ExactCohIntfour}{1.81}
\newcommand{\ExactDiagInteight}{2.70}
\newcommand{\ExactDiagIntfour}{2.68}
\newcommand{\ExactDiagSharedInteight}{1.83}
\newcommand{\ExactDiagSharedIntfour}{0.56}
\newcommand{\ExactGate}{12}
\newcommand{\ExactInteightHi}{1.20}
\newcommand{\ExactInteightLo}{0.45}
\newcommand{\ExactKappaInteight}{1.65}
\newcommand{\ExactKappaIntfour}{4.09}
\newcommand{\ExactLateHi}{1.18}
\newcommand{\ExactLateLo}{0.94}
\newcommand{\ExactLinAllInteight}{1.03}
\newcommand{\ExactLinAllIntfour}{0.59}
\newcommand{\ExactLinLoopInteight}{0.98}

\newcommand{\ExactMidHi}{4.6}
\newcommand{\ExactMidLo}{0.8}
\newcommand{\ExactRInteight}{1.12}

\newcommand{\ExplTopShareRfourhgonetwoeight}{91}

\newcommand{\FifteenCriteria}{4}

\newcommand{\FifteenPassed}{3}
\newcommand{\FifteenPooled}{312}

\newcommand{\FifthCriteria}{13}
\newcommand{\FifthFailed}{7}

\newcommand{\FisherFinalV}{64}
\newcommand{\FisherJvpAggr}{0.56}
\newcommand{\FisherJvpFour}{1.04}
\newcommand{\FisherJvpHi}{0.80}
\newcommand{\FisherJvpLo}{0.39}

\newcommand{\FisherV}{94}

\newcommand{\FourteenAccExact}{43.0}

\newcommand{\FourteenAgreeFisher}{84.8}
\newcommand{\FourteenAgreeGuard}{83.0}

\newcommand{\FourteenCriteria}{5}

\newcommand{\FourteenFisherOnly}{8}

\newcommand{\FourteenGuardOnly}{6}

\newcommand{\FourteenParse}{112}
\newcommand{\FourteenPassed}{2}
\newcommand{\FourthCriteria}{17}
\newcommand{\FourthFailed}{7}

\newcommand{\FpeightHi}{1.19}
\newcommand{\FpeightLo}{1.02}
\newcommand{\FpeightPoints}{11}
\newcommand{\FpeightTorchWriteA}{1.06}

\newcommand{\FpeightTorchWriteC}{1.01}
\newcommand{\GaussHi}{3.0}
\newcommand{\GaussLaw}{2.80}
\newcommand{\GaussLo}{2.3}

\newcommand{\GsmN}{1319}

\newcommand{\HugCoda}{2}
\newcommand{\HugCore}{4}
\newcommand{\HugDiagDfourall}{1.14}
\newcommand{\HugDiagDfourallsr}{1.04}

\newcommand{\HugEOneRows}{1{,}529}
\newcommand{\HugEOneSeqs}{16}
\newcommand{\HugETwoBatches}{6}
\newcommand{\HugETwoOomGB}{97.3}
\newcommand{\HugETwoPeakGB}{72.8}
\newcommand{\HugETwoRows}{787}
\newcommand{\HugETwoSeqs}{8}
\newcommand{\HugFinalShare}{35.6}
\newcommand{\HugHeadDim}{96}
\newcommand{\HugHeads}{55}

\newcommand{\HugPrelude}{2}
\newcommand{\HugR}{32}

\newcommand{\HugScaleShallowHi}{1.00}
\newcommand{\HugScaleShallowLo}{0.87}

\newcommand{\HugScaleThirtyTwoHalf}{0.82}
\newcommand{\HugScaleThirtyTwoQuarter}{1.12}
\newcommand{\HugScaleThirtyTwoTwo}{0.29}

\newcommand{\KappaBatchAbove}{6}
\newcommand{\KappaBatchHi}{4.48}
\newcommand{\KappaBatchLo}{1.65}

\newcommand{\KappaFour}{1.06}
\newcommand{\KappaGateChecked}{36}
\newcommand{\KappaGram}{2.73}
\newcommand{\KappaGrowth}{3.44}
\newcommand{\KappaGrowthNew}{2.53}
\newcommand{\KappaHi}{1.34}
\newcommand{\KappaLo}{1.28}
\newcommand{\KappaMixedEight}{1.37}
\newcommand{\KappaMixedFour}{1.02}
\newcommand{\KappaMixedSixteen}{1.69}
\newcommand{\KappaMixedThirtyTwo}{4.79}
\newcommand{\KappaOffTotal}{1.72}

\newcommand{\KappaSE}{0.21}
\newcommand{\KappaSigmaGap}{1.0}

\newcommand{\KappaSpreadMax}{1.8}
\newcommand{\KappaSpreadNew}{1.6}
\newcommand{\KappaSubAll}{2.04, 2.92, 2.46, 3.25}

\newcommand{\KappaThirtyTwo}{2.68}
\newcommand{\KappaThirtyTwoInteight}{2.70}
\newcommand{\KappaThirtyTwoOld}{3.66}
\newcommand{\KurtPlainK}{9.5}
\newcommand{\KurtPlainV}{6.4}
\newcommand{\KurtRotK}{3.1}
\newcommand{\KurtRotV}{3.7}

\newcommand{\KvKiBToken}{768}

\newcommand{\LagFive}{0.05}
\newcommand{\LagFour}{0.29}
\newcommand{\LagOne}{0.23}
\newcommand{\LagOneMax}{0.44}
\newcommand{\LagSix}{0.22}

\newcommand{\LagThree}{0.08}
\newcommand{\LagTwo}{0.43}
\newcommand{\LagTwoMax}{0.78}
\newcommand{\LateAdjHi}{0.22}
\newcommand{\LateAdjLo}{0.04}
\newcommand{\LateFisherShare}{69}
\newcommand{\LatePairs}{0.068}
\newcommand{\LatePairsNew}{0.061}
\newcommand{\LedgerBatches}{21}
\newcommand{\LedgerCriteria}{176}
\newcommand{\LedgerFail}{60}
\newcommand{\LedgerFalsified}{5}
\newcommand{\LedgerGates}{22}
\newcommand{\LedgerPass}{116}
\newcommand{\LinThirtyTwoFpeight}{0.69}
\newcommand{\LinThirtyTwoInteight}{1.03}
\newcommand{\LinThirtyTwoIntfour}{0.59}
\newcommand{\LocALayer}{6}
\newcommand{\LocARatio}{43}
\newcommand{\LocAShare}{92}

\newcommand{\LocCases}{10}
\newcommand{\LocEarlyCells}{10}
\newcommand{\LocFalsify}{8}
\newcommand{\LocLayerSixCount}{4}
\newcommand{\LocMaxSpread}{6}
\newcommand{\LocSeqs}{5}
\newcommand{\LocShareThreeHi}{2.56}
\newcommand{\LocShareThreeLo}{0.04}
\newcommand{\LocTopHalf}{4}

\newcommand{\MathFinalShare}{63.6}
\newcommand{\MathMedianInBand}{21}

\newcommand{\MathPlanned}{32}
\newcommand{\MathRows}{10{,}786}
\newcommand{\MathSeqs}{24}
\newcommand{\MathSetDiff}{1}
\newcommand{\MathVShare}{95.7}
\newcommand{\MeasFinalDfourhall}{69}

\newcommand{\MeasKDfourhall}{89}
\newcommand{\MeasKFeightall}{94}

\newcommand{\MedianInBand}{19}

\newcommand{\MidLagOne}{0.25}
\newcommand{\MidLagTwo}{0.62}
\newcommand{\MonoSteps}{465}
\newcommand{\MonoViol}{168}

\newcommand{\NineBelowOneB}{2}
\newcommand{\NineChains}{12}
\newcommand{\NineChainsB}{12}
\newcommand{\NineIntMidA}{0.55}
\newcommand{\NineIntMidB}{0.45}
\newcommand{\NineLateCov}{0.71}
\newcommand{\NineLateEnr}{0.68}
\newcommand{\NineMidA}{2.63}
\newcommand{\NineMidB}{3.07}

\newcommand{\NineOddA}{0.43}
\newcommand{\NineOddB}{0.52}

\newcommand{\NineShrMidA}{0.25}
\newcommand{\NineShrMidB}{0.25}
\newcommand{\NineVarA}{0.37}
\newcommand{\NineVarB}{0.90}
\newcommand{\NineteenCriteria}{2}
\newcommand{\NineteenP}{0.025}
\newcommand{\NineteenPassed}{1}
\newcommand{\NineteenQ}{0.047}

\newcommand{\NineteenU}{106}
\newcommand{\NineteenV}{138}
\newcommand{\NineteenX}{150}
\newcommand{\NineteenY}{117}

\newcommand{\NonsplitHi}{1.00}
\newcommand{\NonsplitLo}{0.91}
\newcommand{\NumLayers}{24}
\newcommand{\NumLoops}{4}
\newcommand{\OtherScaleMedHi}{1.06}
\newcommand{\OtherScaleMedLo}{1.00}
\newcommand{\OuroFinalShareExpl}{67}
\newcommand{\OuroResidRatio}{10}
\newcommand{\OwnAggrDfourall}{1.58}

\newcommand{\OwnAggrRtwoh}{4.12}
\newcommand{\OwnArms}{8}
\newcommand{\OwnCap}{224}
\newcommand{\OwnConvKFirst}{0.54}
\newcommand{\OwnConvKLast}{0.014}
\newcommand{\OwnConvKSixteen}{0.044}
\newcommand{\OwnDiagDfourall}{1.12}
\newcommand{\OwnDiagDfourallsr}{0.96}
\newcommand{\OwnFinalShare}{40.8}
\newcommand{\OwnFirstTwenty}{2.0}
\newcommand{\OwnGen}{80}
\newcommand{\OwnGenMedian}{87}

\newcommand{\OwnInBand}{7}
\newcommand{\OwnKappaBatchHi}{2.79}
\newcommand{\OwnKappaBatchLo}{1.96}

\newcommand{\OwnKappaFour}{1.079}

\newcommand{\OwnKappaRelSE}{5.6}
\newcommand{\OwnKappaSE}{0.14}
\newcommand{\OwnKappaSeqs}{8}
\newcommand{\OwnKappaSpread}{1.1}
\newcommand{\OwnKappaThirtyTwo}{2.43}
\newcommand{\OwnKlRowBfsixteen}{$6.5\times10^{-7}$}
\newcommand{\OwnKlRowDtwoh}{$2.4\times10^{-2}$}
\newcommand{\OwnKlRowRfourh}{$7.6\times10^{-7}$}
\newcommand{\OwnKlRowRtwoh}{$7.0\times10^{-5}$}
\newcommand{\OwnLastFive}{89}
\newcommand{\OwnLength}{6}
\newcommand{\OwnMedDtwoh}{1.26}
\newcommand{\OwnMedHi}{1.01}
\newcommand{\OwnMedLo}{0.97}
\newcommand{\OwnMismatch}{0}
\newcommand{\OwnResidRatio}{346}
\newcommand{\OwnRows}{6{,}674}
\newcommand{\OwnSeqs}{64}
\newcommand{\OwnSrExp}{0.96}
\newcommand{\OwnSrRn}{1.68}
\newcommand{\OwnStop}{74}
\newcommand{\OwnTfourA}{1.21}
\newcommand{\OwnTfourAMiss}{0.01}
\newcommand{\OwnVShare}{58}
\newcommand{\POneDSlope}{1.24}
\newcommand{\POnePass}{7}

\newcommand{\PrDev}{0.5}

\newcommand{\PrTfourAFals}{1.2}

\newcommand{\PrTfourC}{30}

\newcommand{\RaggedDeltaHalf}{-1.97}

\newcommand{\ResidReadPct}{124}
\newcommand{\ResidStorePct}{49}

\newcommand{\ScaleDfourhallHalf}{1.00}
\newcommand{\ScaleDfourhallQuarter}{1.02}
\newcommand{\ScaleDfourhallTwo}{1.05}
\newcommand{\ScaleFeightallHalf}{1.02}
\newcommand{\ScaleFeightallQuarter}{1.03}
\newcommand{\ScaleFeightallTwo}{0.99}
\newcommand{\ScaleRtwohHalf}{0.23}
\newcommand{\ScaleRtwohQuarter}{0.25}
\newcommand{\ScaleRtwohTwo}{2.59}
\newcommand{\ScrChains}{6}
\newcommand{\ScrPlainKRatio}{4.51}

\newcommand{\SevAltExp}{3}
\newcommand{\SevAltPtwo}{3}
\newcommand{\SevCovHi}{19}
\newcommand{\SevCovLo}{6}

\newcommand{\SevDeterm}{48}
\newcommand{\SevDetermOf}{48}

\newcommand{\SevEvenRatio}{1.23}

\newcommand{\SevLateHi}{87}
\newcommand{\SevLateLo}{43}
\newcommand{\SevRhoExpHi}{$0.07$}
\newcommand{\SevRhoExpLo}{$-0.02$}
\newcommand{\SevRhoPtwoHi}{0.18}
\newcommand{\SevRhoPtwoLo}{0.11}
\newcommand{\SevStrongEven}{0.80}
\newcommand{\SevStrongKappa}{0.57}
\newcommand{\SevStrongOdd}{$-0.89$}
\newcommand{\SeventeenCriteria}{4}
\newcommand{\SeventeenFisherChanges}{119}
\newcommand{\SeventeenFormats}{10}

\newcommand{\SeventeenGuardChanges}{136}
\newcommand{\SeventeenKfourChanges}{243}
\newcommand{\SeventeenKfourDelta}{$-1.1$}
\newcommand{\SeventeenKlTwentyfiveAcc}{56.0}
\newcommand{\SeventeenKlTwentyfiveChanges}{235}

\newcommand{\SeventeenKlTwentyfiveKtwo}{21}
\newcommand{\SeventeenKlTwentyfiveP}{5\times10^{-5}}

\newcommand{\SeventeenKuX}{102}
\newcommand{\SeventeenKuY}{65}
\newcommand{\SeventeenPassed}{2}
\newcommand{\SeventeenRho}{0.82}
\newcommand{\SeventhCriteria}{4}
\newcommand{\SixConRatioIntfour}{1.6}
\newcommand{\SixConRatioShrink}{2.9}
\newcommand{\SixConReHi}{$-1.12$}
\newcommand{\SixConReLo}{$-1.03$}
\newcommand{\SixExpReHi}{1.25}
\newcommand{\SixExpReLo}{1.02}
\newcommand{\SixMedConIntfour}{4.34}
\newcommand{\SixMedConShrink}{2.85}
\newcommand{\SixMedExpIntfour}{2.64}
\newcommand{\SixMedExpShrink}{0.97}

\newcommand{\SixRepl}{0.84}
\newcommand{\SixRhoIntfour}{$-0.18$}
\newcommand{\SixRhoShrink}{$-0.36$}
\newcommand{\SixSeqs}{16}
\newcommand{\SixShrinkLinHi}{1.58}
\newcommand{\SixShrinkLinLo}{1.01}
\newcommand{\SixteenAccExact}{60.4}
\newcommand{\SixteenAccInt}{59.3}
\newcommand{\SixteenAccP}{0.12}

\newcommand{\SixteenCriteria}{6}

\newcommand{\SixteenN}{1119}
\newcommand{\SixteenPassed}{4}
\newcommand{\SixthCriteria}{5}
\newcommand{\SpecAmplify}{7}
\newcommand{\SpecCells}{32}
\newcommand{\SpecConverged}{10}
\newcommand{\SpecKrylov}{24}
\newcommand{\SpecLateContract}{11/12}
\newcommand{\SpecLateNeg}{58}
\newcommand{\SpecMidNeg}{50}
\newcommand{\SpecOtherNegHi}{$-0.82$}
\newcommand{\SpecOtherNegLo}{$-1.10$}
\newcommand{\SpecOtherNegMid}{10/15}
\newcommand{\SpecSeqs}{4}
\newcommand{\SpecSwitchConv}{4}
\newcommand{\SpecSwitchHi}{1.23}
\newcommand{\SpecSwitchLate}{$-0.95$}
\newcommand{\SpecSwitchLo}{1.04}
\newcommand{\SpecSwitchMidCells}{5}
\newcommand{\SplitAOneMin}{$-0.80$}
\newcommand{\SplitATwoMax}{$+0.73$}
\newcommand{\SplitAdjShare}{65}
\newcommand{\SplitAlign}{0.13}

\newcommand{\SplitCross}{13.9}
\newcommand{\SplitCrossSr}{2.5}

\newcommand{\SplitMidAOne}{$-0.38$}
\newcommand{\SplitMidATwo}{$+0.31$}
\newcommand{\SplitMidX}{$-0.28$}
\newcommand{\SplitRho}{0.76}
\newcommand{\SplitRhoJ}{0.87}
\newcommand{\SplitRhoJSr}{0.02}

\newcommand{\SplitSeqs}{10}
\newcommand{\SplitX}{0.09}
\newcommand{\SrExpConf}{1.05}
\newcommand{\SrExpHug}{1.04}
\newcommand{\SrExpMath}{1.09}
\newcommand{\SrOverExpected}{0.85}
\newcommand{\SrOverRnMeas}{1.72}
\newcommand{\SrOverRnPred}{2.00}
\newcommand{\SrRnConf}{1.95}
\newcommand{\SrRnHug}{1.76}
\newcommand{\SrRnMath}{2.06}
\newcommand{\StatCellsTotal}{192}
\newcommand{\StatConvCells}{177}
\newcommand{\StatExpanding}{9}
\newcommand{\StatExpandingPct}{14}
\newcommand{\StatKrylov}{40}
\newcommand{\StatMedExp}{1.01}
\newcommand{\StatMedNon}{1.01}
\newcommand{\StatRank}{2}
\newcommand{\StatRho}{$0.21$}
\newcommand{\StatRhoAbs}{$-0.08$}
\newcommand{\StatSeqs}{64}
\newcommand{\StatSwitchRe}{1.23}

\newcommand{\StatTopRe}{1.25}
\newcommand{\SwitchControls}{2}

\newcommand{\SwitchLateLinHi}{1.05}
\newcommand{\SwitchLateLinLo}{0.84}

\newcommand{\SwitchLayerJvp}{0.113}
\newcommand{\SwitchLayerKl}{5.95}
\newcommand{\SwitchLoop}{8}
\newcommand{\SwitchLoopJvp}{0.0021}
\newcommand{\SwitchLoopKl}{1.83}

\newcommand{\SwitchMargin}{2.55}
\newcommand{\SwitchMarginMedian}{3.36}
\newcommand{\SwitchMarginPct}{38}
\newcommand{\SwitchOtherLayersMax}{0.03}
\newcommand{\SwitchRestKl}{0.22}
\newcommand{\SwitchRowKl}{6.55}
\newcommand{\SwitchRowShare}{99.0}
\newcommand{\SwitchRows}{85}
\newcommand{\SwitchRtwohEps}{1.5}
\newcommand{\SwitchRtwohRatio}{27}
\newcommand{\SwitchWholeJvp}{0.221}
\newcommand{\SwitchWholeKl}{6.78}

\newcommand{\TSevenCases}{24}
\newcommand{\TSevenLayerSix}{3}
\newcommand{\TSevenLoopsMin}{99.8}
\newcommand{\TSevenMaxLayers}{7}
\newcommand{\TSevenNewSeqs}{12}
\newcommand{\TSevenScored}{19}
\newcommand{\TSevenTop}{0.76}
\newcommand{\TSevenTopHi}{0.92}
\newcommand{\TSevenTopLo}{0.56}
\newcommand{\TailAlphaK}{2.8}

\newcommand{\TenCriteria}{14}

\newcommand{\TenGsm}{136}

\newcommand{\TenMath}{24}

\newcommand{\TenPassed}{8}

\newcommand{\TenThirtyoneAgg}{5.18}

\newcommand{\TenThirtyoneMed}{0.60}

\newcommand{\TenTopLinHi}{314}
\newcommand{\TenTopLinLo}{151}

\newcommand{\TestsDedupCuda}{5}

\newcommand{\TfourA}{1.13}
\newcommand{\ThirdCriteria}{21}
\newcommand{\ThirteenAccExact}{68.5}
\newcommand{\ThirteenCriteria}{8}

\newcommand{\ThirteenMoved}{9}
\newcommand{\ThirteenMovedPct}{4.5}
\newcommand{\ThirteenN}{200}
\newcommand{\ThirteenPassed}{5}

\newcommand{\ThirteenSwitchers}{19}
\newcommand{\TopJump}{54}
\newcommand{\TopRatioDfourall}{31}
\newcommand{\TopRatioRtwoh}{41}
\newcommand{\TopScaleHalf}{2.26}
\newcommand{\TopScaleQuarter}{1.11}
\newcommand{\TopScaleTwo}{9.08}
\newcommand{\TopShareDfourall}{36}
\newcommand{\TopShareRtwoh}{73}

\newcommand{\TopTokens}{85}
\newcommand{\TrimAggrDfourall}{1.02}
\newcommand{\TrimAggrRtwoh}{1.07}

\newcommand{\TwelveCriteria}{6}
\newcommand{\TwelveGenShare}{98}

\newcommand{\TwelveGuardMath}{0.85}
\newcommand{\TwelveGuardMathWorse}{18}
\newcommand{\TwelveGuardTwentyfive}{0.90}

\newcommand{\TwelveHeadShare}{96}
\newcommand{\TwelveMathCellShare}{80}
\newcommand{\TwelveMathHeadShare}{96}
\newcommand{\TwelvePassed}{5}
\newcommand{\TwelvePosZero}{$4\times10^{-9}$}

\newcommand{\TwelveRowProbHi}{0.988}
\newcommand{\TwelveRowProbLo}{0.975}

\newcommand{\TwelveSwitchSeqs}{9}

\newcommand{\TwentyCalBatches}{13}
\newcommand{\TwentyCalChains}{16}
\newcommand{\TwentyCriteria}{8}

\newcommand{\TwentyFbUniThirtythree}{1.08}
\newcommand{\TwentyHW}{an RTX PRO 6000}
\newcommand{\TwentyHWName}{RTX PRO 6000}
\newcommand{\TwentyKlFbAgg}{1.065}
\newcommand{\TwentyKlFbBetter}{38}
\newcommand{\TwentyKlFbHi}{1.29}
\newcommand{\TwentyKlFbLo}{0.13}
\newcommand{\TwentyKlFbMed}{0.975}
\newcommand{\TwentyKlFbP}{0.084}
\newcommand{\TwentyKlFbWorseN}{26}

\newcommand{\TwentyKlUniHi}{0.58}
\newcommand{\TwentyKlUniLo}{0.06}
\newcommand{\TwentyOutCell}{values of core layer 1 in loops 1--4}
\newcommand{\TwentyOutRatio}{7.2}

\newcommand{\TwentyPassed}{6}

\newcommand{\TwentyRankFour}{0.982}
\newcommand{\TwentyRankTwo}{0.987}
\newcommand{\TwentyReproAgg}{1.0002}
\newcommand{\TwentyReproRho}{0.9998}
\newcommand{\TwentySeqs}{64}
\newcommand{\TwentyTopShareHi}{88}
\newcommand{\TwentyTopShareLo}{86}
\newcommand{\WeightGBStep}{10.1}
\newcommand{\XdevArms}{15}
\newcommand{\XdevHi}{1.01}
\newcommand{\XdevLo}{0.92}

\newcommand{\TwentyoneCalBatches}{8}
\newcommand{\TwentyoneCalBatchesTotal}{22}
\newcommand{\TwentyoneCalChains}{26}
\newcommand{\TwentyoneCriteria}{11}

\newcommand{\TwentyoneFisherIntAgg}{0.53}
\newcommand{\TwentyoneFisherV}{96}
\newcommand{\TwentyoneKlFisherAgg}{1.134}
\newcommand{\TwentyoneKlFisherFortyeight}{1.09}
\newcommand{\TwentyoneKlFisherGsm}{1.13}
\newcommand{\TwentyoneKlFisherIns}{1.10}
\newcommand{\TwentyoneKlFisherMed}{1.08}
\newcommand{\TwentyoneKlFisherWorse}{79}

\newcommand{\TwentyoneKlIntAgg}{0.60}

\newcommand{\TwentyoneLinAgg}{0.95}
\newcommand{\TwentyoneLinMed}{0.96}
\newcommand{\TwentyoneMath}{175}
\newcommand{\TwentyoneMeasK}{88}
\newcommand{\TwentyoneMedRatio}{0.94}
\newcommand{\TwentyoneOutCell}{keys of core layer 18 in loop 3}
\newcommand{\TwentyoneOutRankC}{4}
\newcommand{\TwentyoneOutRankD}{174}
\newcommand{\TwentyoneOutRatio}{125}

\newcommand{\TwentyonePassed}{8}
\newcommand{\TwentyoneSeqs}{64}

\newcommand{\PairedN}{1119}
\newcommand{\PairedGenerations}{3357}
\newcommand{\PairedExactCorrect}{762}
\newcommand{\PairedKfourCorrect}{755}
\newcommand{\PairedKlCorrect}{727}
\newcommand{\PairedKlKfourDelta}{-2.50}
\newcommand{\PairedKlKfourLo}{-4.63}
\newcommand{\PairedKlKfourHi}{-0.38}

\newcommand{\ScoreTestStrictDelta}{-3.31}

\newcommand{\ScoreTestStrictExactUnparsed}{266}

\newcommand{\ScoreTestStrictUniformUnparsed}{227}

\newcommand{\ScoreTestStrictAllocUnparsed}{256}

\newcommand{\ScoreTestFlexibleDelta}{-0.18}
\newcommand{\ScoreTestFlexibleLo}{-2.37}
\newcommand{\ScoreTestFlexibleHi}{2.01}

\newcommand{\ScoreTestFlexibleExactUnparsed}{0}

\newcommand{\ScoreTestFlexibleUniformUnparsed}{0}
\newcommand{\ScoreTestFlexibleUniformCorrect}{818}
\newcommand{\ScoreTestFlexibleAllocUnparsed}{1}
\newcommand{\ScoreTestFlexibleAllocCorrect}{816}
\newcommand{\ScoreTestFlexibleAllocExactDelta}{-5.27}
\newcommand{\ScoreTestFlexibleAllocExactLo}{-7.50}
\newcommand{\ScoreTestFlexibleAllocExactHi}{-3.04}
\newcommand{\ScoreTestFlexibleUniformExactDelta}{-5.09}
\newcommand{\ScoreTestFlexibleUniformExactLo}{-7.32}
\newcommand{\ScoreTestFlexibleUniformExactHi}{-2.86}
\newcommand{\ScoreTrainStrictDelta}{-2.50}

\newcommand{\ScoreTrainStrictExactUnparsed}{207}

\newcommand{\ScoreTrainStrictUniformUnparsed}{175}

\newcommand{\ScoreTrainStrictAllocUnparsed}{214}

\newcommand{\ScoreTrainFlexibleDelta}{0.00}
\newcommand{\ScoreTrainFlexibleLo}{-1.92}
\newcommand{\ScoreTrainFlexibleHi}{1.92}

\newcommand{\ScoreTrainFlexibleExactUnparsed}{0}

\newcommand{\ScoreTrainFlexibleUniformUnparsed}{0}

\newcommand{\ScoreTrainFlexibleAllocUnparsed}{0}
\newcommand{\ScoreTrainFlexibleAllocCorrect}{883}
\newcommand{\ScoreTrainFlexibleAllocExactDelta}{-3.40}
\newcommand{\ScoreTrainFlexibleAllocExactLo}{-5.32}
\newcommand{\ScoreTrainFlexibleAllocExactHi}{-1.47}

\newcommand{\ScoreTrainFlexibleUniformExactLo}{-5.27}
\newcommand{\ScoreTrainFlexibleUniformExactHi}{-1.52}

\newcommand{\AllRows}{17904}
\newcommand{\AllGuardBytes}{26.37}
\newcommand{\AllFisherGuardBytes}{31.32}
\newcommand{\AllTestMeasuredTwentyfiveStrictDelta}{-3.31}
\newcommand{\AllTestMeasuredTwentyfiveStrictLo}{-5.56}
\newcommand{\AllTestMeasuredTwentyfiveStrictHi}{-1.05}
\newcommand{\AllTestMeasuredTwentyfiveFlexibleDelta}{-0.18}
\newcommand{\AllTestMeasuredTwentyfiveFlexibleLo}{-2.37}
\newcommand{\AllTestMeasuredTwentyfiveFlexibleHi}{2.01}
\newcommand{\AllTestMeasuredTwentyfiveInteractionDelta}{3.13}
\newcommand{\AllTestMeasuredTwentyfiveInteractionLo}{1.56}
\newcommand{\AllTestMeasuredTwentyfiveInteractionHi}{4.69}
\newcommand{\AllTestMeasuredThirtyoneStrictDelta}{0.98}
\newcommand{\AllTestMeasuredThirtyoneStrictLo}{-0.71}
\newcommand{\AllTestMeasuredThirtyoneStrictHi}{2.67}
\newcommand{\AllTestMeasuredThirtyoneFlexibleDelta}{1.70}
\newcommand{\AllTestMeasuredThirtyoneFlexibleLo}{0.09}
\newcommand{\AllTestMeasuredThirtyoneFlexibleHi}{3.31}

\newcommand{\AllTestMeasuredFortyeightStrictDelta}{-0.27}
\newcommand{\AllTestMeasuredFortyeightStrictLo}{-1.72}
\newcommand{\AllTestMeasuredFortyeightStrictHi}{1.19}
\newcommand{\AllTestMeasuredFortyeightFlexibleDelta}{0.00}
\newcommand{\AllTestMeasuredFortyeightFlexibleLo}{-1.44}
\newcommand{\AllTestMeasuredFortyeightFlexibleHi}{1.44}

\newcommand{\AllTestFisherThirtyoneStrictDelta}{1.88}

\newcommand{\AllTestFisherThirtyoneFlexibleDelta}{2.23}

\newcommand{\AllTestMeasuredGuardFlexibleDelta}{2.59}
\newcommand{\AllTestMeasuredGuardFlexibleLo}{0.49}
\newcommand{\AllTestMeasuredGuardFlexibleHi}{4.70}

\newcommand{\AllTrainMeasuredTwentyfiveInteractionDelta}{2.50}
\newcommand{\AllTrainMeasuredTwentyfiveInteractionLo}{1.00}
\newcommand{\AllTrainMeasuredTwentyfiveInteractionHi}{4.00}

\newcommand{\AllTrainMeasuredGuardFlexibleDelta}{1.16}
\newcommand{\AllTrainMeasuredGuardFlexibleLo}{-0.68}
\newcommand{\AllTrainMeasuredGuardFlexibleHi}{3.01}

\newcommand{\MathScoreN}{200}
\newcommand{\MathScoreOutputs}{1800}
\newcommand{\MathScoreExactString}{43.0}
\newcommand{\MathScoreExactSymbolic}{50.0}

\newcommand{\MathScoreMinCap}{40.5}
\newcommand{\MathScoreMaxCap}{46.5}

\newcommand{\CheckpointMathN}{175}
\newcommand{\CheckpointCalN}{26}
\newcommand{\CheckpointCalBatches}{8}
\newcommand{\CheckpointCalMin}{129.1}
\newcommand{\CheckpointFisherV}{96.5}
\newcommand{\CheckpointMeasuredK}{88.4}
\newcommand{\CheckpointKlUniform}{0.596}
\newcommand{\CheckpointKlFisher}{1.134}
\newcommand{\CheckpointFisherUniform}{0.525}

\newcommand{\PackedNativeTests}{36}
\newcommand{\PackedAttentionCalls}{4320}
\newcommand{\PackedMaxAttentionError}{2.38\times 10^{-6}}
\newcommand{\PackedMaxAttentionScaled}{0.040}
\newcommand{\PackedRelativeRange}{1.56--1.64}
\newcommand{\PackedIntMismatch}{1}
\newcommand{\PackedIntBFChanges}{887}
\newcommand{\PackedKLBFChanges}{872}
\newcommand{\PackedFisherBFChanges}{838}
\newcommand{\PackedOracleMax}{9.67\times 10^{-7}}


\newcommand{\KL}{\mathrm{KL}}
\newcommand{\Var}{\operatorname{Var}}
\newcolumntype{P}[1]{>{\raggedright\arraybackslash}p{#1}}
\definecolor{kcol}{HTML}{1F5FA8}
\definecolor{vcol}{HTML}{D9730D}
\definecolor{wcol}{HTML}{2E8B57}
\definecolor{gcol}{HTML}{8A8A8A}

\usepackage{float}
\mlsystitlerunning{KV-Cache Error in Looped Language Models}

\begin{document}

\twocolumn[
\mlsystitle{Where KV-Cache Error Lives in Looped Language Models: A Diagnostic Study of Ouro and Huginn}

\begin{mlsysauthorlist}
\mlsysauthor{Dezhen Lu}{tue}
\mlsysauthor{Runguo Li}{uiuc}
\mlsysauthor{Zixiang Ni}{xjtu}
\mlsysauthor{Yuchen Tang}{zstu}
\mlsysauthor{Xingquan Kong}{swpu}
\mlsysauthor{Jiaxing Yang}{fudan}
\mlsysauthor{Jiarui Guo}{nankai}
\mlsysauthor{Lingxiang Gao}{csuft}
\end{mlsysauthorlist}
\mlsysaffiliation{tue}{University of T\"ubingen, T\"ubingen, Germany}
\mlsysaffiliation{uiuc}{University of Illinois Urbana-Champaign, Urbana, IL, USA}
\mlsysaffiliation{xjtu}{Xi'an Jiaotong University, Xi'an, China}
\mlsysaffiliation{zstu}{Zhejiang Sci-Tech University, China}
\mlsysaffiliation{swpu}{Southwest Petroleum University, China}
\mlsysaffiliation{fudan}{Fudan University, Shanghai, China}
\mlsysaffiliation{nankai}{Nankai University, Tianjin, China}
\mlsysaffiliation{csuft}{Swan College, Central South University of Forestry and Technology, Changsha, China}
\mlsyscorrespondingauthor{Dezhen Lu}{dezhen.lu@student.uni-tuebingen.de}

\mlsyskeywords{looped transformers, recurrent depth, KV cache, quantization, Fisher information, pre-registration}

\vskip 0.3in

\begin{abstract}
A looped language model applies one block of layers $R$ times per token and keeps a key--value (KV) plane for each
loop. We study where storage error changes the next-token distribution and how shared-weight error accumulates, using
Ouro-1.4B ($R=\NumLoops$) on GSM8K and MATH and Huginn-0125 ($R=\HugR$) on GSM8K-derived chains, with dense, simulated quantization.
Per unit of error, values carry \FisherV\% of the diagonal Fisher mass, but measured KL is dominated by keys with larger
quantization errors. Local second-order prediction follows the measured KL of typical sequences, including \ConfSeqs{}
held-out chains, but misses a few confident rows, and a local second-order factor for shared weights of
\ExactKappaInteight--\ExactKappaIntfour{} at $R=32$ depends on the error direction. A cell-level allocator lowers one-step
KL at 25--48\% of BF16 bytes, while paired task-accuracy comparisons depend on budget and the 25\% gap against K4V2
shrinks to near zero under a post hoc answer-extraction check. A later Ouro-2.6B test retains the key/value split but
falsifies measured-cost allocation's advantage over Fisher at 31\%; across all \LedgerBatches{} registrations,
\LedgerFail{} of \LedgerCriteria{} criteria failed.
\end{abstract}
]

\printAffiliationsAndNotice{}
\ifdefined\ANONYMOUS
\hypersetup{pdftitle={Where KV-Cache Error Lives in Looped Language Models: A Diagnostic Study of Ouro and Huginn}, pdfauthor={Anonymous Submission}, pdfsubject={MLSys 2027}}
\else
\hypersetup{pdftitle={Where KV-Cache Error Lives in Looped Language Models: A Diagnostic Study of Ouro and Huginn}, pdfauthor={Dezhen Lu, Runguo Li, Zixiang Ni, Yuchen Tang, Xingquan Kong, Jiaxing Yang, Jiarui Guo, Lingxiang Gao}}
\fi

\section{Introduction}
\label{sec:intro}

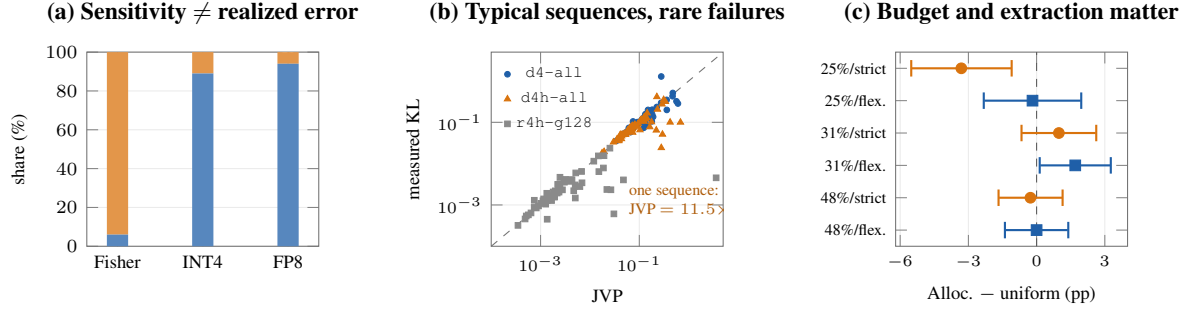
\begin{figure*}[t]
\centering
\pgfmathsetmacro{\kfisher}{100-\FisherV}
\pgfmathsetmacro{\vintfour}{100-\MeasKDfourhall}
\pgfmathsetmacro{\vfeight}{100-\MeasKFeightall}
\begin{tikzpicture}[font=\scriptsize]
\begin{axis}[
  width=4.65cm, height=4.15cm,
  ybar stacked, bar width=8pt,
  ymin=0, ymax=100,
  symbolic x coords={Fisher,INT4,FP8}, xtick=data,
  xticklabels={Fisher,INT4,FP8},
  ylabel={share (\%)}, ylabel style={font=\scriptsize},
  tick label style={font=\scriptsize},
  title={(a) Sensitivity $\neq$ realized error},
  title style={font=\small\bfseries,yshift=1pt},
  axis line style={draw=black!55},
  ymajorgrids, grid style={black!8},
  enlarge x limits=0.18,
]
\addplot[fill=kcol!75,draw=none] coordinates {
  (Fisher,\kfisher) (INT4,\MeasKDfourhall) (FP8,\MeasKFeightall)};
\addplot[fill=vcol!75,draw=none] coordinates {
  (Fisher,\FisherV) (INT4,\vintfour) (FP8,\vfeight)};
\node[font=\tiny\bfseries,text=white] at (axis cs:Fisher,3) {K};
\node[font=\tiny\bfseries,text=white] at (axis cs:Fisher,53) {V};
\node[font=\tiny\bfseries,text=white] at (axis cs:INT4,44) {K};
\node[font=\tiny\bfseries,text=white] at (axis cs:INT4,95) {V};
\node[font=\tiny\bfseries,text=white] at (axis cs:FP8,47) {K};
\node[font=\tiny\bfseries,text=white] at (axis cs:FP8,97) {V};
\end{axis}

\begin{loglogaxis}[
  at={(5.35cm,0)}, anchor=south west,
  width=4.65cm, height=4.15cm,
  xmin=1e-4,xmax=5,ymin=1e-4,ymax=5,
  xlabel={JVP},ylabel={measured KL},
  xlabel style={font=\scriptsize},ylabel style={font=\scriptsize},
  tick label style={font=\tiny},
  title={(b) Typical sequences, rare failures},
  title style={font=\small\bfseries,yshift=1pt},
  legend style={draw=none,fill=none,font=\tiny,at={(0.02,0.98)},anchor=north west},
  axis line style={draw=black!55},
  grid=major, grid style={black!8},
]
\addplot[domain=1e-4:5,samples=2,dashed,black!55,forget plot] {x};
\addplot[only marks,mark=*,mark size=1.05pt,kcol]
  table[x=jvp,y=kl,restrict expr to domain={\thisrow{arm}}{0:0}] {figs/scatter.dat};
\addplot[only marks,mark=triangle*,mark size=1.35pt,vcol]
  table[x=jvp,y=kl,restrict expr to domain={\thisrow{arm}}{1:1}] {figs/scatter.dat};
\addplot[only marks,mark=square*,mark size=1.05pt,gcol]
  table[x=jvp,y=kl,restrict expr to domain={\thisrow{arm}}{2:2}] {figs/scatter.dat};
\legend{\texttt{d4-all},\texttt{d4h-all},\texttt{r4h-g128}}
\node[anchor=west,font=\tiny,text=vcol!80!black,align=left] at (axis cs:4e-2,1.2e-3)
  {one sequence:\\JVP $=11.5\times$ KL};
\end{loglogaxis}

\begin{axis}[
  at={(10.70cm,0)}, anchor=south west,
  width=4.65cm, height=4.15cm,
  xmin=-6.2, xmax=4, ymin=0.5, ymax=6.5,
  xtick={-6,-3,0,3}, ytick={1,2,3,4,5,6},
  yticklabels={48\%/flex.,48\%/strict,31\%/flex.,31\%/strict,25\%/flex.,25\%/strict},
  xlabel={Alloc. $-$ uniform (pp)}, xlabel style={font=\scriptsize},
  tick label style={font=\tiny},
  title={(c) Budget and extraction matter},
  title style={font=\small\bfseries,yshift=1pt},
  axis line style={draw=black!55},
  xmajorgrids, grid style={black!8},
]
\draw[dashed,black!65] (axis cs:0,0.5) -- (axis cs:0,6.5);
\addplot[only marks,mark=*,mark size=2pt,vcol] coordinates
  {(\AllTestMeasuredTwentyfiveStrictDelta,6) (\AllTestMeasuredThirtyoneStrictDelta,4) (\AllTestMeasuredFortyeightStrictDelta,2)};
\addplot[only marks,mark=square*,mark size=2pt,kcol] coordinates
  {(\AllTestMeasuredTwentyfiveFlexibleDelta,5) (\AllTestMeasuredThirtyoneFlexibleDelta,3) (\AllTestMeasuredFortyeightFlexibleDelta,1)};
\draw[|-|,thick,vcol] (axis cs:\AllTestMeasuredTwentyfiveStrictLo,6) -- (axis cs:\AllTestMeasuredTwentyfiveStrictHi,6);
\draw[|-|,thick,vcol] (axis cs:\AllTestMeasuredThirtyoneStrictLo,4) -- (axis cs:\AllTestMeasuredThirtyoneStrictHi,4);
\draw[|-|,thick,vcol] (axis cs:\AllTestMeasuredFortyeightStrictLo,2) -- (axis cs:\AllTestMeasuredFortyeightStrictHi,2);
\draw[|-|,thick,kcol] (axis cs:\AllTestMeasuredTwentyfiveFlexibleLo,5) -- (axis cs:\AllTestMeasuredTwentyfiveFlexibleHi,5);
\draw[|-|,thick,kcol] (axis cs:\AllTestMeasuredThirtyoneFlexibleLo,3) -- (axis cs:\AllTestMeasuredThirtyoneFlexibleHi,3);
\draw[|-|,thick,kcol] (axis cs:\AllTestMeasuredFortyeightFlexibleLo,1) -- (axis cs:\AllTestMeasuredFortyeightFlexibleHi,1);
\end{axis}
\end{tikzpicture}
\caption{Experimental overview on Ouro. (a) Share of diagonal Fisher mass carried by values (orange) versus share of measured
INT4 and FP8 KL carried by keys (blue), on \EOneSeqs{} GSM8K chains. (b) Per-sequence measured KL against the local
second-order prediction; most sequences lie near the diagonal. (c) Accuracy difference, measured-cost allocation minus
its uniform baseline, on the same 1119 GSM8K test problems at three budgets. Orange circles: strict extraction;
blue squares: flexible extraction; bars: post hoc paired 95\% intervals. Exact storage uses FP32; BF16 defines the byte budget.
Dashed lines: equality in (b), zero difference in (c).}
\label{fig:overview}
\end{figure*}

A looped (recurrent-depth) language model applies a shared stack of layers several times per token
\cite{universal,huginn,ouro}. Inference engines for these models keep one KV plane per loop so that a token at depth
$d$ attends to the depth-$d$ keys and values of earlier tokens \cite{cdb}; with early exit, a token that stops at
depth $e<R$ copies its depth-$e$ entries into the deeper planes or the engine reads them through an alias
\cite{raggedkv}. Either way the cache holds $R$ planes per token. For Ouro-1.4B in BF16 that is \KvKiBToken{}\,KiB per
token, and a decode step reads about \WeightGBStep\,GB of weights, so KV reads overtake weight reads once the resident
context exceeds about \CrossoverKTok{}k tokens across a batch.

Prior work reduces this cache with residual coding against the final loop's plane \cite{residualquant}, cross-loop
latent compression \cite{lla}, sparse late-loop updates \cite{flashloop} and general KV quantizers
\cite{kivi,kvquant,quarot}. To choose among them, one needs to know which planes and which of keys and values carry the
error, whether a cheap sensitivity estimate ranks formats correctly, and whether reusing one quantized weight $R$ times
compounds its error. The measurements below examine these questions on identical rows. Figure~\ref{fig:overview}
summarizes the main outcome: sensitivity and realized error favor different cache components, typical sequences follow
the local second-order map, and lower one-step KL does not guarantee task accuracy. Criteria and input tables were
frozen before each GPU job; many predictions failed, and \cref{sec:ledger} records the verdicts.

\textbf{Contributions.}
\vspace{-0.6\baselineskip}
\begin{itemize}\setlength{\itemsep}{1pt}\setlength{\parskip}{0pt}
\item A row-matched protocol for comparing measured KL, local second-order prediction and diagonal Fisher on the same
  cache perturbations, together with a sensitivity map of the multi-plane cache (\cref{sec:method,sec:where}).
\item Evidence that second-order prediction works for typical sequences but fails on a small number of confident rows,
  and a localization of that failure to early layers (\cref{sec:linear}).
\item Tests of seven statistical and dynamical predictions on held-out chains, a second distribution and a second
  model (\cref{sec:theory,sec:weights}).
\item A cell-level allocation study linking Ouro-1.4B's Fisher failure to a 2-bit key cell post hoc, with mixed task results.
  Both allocator families protect the second Ouro checkpoint's outlier, where measured costs lose to Fisher
  (\cref{sec:repr}).
\end{itemize}

\textbf{Scope.} Allocation uses dense planes and quantization simulated in FP32. A separate batch-one L40S pilot times a
physical packed reader, but its full-model numerical gate fails; quality-qualified serving performance and long-context
capacity remain unestablished. Task accuracy concerns Ouro-1.4B; Huginn supplies KV and shared-weight diagnostics, and
Ouro-2.6B adds a teacher-forced checkpoint test. Cache-design relevance is indirect.

\section{Three predictors on identical rows}
\label{sec:method}

\textbf{Model and cache.} Ouro-1.4B has \NumLayers{} layers executed $R=\NumLoops$ times; every loop ends with a
normalization whose output feeds the next loop and the exit gate \cite{ouro}. The cache stores, for every token, layer
and loop, one key and one value vector per KV head. We write $X$ for the stored planes and index a plane by its loop
$\ell\in\{1,\dots,R\}$. Huginn-0125 runs a prelude of \HugPrelude{} layers once, a core of \HugCore{} layers $R=\HugR$
times from a random initial state, and a coda of \HugCoda{} layers \cite{huginn}; we quantize its core planes and replace
the random state by a deterministic draw seeded by the token ids. This changes the model's default inference behavior,
so the Huginn results describe this deterministic variant.

\textbf{Protocol.} We teacher-force fixed greedy chains. For every scored row $t$ the model runs all $R$ loops, reads
strict-prefix keys and values from the stored planes, and uses its freshly computed key and value on the diagonal,
exactly as a decode step does. A storage format maps the planes to $X+\Delta$; the row distribution changes from
$p_t=\mathrm{softmax}(z_t(X))$ to $q_t$, and we report $\sum_t\KL(p_t\,\|\,q_t)$. Arithmetic is the inference engine's
layer code in FP32 with TF32 disabled; only the paged cache is replaced by dense planes.

\textbf{Three predictors.} With $J_t=\partial z_t/\partial X$,
\begin{equation}
\KL(p_t\|q_t)=\tfrac12\,\Delta^{\!\top}J_t^{\!\top}(\mathrm{diag}\,p_t-p_tp_t^{\!\top})J_t\Delta+O(\|\Delta\|^3).
\end{equation}
(i)~The \emph{exact second order} $\tfrac12\sum_t\Var_{p_t}(J_t\Delta)$ needs one forward-mode JVP per format.
(ii)~The \emph{diagonal Fisher} $\tfrac12\sum_iF_{ii}\Delta_i^2$ uses \EOneDraws{} sampled-label draws per row; the
argmax-label Fisher is biased and our tests reject it. (iii)~The \emph{measured} KL. (i) versus (ii) isolates
off-diagonal terms, (iii) versus (i) nonlinearity.

\textbf{Weight sharing.} We give each loop its own leaf copy $W_\ell$ of a weight, so one backward pass yields every
loop's gradient, and compare
\begin{equation}
\kappa=\frac{\sum_iE[(\sum_\ell g_{\ell,i})^2]\Delta W_i^2}{\sum_\ell\sum_iE[g_{\ell,i}^2]\Delta W_i^2},\qquad
\kappa_2=\frac{\mathrm{JVP}_\text{all}}{\sum_\ell\mathrm{JVP}_\ell},
\label{eq:kappa}
\end{equation}
near 1 if the $R$ uses add incoherently and near $R$ if they add coherently. Here $\kappa_2$ is the ratio of local
second-order terms at the unquantized model, not the realized coherence of a finite quantization error;
$\mathrm{JVP}_\ell$ denotes the scalar score $\tfrac12\sum_t\Var_{p_t}((\partial z_t/\partial W_\ell)\Delta W)$ with
$\Delta W$ applied in loop $\ell$ only, and $\mathrm{JVP}_\text{all}$ applies it in every loop.

\textbf{Registration and devices.} For each batch, the analysis code and criteria entered a frozen input table before
the run; the job checked that table, ran CPU self-tests and archived the outputs with checksums (supplementary
Appendix~\ref{app:batches}). The registered analysis batches used single GH200 GPUs except the twentieth registration
(\TwentyHWName), where a uniform INT4 control reproduced the GH200 per-chain KL (aggregate ratio \TwentyReproAgg).
The separate L40S follow-ups appear in Appendices~\ref{app:l40s-qualification} and \ref{app:l40s-packed}.
Earlier H100 results and analyses added after a run are marked as such. The batches were run sequentially. Some later
criteria therefore use earlier outcomes; we identify those analyses as confirmations rather than prospective tests.
Here ``registered'' means that the criteria were written to a frozen file before the corresponding run and checked
against its outputs; it is an audit trail, not a guarantee against post hoc choices, which we label where they occur.

\textbf{Statistics.} Confidence intervals for the shared-weight results are batch bootstraps (\cref{tab:e2,tab:kappa}).
Answer-change counts in free generation are sign tests on discordant problems, one-sided where the registration
specifies it. We report these p-values descriptively. The registered batches are not corrected for multiplicity across
criteria, and we do not treat any single p-value as the main evidence. GSM8K uses the registered strict-match rule.
A post hoc audit reuses the saved task's flexible-extract rule, which selects the last numeric match; it tests
format sensitivity, not mathematical correctness. Direct format comparisons and changes in those differences use
paired per-problem intervals (supplementary Appendix~\ref{app:scoring-robustness}).
These intervals condition on the tested checkpoint, prompts and allocated format; they do not propagate uncertainty
from selecting or resampling the calibration chains and recomputing the allocation.

\section{Values carry the sensitivity, keys carry the error}
\label{sec:where}

On \EOneSeqs{} GSM8K chains (\EOneRows{} scored rows), values carry \FisherV\% of the diagonal Fisher mass of the stored
cache and the final loop's value plane alone \FisherFinalV\% (\cref{tab:sens}). Measured KL splits the other way:
quantizing only keys or only values with direct INT4 attributes \MeasKDfourhall\% of the KL to keys (\MeasKFeightall\%
for FP8), and the final plane contributes \MeasFinalDfourhall\%. The diagonal prediction attributes
\DiagKDfourhall\% to keys, so the reversal comes from error sizes, not from a failure of the map: per element, the keys'
quantization error is much larger. For the \XdevArms{} formats measured in both studies, the GH200 KL divided by the
earlier H100 value lies in \XdevLo--\XdevHi.

\begin{table}[t]
\centering\small
\caption{Share of diagonal Fisher mass of Ouro's stored KV cache by loop plane and key/value (GH200,
\EOneSeqs{} chains): sensitivity per unit of error, independent of format.}
\label{tab:sens}
\begin{tabular}{lrrrr}
\toprule
 & Plane 1 & Plane 2 & Plane 3 & Plane 4 \\
\midrule
K & 0.6\% & 0.8\% & 0.5\% & 3.7\% \\
V & 7.0\% & 10.9\% & 12.8\% & 63.6\% \\
 
\bottomrule
\end{tabular}
\end{table}

\section{Second order is right for typical sequences, not in aggregate}
\label{sec:linear}

\begin{table*}[t]
\centering\small\setlength{\tabcolsep}{4pt}
\caption{One-step KL of \EOneArms{} KV formats on Ouro (GH200, \EOneSeqs{} chains, \EOneRows{} rows). Aggregate ratios
are sums over sequences, as registered; $^\dagger$medians of per-sequence ratios are post hoc. Last column:
reproduction against the earlier H100 measurement where it exists. \texttt{r*}: residual against the final plane;
\texttt{d*}: direct; \texttt{h}: Hadamard rotation; \texttt{f8}: FP8; \texttt{-g$n$}: group size.}
\label{tab:e1}
\begin{tabular}{lrrrrrr}
\toprule
Format & KL / row & top-1 & meas./JVP (aggr.) & meas./JVP (med.)$^\dagger$ & meas./diag (med.)$^\dagger$ & GH200/H100 \\
\midrule
\texttt{bf16} & 9.8e-07 & 0.9998 & 1.03 & 1.00 & 1.03 & 1.00 \\
\texttt{f8all} & 1.6e-04 & 0.9965 & 0.91 & 0.99 & 1.04 & -- \\
\texttt{f8r4h} & 1.4e-04 & 0.9971 & 0.84 & 0.99 & 1.02 & 1.00 \\
\texttt{r4h} & 2.9e-05 & 0.9995 & 0.39 & 0.96 & 1.29 & 0.92 \\
\texttt{r4} & 3.1e-05 & 0.9994 & 0.10 & 0.97 & 1.28 & 1.01 \\
\texttt{r4h-g128} & 3.1e-05 & 0.9992 & 0.05 & 0.96 & 1.32 & 0.99 \\
\texttt{r4h-nols} & 8.0e-05 & 0.9988 & 0.11 & 0.95 & 1.25 & 1.00 \\
\texttt{r442h} & 9.2e-05 & 0.9988 & 0.57 & 1.01 & 1.14 & 0.97 \\
\texttt{d4} & 4.2e-04 & 0.9973 & 0.95 & 1.00 & 1.18 & -- \\
\texttt{d4h} & 2.7e-04 & 0.9973 & 0.53 & 0.96 & 1.21 & 0.99 \\
\texttt{d4h-g128} & 3.3e-04 & 0.9968 & 0.62 & 0.92 & 1.12 & -- \\
\texttt{d4h-all} & 8.7e-04 & 0.9937 & 0.79 & 1.01 & 1.07 & -- \\
\texttt{d4-all} & 1.5e-03 & 0.9932 & 1.02 & 1.02 & 1.06 & -- \\
\texttt{d4-all-sr} & 2.5e-03 & 0.9883 & 0.91 & 1.01 & 1.07 & -- \\
\texttt{r2} & 6.6e-04 & 0.9955 & 0.37 & 1.02 & 1.26 & 1.00 \\
\texttt{r2h} & 2.1e-03 & 0.9961 & 2.40 & 0.91 & 1.23 & 1.00 \\
\texttt{r2h-g16} & 2.9e-04 & 0.9974 & 0.13 & 1.00 & 1.20 & 1.00 \\
\texttt{r2h-g128} & 3.8e-03 & 0.9949 & 1.11 & 0.83 & 1.31 & 1.00 \\
\texttt{r2h-nols} & 1.7e-02 & 0.9896 & 2.27 & 1.31 & 2.00 & 1.01 \\
\texttt{rqpaper} & 2.6e-03 & 0.9937 & 1.61 & 0.99 & 1.14 & 1.00 \\
\texttt{d2h} & 1.7e-02 & 0.9815 & 1.87 & 1.28 & 1.83 & 1.00 \\
 
\bottomrule
\end{tabular}
\end{table*}

\textbf{Registered outcome.} The aggregate measured/JVP ratio falls in $[0.8,1.25]$ for only \POnePass{} of
\EOneArms{} formats, and the log--log slope of measured KL on the diagonal prediction is \POneDSlope{} (\cref{tab:e1};
criteria in \cref{tab:prereg}). Additivity of plane and K/V subsets, the rounding-temperature test and the numerical
floors pass. For formats in the linear band, KL scales as $\epsilon^2$ under $\Delta\to\epsilon\Delta$.

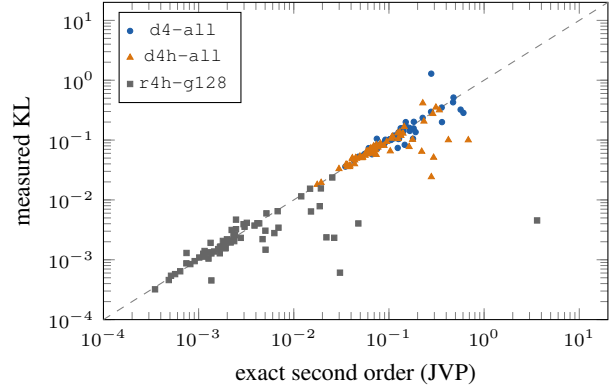
\begin{figure}[t]
\centering
\begin{tikzpicture}
\begin{loglogaxis}[width=\columnwidth,height=0.70\columnwidth,
  xlabel={exact second order (JVP)}, ylabel={measured KL},
  legend pos=north west, legend style={font=\scriptsize}, tick label style={font=\scriptsize},
  label style={font=\small}, xmin=1e-4, xmax=20, ymin=1e-4, ymax=20]
\addplot[only marks,mark=*,mark size=1pt,kcol] table[x=jvp,y=kl,restrict expr to domain={\thisrow{arm}}{0:0}] {figs/scatter.dat};
\addplot[only marks,mark=triangle*,mark size=1.3pt,vcol] table[x=jvp,y=kl,restrict expr to domain={\thisrow{arm}}{1:1}] {figs/scatter.dat};
\addplot[only marks,mark=square*,mark size=1pt,black!60] table[x=jvp,y=kl,restrict expr to domain={\thisrow{arm}}{2:2}] {figs/scatter.dat};
\addplot[domain=1e-4:20,samples=2,dashed,gray] {x};
\legend{\texttt{d4-all},\texttt{d4h-all},\texttt{r4h-g128}}
\end{loglogaxis}
\end{tikzpicture}
\caption{Per-sequence measured one-step KL against the exact second order (Ouro, GH200). Dashed: equality. Most
sequences lie on the diagonal; a few lie far below it.}
\label{fig:scatter}
\end{figure}

\textbf{Typical sequences.} The aggregate is dominated by a few sequences. The median per-sequence ratio lies in
$[0.8,1.25]$ for \MedianInBand{} of \EOneArms{} formats (\cref{fig:scatter}). The second batch made this a registered
criterion and tested it on \ConfSeqs{} chains not used here: \ConfMedianInBand{} of \ConfArms{} medians fall in the
band, and \MathMedianInBand{} of \ConfArms{} on MATH-500.

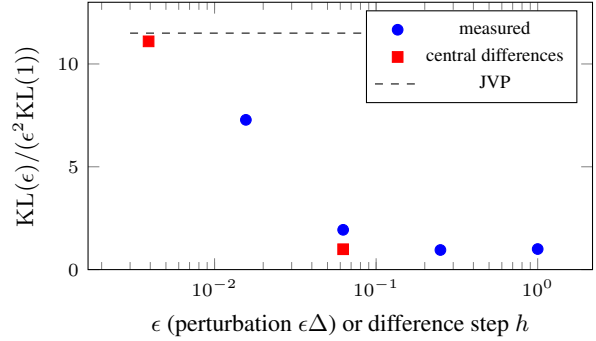
\begin{figure}[t]
\centering
\begin{tikzpicture}
\begin{semilogxaxis}[width=\columnwidth,height=0.62\columnwidth,
  xlabel={$\epsilon$ (perturbation $\epsilon\Delta$) or difference step $h$},
  ylabel={$\KL(\epsilon)/(\epsilon^2\KL(1))$}, ymin=0, ymax=13,
  legend pos=north east, legend style={font=\scriptsize}, tick label style={font=\scriptsize},
  label style={font=\small}]
\addplot[mark=*,blue,only marks] table[x=eps,y=ratio] {figs/sweep.dat};
\addplot[only marks,mark=square*,red] coordinates {(0.00390625,\DiagFdSmall) (0.0625,\DiagFdLarge)};
\addplot[dashed,black,domain=0.003:1.2,samples=2] {\DiagJvpOverKl};
\legend{measured, central differences, JVP}
\end{semilogxaxis}
\end{tikzpicture}
\caption{The dominant sequence for \texttt{d4h-all} (post hoc, CPU). Measured KL approaches the JVP only for
perturbations far smaller than the quantization error.}
\label{fig:sweep}
\end{figure}

\textbf{Why the aggregate fails (post hoc).} On the sequence with the largest discrepancy for \texttt{d4h-all}, the JVP
predicts \DiagJvpOverKl{} times the measured KL, and it is the correct small-perturbation limit: $\KL(\epsilon)/\epsilon^2$
approaches it only for perturbations far below the quantization error (\cref{fig:sweep}). One row carries
\DiagRowShare\% of the second-order term; there the model is confident (top probability \DiagTopP), and linearizing the
network while keeping the softmax exact makes the prediction worse, so the saturation is inside the network. A
re-registered localization on \TSevenNewSeqs{} new sequences finds it in a few layers of the first two loops: in all
\TSevenScored{} scored cases loops 1 and 2 hold at least \TSevenLoopsMin\% of the excess (supplementary Appendix~\ref{app:theory}).

These results support local second-order prediction on typical sequences of the tested formats. Aggregate ratios
remain necessary to expose concentrated errors; reporting only medians would hide them. We therefore report both,
because the linear radius differs between rows by orders of magnitude.

\section{Tests of statistical and dynamical predictions}
\label{sec:theory}

After the first batch we wrote down seven predictions, each with a source theory, a numerical prediction and a
falsification condition, and tested them on \ConfSeqs{} held-out GSM8K chains, \MathSeqs{} MATH-500 chains and Huginn
(\cref{tab:theory}; full derivations and measurements in supplementary Appendix~\ref{app:theory}).

\begin{table*}[t]
\centering\footnotesize\setlength{\tabcolsep}{3pt}
\caption{Predictions of the second registration (GH200). T1--T3 and T6--T7 on Ouro, T3--T5 also or only on Huginn.
``Fail (not falsified)'': the composite criterion failed but the stated falsification condition did not occur. The
post hoc column is not part of any verdict.}
\label{tab:theory}
\begin{tabular}{lP{2.2cm}P{4.9cm}P{4.2cm}P{1.9cm}P{2.3cm}}
\toprule
 & Source theory & Registered prediction & Measured & Verdict & Post hoc reading \\
\midrule
T1 & extreme values, stable laws & heavy-tailed formats: aggregate outside [0.8, 1.25], \texttt{r4h-g128} deviation $\ge$0.5; $\ge$18/24 medians in band; Hill $\hat\alpha$(\texttt{r4h-g128})$<1$ & aggregate 0.79 (dev.\ 0.21); 24/24 medians; $\hat\alpha$ = 1.27 & fail (not falsified) & tail $1<\alpha<2$, not $\alpha<1$ \\
T2 & extreme values of group maxima; rotation as Gaussianization & rotated ratios 1.41, 1.61 $\pm$25\%; unrotated [3.4, 5.4]; Gaussian law 2.8 excluded & 5/5 in band; unrotated 3.84 (MATH 4.37) & pass & -- \\
T3 & quenched vs annealed disorder & SR/RN in [1.5, 2.5]; SR / expected-error prediction in [0.6, 1.4] & 1.95 / 1.05 (MATH 2.06 / 1.09; Huginn 1.76 / 1.04) & pass & -- \\
T4 & fixed-point (equilibrium) dynamics & (a) RN/SR diagonal underestimate $\ge$1.5; (b) plane change monotone, $<$0.1 after loop 16; (c) final-plane Fisher share $<$30\% & (a) 1.13; (b) 0.54$\to$0.014; (c) 35.6\% & (a) falsified, (b) pass, (c) fail & third batch: errors correlated, Jacobians misaligned \\
T5 & correlation length of loop gradients & (a) $\kappa$ non-decreasing, $\kappa(32)/\kappa(4)\ge$1.5; (b) format spread $<$10\%; (c) correlation decreasing along the diagonal, late pairs $\ge$ early; (d) Fisher-free ratio within 30\% & (a) 3.44; (b) $\le$1.8\%; (c) 168/465 rising steps, late 0.068 vs early 0.26; (d) 6/6 & (a,b,d) pass, (c) fail & 32 draws: $\kappa(32)$ 2.68; middle loops; parity \\
T6 & distribution independence (MATH-500) & V share $\ge$84\%, final plane in [57, 77]\% & 95.7\%, 63.6\% & pass & -- \\
T7 & saturation localized in few layers & top 3 layers hold $\ge$50\% of JVP; excess not spread over $\ge$8 layers & top-3 share 0.04--2.56; at most 6 layers & fail (not falsified) & metric ill-posed; excess share confirmed in third batch \\
 
\bottomrule
\end{tabular}
\end{table*}

\textbf{What the statistical models get right.} Per-group scaling sets the error by each group's range, which grows like
$\sqrt{2\ln g}$ for a Gaussian tail and like $g^{1/\alpha}$ for a power-law tail. Measured KL follows the exact error
ratios of both branches on both distributions: heavy-tailed without rotation (keys' kurtosis \KurtPlainK), near-Gaussian
after a Hadamard rotation (T2, \cref{tab:groups}). Round-to-nearest has mean square error $\tfrac1{12}$ of a step
squared and stochastic rounding $\tfrac16$; stochastic rounding measures \SrRnConf{} times round-to-nearest on Ouro and
\OwnSrRn{} on Huginn's own chains (T3). The aggregate's tail is heavy, but its index lies between 1 and 2, not below 1
(T1).

\begin{table}[t]
\centering\small\setlength{\tabcolsep}{3pt}
\caption{Group-size scaling of median one-step KL (Ouro, GH200, held-out GSM8K and MATH-500 chains). rot.: Hadamard
rotation; res.: residual. Predictions are exact quantization
error ratios from CPU; the unrotated band was pre-registered around the key-weighted ratio. The Gaussian law is
the falsification alternative (band [\GaussLo, \GaussHi]).}
\label{tab:groups}
\begin{tabular}{lrrr}
\toprule
KL ratio, group size $g$ & Predicted & GSM8K & MATH \\
\midrule
rot.\ INT4, 32/16 & 1.41 $\pm$25\% & 1.41 & 1.32 \\
rot.\ INT4, 128/32 & 1.61 $\pm$25\% & 1.67 & 1.72 \\
rot.\ INT2 res., 32/16 & 1.41 $\pm$25\% & 1.44 & 1.68 \\
rot.\ INT2 res., 128/32 & 1.61 $\pm$25\% & 1.88 & 1.51 \\
unrot.\ INT4, 128/16 & [3.4, 5.4] & 3.84 & 4.37 \\
Gaussian law, 128/16 & 2.80 & -- & -- \\
 
\bottomrule
\end{tabular}
\end{table}

\textbf{What the fixed-point picture gets wrong.} Huginn is trained to approach a fixed point over its loops \cite{deq},
and its planes do converge (T4b). Adjacent converged planes should then receive nearly equal round-to-nearest errors
whose cross terms add, so the diagonal prediction should underestimate it. It barely does (ratio \TfourA, falsification
threshold \PrTfourAFals). A third registration separated the two reasons on Huginn's own chains: the errors of adjacent
late planes are strongly correlated ($\rho^J=\SplitRhoJ$), but the output responds to adjacent planes along nearly
orthogonal directions (alignment \SplitAlign), so cross terms make up only \SplitCross\% of the second-order term. On
Huginn's own chains one sequence switches on above the quantization scale; its excess sits in one row, one layer and a
band of middle loops (supplementary Appendix~\ref{app:huginn}).

\section{Shared-weight error accumulates through the middle loops}
\label{sec:weights}

\begin{table*}[t]
\centering\small
\caption{Quantizing Ouro's shared core weights (GH200, \ETwoSeqs{} chains, \ETwoBatches{} batches per depth). KL is
summed over scored rows; intervals are batch bootstraps.}
\label{tab:e2}
\begin{tabular}{lrrrrr}
\toprule
Format & KL, $R{=}4$ & $\kappa$ [95\% CI] & $\KL_\text{all}/\sum_\ell\KL_\ell$ [95\% CI] & meas./JVP, $R{=}4$ & meas./JVP, $R{=}1$ \\
\midrule
\texttt{int8\_ch} & 1.03 & 1.28 [1.22, 1.35] & 1.13 [1.03, 1.22] & 0.84 & 1.07 \\
\texttt{fp8\_block128} & 4.42 & 1.33 [1.26, 1.40] & 1.36 [1.24, 1.51] & 0.88 & 1.02 \\
\texttt{fp8\_channel} & 5.59 & 1.33 [1.26, 1.40] & 1.33 [1.11, 1.49] & 0.96 & 1.06 \\
\texttt{fp8\_tensor} & 3.29 & 1.32 [1.26, 1.40] & 1.16 [1.07, 1.27] & 0.91 & 0.99 \\
\texttt{int4\_g128} & 109 & 1.33 [1.27, 1.41] & 1.39 [1.26, 1.52] & 0.68 & 3.45 \\
\texttt{int4\_g128\_sr} & 307 & 1.30 [1.24, 1.38] & 1.72 [1.51, 1.95] & 0.77 & 6.28 \\
\texttt{int3\_g128} & 4.18e+03 & 1.34 [1.28, 1.42] & 3.81 [3.37, 4.29] & 5.21 & 7.42 \\
 
\bottomrule
\end{tabular}
\end{table*}

\textbf{Ouro.} At $R=4$, $\kappa$ is \KappaLo--\KappaHi{} for all seven formats (\cref{tab:e2}): the four uses' gradients
are mildly correlated, and $\kappa$ is a property of that geometry rather than of the format. The Fisher-free ratio agrees
at \CohLo--\CohHi{} in or near the linear regime. The registered amplification hypothesis, that sharing makes linear
response fail at a finer format at $R=4$ than at $R=1$, is not supported.

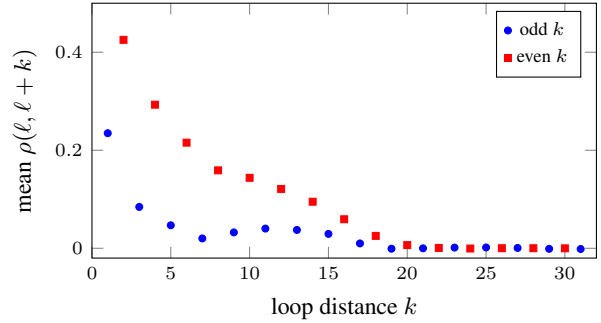
\begin{figure}[t]
\centering
\begin{tikzpicture}
\begin{axis}[width=\columnwidth,height=0.6\columnwidth,
  xlabel={loop distance $k$}, ylabel={mean $\rho(\ell,\ell+k)$}, ymin=-0.02, ymax=0.5, xmin=0, xmax=32,
  legend pos=north east, legend style={font=\scriptsize}, tick label style={font=\scriptsize},
  label style={font=\small}]
\addplot[only marks,mark=*,mark size=1.3pt,blue] table[x=lag,y=rho,restrict expr to domain={\thisrow{even}}{0:0}] {figs/lag.dat};
\addplot[only marks,mark=square*,mark size=1.3pt,red] table[x=lag,y=rho,restrict expr to domain={\thisrow{even}}{1:1}] {figs/lag.dat};
\legend{odd $k$, even $k$}
\end{axis}
\end{tikzpicture}
\caption{Correlation of Fisher-weighted loop gradients against loop distance, Huginn at $R=32$, INT4 (GH200,
\DrawsNew{} Fisher draws, from a count sketch of the per-loop gradients). The parity alternation is a post hoc
observation.}
\label{fig:lag}
\end{figure}

\textbf{Huginn.} The diagonal $\kappa$ grows from \KappaFour{} at $R=4$ to $\KappaThirtyTwo\pm\KappaSE$ at $R=32$
(\DrawsNew{} Fisher draws, jackknife; \cref{tab:kappa}), with formats within \KappaSpreadNew\% and $\kappa(32)=
\OwnKappaThirtyTwo\pm\OwnKappaSE$ on Huginn's own chains; with \DrawsOld{} draws the second registration had overstated
it as \KappaThirtyTwoOld, and disjoint \DrawsOld-draw subsets scatter as \KappaSubAll. Fisher-weighted loop gradients
correlate more at even than at odd loop distances (\cref{fig:lag}, post hoc). The local second-order ratio
(\cref{eq:kappa}) separates the formats that the diagonal Fisher averages together: $\kappa_2$ is \ExactKappaIntfour{} for INT4 and
\ExactKappaInteight{} for INT8, where the diagonal values are \ExactDiagIntfour{} and \ExactDiagInteight. INT8 is linear, and its measured
coherence $\KL_\text{all}/\sum_\ell\KL_\ell=\ExactCohInteight$ agrees. Synthetic error directions (stochastic rounding,
group-mean bias, shrinkage) all give \DirLo--\DirHi, and a chain's $\kappa_2$ is largely shared across
directions (rank correlation \SixRepl{} on \SixSeqs{} chains). That it rises with an expanding middle-loop map was
registered and falsified.

\begin{table}[t]
\centering\footnotesize\setlength{\tabcolsep}{2.2pt}
\caption{Shared-weight coherence on Huginn (GH200, \HugETwoSeqs{} chains, \HugETwoBatches{} batches per depth).
$R=4$ and $32$ from the third batch, $R=8$ and $16$ from the second; Fisher draws per batch as listed. The
\DrawsOld-draw rows carry the larger sampling error of the second batch. coh.: the
Fisher-free ratio $\KL_\text{all}/\sum_\ell\KL_\ell$ (INT4), measured for $R\le8$ in the second batch and at $R=32$ in the
fourth. Last column: INT8 / FP8 / INT4.}
\label{tab:kappa}
\begin{tabular}{rrrrrrrc}
\toprule
 & & \multicolumn{3}{c}{$\kappa$} & & & \\
\cmidrule(lr){3-5}
$R$ & draws & INT8 & FP8 & INT4 & spread & coh. & meas./JVP \\
\midrule
4 & 32 & 1.068 & 1.062 & 1.062 & 0.6\% & 1.04 & 1.00 / 1.00 / 0.99 \\
8 & 8 & 1.201 & 1.194 & 1.191 & 0.8\% & 1.26 & 1.00 / 1.00 / 0.94 \\
16 & 8 & 1.392 & 1.384 & 1.379 & 0.9\% & -- & 1.00 / 0.81 / 0.97 \\
32 & 32 & 2.697 & 2.655 & 2.685 & 1.6\% & 1.81 & 1.03 / 0.69 / 0.59 \\
 
\bottomrule
\end{tabular}
\end{table}

\textbf{Where the coherence sits.} Because the JVP is a quadratic form in the coefficients $a_\ell$ of the tangent
$\sum_\ell a_\ell\Delta W_\ell$, JVPs of four loop blocks (1--8, 9--24, 25--28, 29--32), of their six pairwise unions and
of the alternating signs $(-1)^\ell$ split the cross term $\mathrm{JVP}_\text{all}-\sum_\ell\mathrm{JVP}_\ell$ exactly:
into within- and between-block parts, and into even- and odd-lag sums. On \SixSeqs{} chains the split closes to
$\EightClose$ of $\mathrm{JVP}_\text{all}$. For shrinkage the late loops carry most of the cross term only because they
carry most of the single-loop mass: relative to equal correlation of all loop pairs they are under-correlated
(\EightLateEnrich$\times$), while pairs within loops 9--24 are \EightMidEnrich$\times$ enriched. For INT4 the middle block
also carries most of the chain-to-chain variation (\EightIntMidCov, against \EightMidCov{} for shrinkage). We registered
these readings and tested them on two fresh samples of \NineChains{} chains each, the second frozen before the first was
read (\cref{tab:ninth}): the middle-block enrichment (\NineMidA{} and \NineMidB) and the direction dependence (INT4
\NineIntMidA{} and \NineIntMidB{} against shrinkage \NineShrMidA{} and \NineShrMidB) replicate. The share of variation
carried by odd-lag cross terms, \EightOddShare{} on the original chains, is \NineOddA{} and \NineOddB{} on the new ones, and
its link to the loop map's spectrum does not replicate. Near the fixed point, loops neither add KV rounding errors
coherently nor align their weight gradients, and pairs of middle loops are the most correlated (\NineMidA--\NineMidB$\times$
their equal-correlation share on the new samples).

\section{Registration ledger}
\label{sec:ledger}

\cref{tab:ledger} lists every registration with its criteria counts. Thresholds of later registrations that used earlier
results say so in the registration; those criteria confirm observed effects rather than predict new ones. Deviations
from the registrations (four code fixes, two of them after out-of-memory failures, and changes of GPU host and type) are listed in
supplementary Appendix~\ref{app:deviations}; none changed a criterion or threshold. The registered batches are not
independent trials, so their count is an auditing device rather than a multiplicity correction.

\begin{table}[!t]
\centering\footnotesize\setlength{\tabcolsep}{2.5pt}
\caption{The \LedgerBatches{} registrations (GH200; the twentieth on \TwentyHWName). Each lettered sub-criterion counts once; a falsified criterion
counts as failed and as falsified; the \LedgerGates{} registered gates (bitwise and minimum-sample checks) all passed and
are not counted. Registration 9 is two independent runs of the same criteria.}
\label{tab:ledger}
\begin{tabular}{rP{4.0cm}rrrl}
\toprule
\# & Question & pass & fail & fals. & Detail \\
\midrule
1 & Ouro KV formats and weights: second-order predictors, subsets, floors & 4 & 4 & 0 & \cref{tab:prereg} \\
2 & seven theory predictions on held-out chains, MATH-500 and Huginn & 7 & 5 & 1 & \cref{tab:theory} \\
3 & Huginn own chains; why coherent errors do not add; $\kappa$ with 32 draws & 17 & 3 & 0 & \cref{tab:third} \\
4 & the switching sequence; exact $\kappa(32)$; loop-map spectrum & 9 & 7 & 0 & \cref{tab:fourth} \\
5 & expanding modes over 64 chains; error directions; converged spectra & 5 & 7 & 0 & \cref{tab:fifth} \\
6 & does $\kappa_2$ rise with an expanding loop map? & 1 & 3 & 1 & \cref{tab:sixth} \\
7 & lag structure of middle-loop cross terms & 2 & 2 & 0 & \cref{tab:seventh} \\
8 & exact block and parity decomposition of the cross term & 3 & 0 & 0 & \cref{tab:eighth} \\
9 & two out-of-sample replications of the eighth batch's post hoc readings & 7 & 3 & 0 & \cref{tab:ninth} \\
10 & bit widths allocated by the sensitivity map, at equal bytes & 8 & 6 & 0 & \cref{tab:tenth} \\
11 & bit widths allocated by measured single-cell KL, on unseen chains & 15 & 2 & 0 & \cref{tab:eleventh} \\
12 & why one cell fails at 2 bits; raising only that cell & 5 & 1 & 0 & \cref{tab:twelfth} \\
13 & free generation: does that cell move answer endings or answers? & 5 & 3 & 1 & \cref{tab:thirteenth} \\
14 & MATH free generation: that cell and answers; a BF16 floor & 2 & 3 & 1 & \cref{tab:fourteenth} \\
15 & free generation at 48\% of BF16 bytes & 3 & 1 & 0 & \cref{tab:fifteenth} \\
16 & free generation on 1119 unused GSM8K problems & 4 & 2 & 0 & \cref{tab:sixteenth} \\
17 & the four largest 200-problem losses on 1119 problems & 2 & 2 & 0 & \cref{tab:seventeenth} \\
18 & the 25\% allocation against K4V2 on fresh training problems & 2 & 0 & 0 & \cref{tab:eighteenth} \\
19 & raising only the 25\% allocation's 2-bit key cells & 1 & 1 & 0 & \cref{tab:nineteenth} \\
20 & measured block costs against diagonal Fisher on Huginn & 6 & 2 & 0 & \cref{tab:twentieth} \\
21 & the Ouro-1.4B readings on a second checkpoint, Ouro-2.6B & 8 & 3 & 1 & \cref{tab:twentyfirst} \\
 
\midrule
 & total & \LedgerPass & \LedgerFail & \LedgerFalsified & \\
\bottomrule
\end{tabular}
\end{table}

\section{Diagnostic formats and cache layout}
\label{sec:repr}

\textbf{Where to spend bits.} The tenth and eleventh registrations turned the sensitivity map into a diagnostic
allocator. Every
(plane, K$|$V, layer) cell gets its own width from \{2, 4, 8, 16\} bits (rotated direct INT with per-group scales, or
BF16). In the format byte model, each plane is read once and nominal bytes read per step equal bytes stored;
the dense FP32 simulation does not realize those traffic savings. A greedy rule frozen before the job spends a budget on the upgrades
with the largest cost reduction per stored bit. The tenth registration took each cell's cost from the diagonal Fisher on
64 calibration chains. The eleventh measured it on the same chains: each cell alone stored at each width, every other
cell exact (\ElevenForwards{} forwards per batch, \ElevenCalMin{} minutes on one GH200). Both were scored against uniform
formats with the same bytes, the eleventh on \ElevenMath{} MATH-500 chains that no registration had used
(\cref{tab:alloc}; criteria in \cref{tab:tenth,tab:eleventh}). With measured costs the allocation lowers aggregate
one-step KL against the uniform format at all three budgets, and at the bytes of INT4 only
\ElevenKlThirtyoneWorse\% of sequences get worse, with KL
within \ElevenLinMax$\times$ of the exact second order. The diagonal-Fisher allocation matches it where no cell drops to 2 bits;
at the bytes of \texttt{d4h} its aggregate (\ElevenFortyeightAggExact) misses the registered 0.30 because one sequence
holds \ElevenOutlierShare\% of its KL.

Where it uses 2-bit cells the diagonal-Fisher allocation fails in the tail: at INT4's bytes its median was
\TenThirtyoneMed{} of INT4's on held-out GSM8K, its aggregate \TenThirtyoneAgg{} there and \ElevenFisherThirtyoneAgg{} on
the new chains.

Post hoc interventions localize a major contributor to one cell. Keys of the first loop's first layer at 2 bits cost
\ElevenCellRatio$\times$ their diagonal-Fisher prediction, \ElevenCellTopShare\% of it in \ElevenCellTopN{} of the
\ElevenCalChains{} chains (the median sequence's ratio is \ElevenCellMedRatio); the diagonal Fisher ranks that
cell's upgrade \ElevenCellRankFisher{}th of 96 key cells, measurement ranks it first. The exact second order misses it
too: the tenth registration's worst sequences had KL/JVP of \TenTopLinLo--\TenTopLinHi, so the response is nonlinear,
not merely off-diagonal. Storing only its 2-bit key cells
(\ElevenKTwoCells{} of them) at 2 bits reproduces \ElevenKeyShareLo--\ElevenKeyShareHi\% of the KL of the tenth
registration's five worst sequences; its \ElevenVTwoCells{} 2-bit value cells reproduce at most \ElevenValShareHi\%. So
the registered rule that forbade 2-bit values failed (\ElevenNoVTwoAgg), and the tenth registration's post hoc
attribution of its tail to value cells was wrong.

The twelfth registration took the cell apart (\cref{tab:twelfth}). One KV head carries \TwelveHeadShare\% of its cost and the keys of generated tokens
\TwelveGenShare\%; the first token carries \TwelvePosZero, so the registered attention-sink account failed. On GSM8K
each of the \TwelveSwitchSeqs{} switching sequences puts over 99\% of its KL into one row: after an answer's two
newlines the reference predicts the next few-shot question's ``Q'' with probability
\TwelveRowProbLo--\TwelveRowProbHi, and with that head's keys at 2 bits it predicts a third newline. On MATH the
flipped rows sit inside the answers, and the cell alone at 2 bits accounts for \TwelveMathCellShare\% of the
allocation's KL. The GSM8K aggregates in \cref{tab:alloc} are therefore dominated by one few-shot-format row, whereas
the dominant MATH rows fall within answers. Neither aggregate directly measures answer correctness. Raising only that cell
to 4 bits (0.2\% more bytes) removes the tail: the diagonal-Fisher allocation
then reaches \TwelveGuardMath{} of INT4's aggregate on the MATH chains and \TwelveGuardTwentyfive{} of K4V2's at 25\%,
though not the measured-cost allocation's \ElevenKlThirtyoneAgg{} and \ElevenKlTwentyfiveAgg.

On Huginn (twentieth
registration, one width per four loops, supplementary \cref{tab:twentieth}) measured block costs again beat uniform
formats at every budget (aggregate \TwentyKlUniLo--\TwentyKlUniHi{} of theirs); against diagonal-Fisher block costs
the result is mixed (\TwentyKlFbAgg{} at INT4's core-cache bytes, median \TwentyKlFbMed; \TwentyKlFbLo--\TwentyKlFbHi{} at the
other budgets). No block cell stands out as Ouro's does (the largest measured/Fisher ratio is \TwentyOutRatio$\times$
the median), the two costs rank the blocks alike (post hoc, $\rho=\TwentyRankTwo$), and one sequence holds
\TwentyTopShareLo--\TwentyTopShareHi\% of the aggregates in which either allocation does badly.

\textbf{Second Ouro checkpoint.} Ouro-2.6B retains the Fisher-value/measured-key split
(\CheckpointFisherV\% / \CheckpointMeasuredK\%). At 31\% bytes, measured-cost allocation's MATH aggregate is
\CheckpointKlUniform{} of uniform INT4's but \CheckpointKlFisher{} of Fisher's, falsifying its registered advantage.
Fisher improves on INT4 (\CheckpointFisherUniform); post hoc, both allocator families keep the outlier cell
at 4 bits or more. Deadlines left \CheckpointCalN/64 measured-calibration and
\CheckpointMathN/176 MATH chains, meeting minimum gates. These teacher-forced 1.4B-generated chains do not measure
2.6B task accuracy (Appendix~\ref{app:checkpoint-v22}).

On Ouro-1.4B's calibration chains, the whole allocation's KL is \ElevenAddKl{} of the sum of isolated-cell costs, which also
approximate the diagonal-Fisher allocation's aggregate there (\ElevenPredFisherAgg{} predicted,
\ElevenInFisherAgg{} measured). At the bytes of uniform INT2 there was nothing to move, because 2 bits is the menu's
floor; the tenth registration missed this.

\textbf{In free generation.} Four registrations generated greedily with each format stored as it would be in a cache
(FP32 compute; \cref{tab:thirteenth,tab:fourteenth,tab:fifteenth,tab:sixteenth}); exact storage reproduced the
teacher-forced chains token for token and did not depend on the batch schedule. Prefill attends exact KV and produces
the first output token before the prompt planes are quantized; the format affects subsequent decode steps.
On GSM8K the diagonal-Fisher allocation moved \ThirteenMovedPct\% of answer endings by one extra newline before the next ``Q:'', short of the
registered 5\% (falsified); post hoc, all \ThirteenMoved{} are among the \ThirteenSwitchers{} sequences that switch
under teacher forcing, the guarded and measured-cost allocations and INT4 moved none, and a moved ending never changed
an answer. On MATH the guard did not improve agreement with exact storage: the guarded and unguarded allocations reproduced
\FourteenAgreeGuard\% and \FourteenAgreeFisher\% of the \FourteenParse{} answers exact storage gives (\FourteenGuardOnly{}
against \FourteenFisherOnly{} discordant problems; falsified). 

A possible explanation is that free generation leaves the teacher-forced path, so the rows that dominate one-step
KL need not recur. Answer-change counts alone do not establish this mechanism.

On the \SixteenN{} previously unused GSM8K test problems, answer-change counts generally grow as bytes fall;
across \SeventeenFormats{} formats, one-step aggregate KL ranks them with $\rho=\SeventeenRho$ (registered).
Under strict-match scoring, exact storage scores \SixteenAccExact\% and uniform INT4 \SixteenAccInt\%
($p=\SixteenAccP$), so the larger loss on the first 200 problems does not replicate
(supplementary \cref{tab:sixteenth,tab:seventeenth}). At 25\% of the bytes the measured-cost allocation does lose accuracy
(\SeventeenKlTwentyfiveAcc\%, $p=\SeventeenKlTwentyfiveP$), more than uniform K4V2 (\SeventeenKfourDelta{} points;
\SeventeenKuX:\SeventeenKuY{} discordant problems, seen post hoc and then registered and replicated on 1119 unused
training problems, \EighteenX:\EighteenY, $p=\EighteenP$, \cref{tab:eighteenth}).
On that replication, the paired accuracy difference is $\PairedKlKfourDelta$ points (post hoc 95\% interval
$[\PairedKlKfourLo,\PairedKlKfourHi]$;
supplementary \cref{tab:paired-accuracy}). The training problems were unused by our earlier experiments, but may
have appeared in model training. Raising the measured-cost allocation's \SeventeenKlTwentyfiveKtwo{} 2-bit key cells to 4 bits uses
\AllGuardBytes\% of BF16 bytes. In the original pooled strict test it recovers part of the gap
(\NineteenX:\NineteenY, $p=\NineteenP$), but still scores below K4V2
(\NineteenU:\NineteenV, $p=\NineteenQ$; supplementary \cref{tab:nineteenth}).
Answer-change counts measure distributional change; accuracy requires paired correctness comparisons.
A non-significant comparison with exact storage does not settle whether an allocation outperforms a uniform format.

\textbf{Answer extraction changes the comparison (post hoc).} Rescoring all saved outputs with the frozen task's
flexible-extract rule changes the allocator-minus-K4V2 gap from $\ScoreTestStrictDelta$ to
$\ScoreTestFlexibleDelta$ points on the test sample, and from $\ScoreTrainStrictDelta$ to
$\ScoreTrainFlexibleDelta$ on the training sample. The flexible-rule paired 95\% intervals are
$[\ScoreTestFlexibleLo,\ScoreTestFlexibleHi]$ and $[\ScoreTrainFlexibleLo,\ScoreTrainFlexibleHi]$;
neither establishes equivalence. Both formats still score below exact under that rule: the allocator's differences are
$\ScoreTestFlexibleAllocExactDelta$ and $\ScoreTrainFlexibleAllocExactDelta$ points, respectively
(supplementary \cref{tab:scoring-robustness}). The strict results remain primary, but do not support an extraction-independent
accuracy advantage for K4V2 or isolate a reasoning failure. The change in the paired gap itself is
$\AllTestMeasuredTwentyfiveInteractionDelta$ points on test and $\AllTrainMeasuredTwentyfiveInteractionDelta$ on training;
both descriptive 95\% intervals exclude zero (supplementary \cref{tab:scoring-all-comparisons}).

\textbf{Direct comparisons across budgets (post hoc).} At 31\% of BF16 bytes, measured-cost allocation minus uniform
INT4 is $\AllTestMeasuredThirtyoneStrictDelta$ points under strict extraction and
$\AllTestMeasuredThirtyoneFlexibleDelta$ under flexible extraction (95\% interval
$[\AllTestMeasuredThirtyoneFlexibleLo,\AllTestMeasuredThirtyoneFlexibleHi]$). At 48\%, the differences are
$\AllTestMeasuredFortyeightStrictDelta$ and $\AllTestMeasuredFortyeightFlexibleDelta$ (\cref{fig:overview}c).
Fisher allocation at 31\% also exceeds uniform INT4 by $\AllTestFisherThirtyoneStrictDelta$ strict and
$\AllTestFisherThirtyoneFlexibleDelta$ flexible points. These unadjusted, retrospective contrasts suggest modest
gains at 31\%; they do not establish a general ranking by KL or prove that the 31\% difference changes with the
extraction rule. The full audit includes every complete arm from the four relevant batches, including guarded
variants with higher byte budgets (supplementary \cref{tab:scoring-all-arms,tab:scoring-all-comparisons}).

\begin{table}[!t]
\centering\footnotesize\setlength{\tabcolsep}{2.6pt}
\caption{Bit allocation by diagonal-Fisher and by measured single-cell costs against uniform formats with the same bytes
(Ouro, GH200; fraction of BF16 bytes). Ratios of per-sequence KL: median, aggregate and sequences that get worse on
\ElevenMath{} unseen MATH-500 chains; median and aggregate on the \TenGsm{} held-out GSM8K chains of the tenth
registration. ``No 2-bit V'': the Fisher rule with a 4-bit floor for values. K$n$V$m$: uniform formats with $n$-bit keys and $m$-bit values; \texttt{d4h}: direct INT4 with Hadamard rotation.}
\label{tab:alloc}
\begin{tabular}{lllrrrrr}
\toprule
 & & & \multicolumn{3}{c}{MATH, unseen} & \multicolumn{2}{c}{GSM8K} \\
\cmidrule(lr){4-6}\cmidrule(lr){7-8}
Bytes & Uniform & Cost & med. & aggr. & worse & med. & aggr. \\
\midrule
25\% & K4V2 & diag.\ Fisher & 1.45 & 4.35 & 61\% & 0.64 & 1.46 \\
 &  & measured & 0.76 & 0.84 & 22\% & 0.58 & 0.64 \\
31\% & INT4 & diag.\ Fisher & 1.27 & 13.6 & 57\% & 0.60 & 5.18 \\
 &  & \quad no 2-bit V & 1.49 & 14.5 & 89\% & 0.93 & 4.63 \\
 &  & measured & 0.66 & 0.69 & 6\% & 0.49 & 0.53 \\
48\% & \texttt{d4h} & diag.\ Fisher & 0.17 & 0.30 & 0.6\% & 0.15 & 0.13 \\
 &  & measured & 0.14 & 0.14 & 0.0\% & 0.13 & 0.11 \\
 
\bottomrule
\end{tabular}
\end{table}

\textbf{Earlier end-task simulations (H100).} In free generation on all \GsmN{} GSM8K test problems \cite{gsm8k}, residual INT4
formats and direct INT4 with rotation stay within the interval width of BF16 (\cref{tab:gsm8k}); the INT2 formats lose
about two points. Residual coding is a capacity format, not a bandwidth one: residual INT4 stores \ResidStorePct\% of the
BF16 bytes but reads \ResidReadPct\% of them per step (\cref{tab:bytes}).

\begin{table}[!t]
\centering\small\setlength{\tabcolsep}{3pt}
\caption{GSM8K, \GsmN{} problems, greedy, FP32 compute with simulated storage (H100). Same text: generations identical
to BF16.}
\label{tab:gsm8k}
\begin{tabular}{lrrrr}
\toprule
Format & Bytes & Acc. (\%) & $\Delta$ vs BF16 (pp) & Same text \\
\midrule
\texttt{fp32} & 200\% & 61.64 & +0.08 [-0.26, +0.41] & 1267 \\
\texttt{bf16} & 100\% & 61.56 & -- & -- \\
\texttt{r4} & 49\% & 61.64 & +0.08 [-0.78, +0.93] & 1053 \\
\texttt{r4h} & 49\% & 62.09 & +0.53 [-0.42, +1.48] & 1086 \\
\texttt{d4h} & 48\% & 61.56 & +0.00 [-1.26, +1.26] & 809 \\
\texttt{f8r4h} & 37\% & 61.26 & -0.30 [-1.38, +0.77] & 930 \\
\texttt{r2} & 40\% & 60.50 & -1.06 [-2.58, +0.45] & 578 \\
\texttt{r2h} & 40\% & 59.44 & -2.12 [-3.59, -0.66] & 622 \\
\texttt{rqpaper} & 22\% & 59.36 & -2.20 [-3.76, -0.64] & 498 \\
 
\bottomrule
\end{tabular}
\end{table}

An earlier H100 kernel study is summarized separately in supplementary Appendix~\ref{app:historical-kernels};
its FP8 throughput measurements do not evaluate the mixed-bit allocator studied here.

\textbf{Exact release of exited planes.} With early exit every plane deeper than a token's exit holds a copy of its
exit plane. Once every position of a page is final, the planes deeper than the page's deepest exit are byte-identical to
it: their table entries can point at its block and their own blocks return to the pool, with no kernel change. Tokens,
exit depths and every per-depth KV view are bitwise equal to the copy layout at page sizes 1 to 16 on CPU and in
\TestsDedupCuda{} GPU cases on GH200 (CUDA graphs, multi-stream finalize). On GSM8K exit traces with 16-token pages the
cache keeps \DedupSeqHalfSixteen{} of the copy layout's blocks over whole sequences and \DedupDecHalfSixteen{} over decode
pages: prompts dominate and mixed exit depths within a page block release. The early-exit policy itself, not the release,
costs accuracy (\RaggedDeltaHalf{} pp at threshold 0.5, H100).

\section{Negative results and limitations}
\label{sec:negative}

The diagonal Fisher is not calibrated for aggregate KL, and group-size ratios fail to predict rankings. Huginn's
fixed-point picture does not imply coherent rounding errors or increasing $\kappa_2$. Residual KV and FP8 without
split-KV decoding do not save read bandwidth. Supplementary Appendix~\ref{app:negative} retains the other negative results.

The guard did not establish a MATH task benefit; a full-sample symbolic audit likewise resolves no allocator advantage
(Appendix~\ref{app:math-scoring}). The 25\% strict-score loss is extraction-sensitive; the 31\% gains are
retrospective, unadjusted and independently unconfirmed. Neither extraction rule adjudicates reasoning, and the
training-split replication is not known to be contamination-free.

Evidence covers two primary checkpoints and a partial second-Ouro diagnostic, fixed mathematics prompts and
greedy decoding with length caps. Prompt
robustness, sampling and matched non-looped controls are untested. Huginn uses a deterministic token-dependent initial
state; its four-loop allocation blocks differ from Ouro's per-loop cells, confounding model-family comparisons.

Isolated-cell costs need not add jointly, and greedy allocation has no global-optimality guarantee. Intervals omit
recalibration variability. Ouro's calibration required \ElevenForwards{} forwards per batch and \ElevenCalMin{} minutes
on one GH200; cost amortization and stability under prompt-distribution shifts are unmeasured.

Nominal budgets include scales and offsets but omit packing, alignment, format indices and workspaces. The FP32
simulation uses exact-KV prefill. A separate L40S physical-cache pilot uses less retained cache but fails its
full-model reader gate; its batch-one timings cannot establish a quality-qualified speedup
(Appendix~\ref{app:l40s-packed}). Capacity and concurrency remain untested. FlashLoop's different packer fails a
strict CPU bitwise check in 16 of 20 synthetic cases (Appendix~\ref{app:l40s-qualification}); this is not a test of
our allocator.

\section{Related work}

Looped and recurrent-depth models include Universal Transformers \cite{universal}, Huginn \cite{huginn}, Ouro
\cite{ouro} and Hyperloop Transformers \cite{hyperloop}. Systems work batches loop depths continuously \cite{cdb},
aliases KV across exit depths \cite{raggedkv}, drafts from shallow loops \cite{loopspec,wavefront,das}, contrasts early
and late readouts \cite{loopcd}, and compresses cross-loop KV \cite{lla,flashloop}. Quantization of looped models has been
studied for weights and activations \cite{loopq,feedbackexposure} and for residual KV \cite{residualquant}. General KV
quantizers \cite{kivi,kvquant} and rotations \cite{quarot} motivate our formats; second-order sensitivity is standard in
weight quantization \cite{hawq,adaround,gptq}, and paged KV management follows \cite{pagedattention}. Our theory tests
borrow tail-index estimation \cite{hill}, the error analysis of stochastic rounding \cite{stochround}, and the
fixed-point view of iterated depth from equilibrium models \cite{deq}.

\section{Conclusion}

Across the studied checkpoints, values dominate Fisher mass while keys dominate measured KL. Second-order prediction
tracks typical sequences, but aggregates and rare rows require direct measurement. Shared-weight amplification also
depends on error direction; at $R=32$ the local ratio ranges from \ExactKappaInteight{} to \ExactKappaIntfour.

Measured costs expose a nonlinear key failure on Ouro-1.4B, yet lose to Fisher on the second checkpoint. Task contrasts
depend on budget and scoring. KL remains a diagnostic; task and serving benefits require separate validation.

\clearpage
\bibliography{refs}
\bibliographystyle{mlsys2025}

\clearpage
\onecolumn
\appendix
\setcounter{table}{0}
\setcounter{figure}{0}
\renewcommand{\thetable}{S\arabic{table}}
\renewcommand{\theHtable}{supp.\arabic{table}}
\renewcommand{\thefigure}{S\arabic{figure}}
\renewcommand{\theHfigure}{supp.\arabic{figure}}
\section*{Supplementary Material}
\section{Measurement design}\label{app:overview}

\begin{figure}[H]
\centering
\resizebox{0.92\textwidth}{!}{
\begin{tikzpicture}[font=\scriptsize, >=Stealth,
  cell/.style={draw=gcol!70, minimum width=3.0mm, minimum height=2.6mm, inner sep=0pt},
  box/.style={draw, rounded corners=1.5pt, inner sep=2.5pt, align=center},
  lab/.style={font=\footnotesize\bfseries}]
\begin{scope}
\node[lab] at (-0.45,3.25) {(a)};
\foreach \l/\y/\s in {1/0.55/12, 2/1.10/12, 3/1.65/12, R/2.60/35} {
  \foreach \t in {0,...,5} {
    \node[cell, fill=kcol!\s] (k\l\t) at (0.45+0.34*\t, \y+0.27) {};
    \node[cell, fill=vcol!\s] (v\l\t) at (0.45+0.34*\t, \y) {};
  }
  \node[anchor=east] at (0.25, \y+0.13) {loop $\l$};
}
\foreach \d in {2.16,2.26,2.36} \fill[black] (1.30,\d) circle (0.55pt);  
\node[anchor=north] at (1.30, 0.32) {prefix positions $\rightarrow$};
\node[draw=red!70!black, very thick, fit=(k20)(v25), inner sep=0.4pt] {};
\node[anchor=west, text width=1.8cm, align=left] at (2.50, 1.45) {stored $X{+}\Delta$:\\ keys \textcolor{kcol}{$\blacksquare$},\\ values \textcolor{vcol}{$\blacksquare$},\\ one plane per loop};
\end{scope}
\begin{scope}[xshift=5.0cm]
\node[lab] at (-0.2,3.25) {(b)};
\node[box] (row) at (0.85,2.55) {decode row $t$\\ reads $X{+}\Delta$};
\node[box, fill=gcol!12] (kl) at (0.85,0.95) {measured\\ $\KL(p_t\,\|\,q_t)$};
\node[box, fill=kcol!10] (jvp) at (3.45,2.20) {exact 2nd order\\ $\tfrac12\Var_{p_t}(J_t\Delta)$};
\node[box, fill=vcol!10] (fi) at (3.45,0.75) {diagonal Fisher\\ $\tfrac12\sum_i F_{ii}\Delta_i^2$};
\draw[->] (row) -- (kl);
\draw[->] (row.east) -- (jvp.west |- row.east);
\draw[->] (row.north) |- (4.95,2.98) |- (fi.east);  
\draw[<->, dashed, gcol] (kl.east) -- node[above, sloped, pos=0.62, text=black] {\tiny nonlinearity} (jvp.south west);
\draw[<->, dashed, gcol] (jvp.south) -- node[right, text=black] {\tiny off-diagonal} (fi.north);
\node[anchor=north, text width=4.5cm, align=center] at (2.15,0.30) {same rows, same stored positions, same diagonal};
\end{scope}
\begin{scope}[xshift=10.65cm]
\node[lab] at (-0.2,3.25) {(c)};
\node[box, fill=wcol!15] (w) at (0.45,1.60) {$W{+}\Delta W$};
\foreach \l/\y in {1/0.55, 2/1.10, 3/1.65, R/2.60} {
  \node[box, minimum width=1.0cm] (L\l) at (2.05, \y) {loop $\l$};
  \draw[->, wcol!70!black] (w.east) -- (L\l.west);
}
\node at (2.05, 2.15) {$\vdots$};
\node[anchor=west, text width=2.5cm, align=left] at (2.75,2.15) {$\kappa_2=\dfrac{\mathrm{JVP}_\text{all}}{\sum_\ell \mathrm{JVP}_\ell}$};
\node[anchor=west, text width=2.5cm, align=left] at (2.75,0.95) {blocks of loops and signs $(-1)^\ell$ give a closed-form split of $\mathrm{JVP}_\text{all}-\sum_\ell\mathrm{JVP}_\ell$};
\end{scope}
\end{tikzpicture}}
\caption{Measurement design. A looped model stores one key and one value plane per loop for every prefix position. We
compare measured next-token KL, its local second-order prediction through a JVP, and diagonal Fisher on identical
rows. For shared weights, loop-specific tangents separate within-loop error from cross-loop coherence.}
\label{fig:design}
\end{figure}
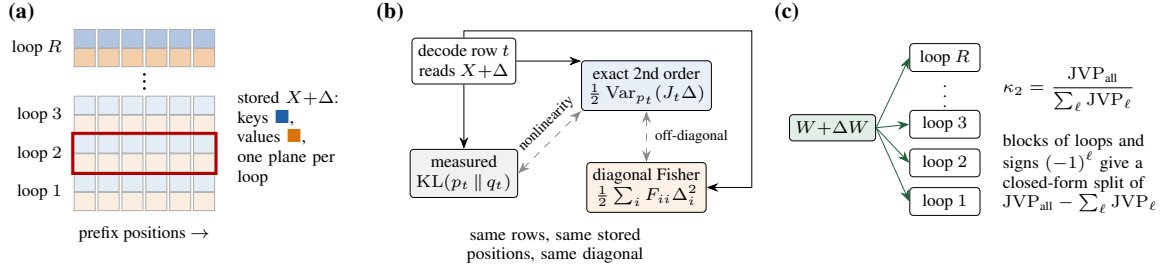

\section{Batches and checks}\label{app:batches}

Decision rules were fixed in analysis code before each batch. The first batch's rules are recoverable from its frozen
code; later batches also retain separate registration documents. Package manifests bind the code, inputs and controls,
and run logs record preflight checksum checks and deterministic CPU self-tests. The released records support content
checks but do not provide an independent timestamp for registration. Batches on the shared cluster also ran
a separate, unscored Nsight Compute pass. Table~\ref{tab:ledger} of the main paper lists all registrations; the
registration tables are Table~\ref{tab:theory} and Tables~\ref{tab:prereg} and~\ref{tab:third}--\ref{tab:ninth}.

\section{First batch: bytes, criteria and the dominant sequence}\label{app:first}

\paragraph{Bytes stored and bytes read.} Formats differ in what they store and in what one decode step reads.
Loop $\ell$ reads plane $\ell$. A residual format that codes planes $1..R{-}1$ against the final plane must read the
reference and the residual for each of those loops; the two reads are a full core apart and do not share cache
lines. Table~\ref{tab:bytes} shows the consequence: the residual INT4 format of the earlier study stores
\ResidStorePct\% of the BF16 bytes but reads \ResidReadPct\% of them per step. Residual coding is a capacity and
transfer format, not a bandwidth format.

\begin{table}[H]
\centering\small
\caption{Stored and per-step read bytes of KV formats, summed over the $R$ planes, from the format definitions
(\S\ref{sec:method}). \texttt{r*}: residual against the final plane; \texttt{d*}: direct; \texttt{h}: Hadamard
rotation; \texttt{f8}: FP8.}
\label{tab:bytes}
\begin{tabular}{lrr}
\toprule
Format & Stored / BF16 & Read per step / BF16 \\
\midrule
\texttt{bf16} & 100\% & 100\% \\
\texttt{f8all} & 51\% & 51\% \\
\texttt{d4h} & 48\% & 48\% \\
\texttt{d4h-all} & 31\% & 31\% \\
\texttt{r4h} & 49\% & 124\% \\
\texttt{f8r4h} & 37\% & 75\% \\
\texttt{r2h} & 40\% & 115\% \\
\texttt{rqpaper} & 22\% & 46\% \\
 
\bottomrule
\end{tabular}
\end{table}

\paragraph{Pre-registered outcome.} Table~\ref{tab:prereg} lists the criteria. The aggregate measured/JVP ratio falls
in $[0.8,1.25]$ for only \POnePass{} of \EOneArms{} formats, the log--log slope of measured KL on the diagonal
prediction is \POneDSlope{}, and the group-size ratios are not predicted. Additivity, the rounding-temperature test
and the floors pass. For the formats in the linear band, KL scales as $\epsilon^2$ under $\Delta\to\epsilon\Delta$:
$\KL(\epsilon)/(\epsilon^2\KL(1))$ at $\epsilon=\tfrac14,\tfrac12,2$ is \ScaleDfourhallQuarter, \ScaleDfourhallHalf,
\ScaleDfourhallTwo{} for \texttt{d4h-all} and \ScaleFeightallQuarter, \ScaleFeightallHalf, \ScaleFeightallTwo{} for
FP8; for residual INT2 (\texttt{r2h}) it is \ScaleRtwohQuarter, \ScaleRtwohHalf, \ScaleRtwohTwo.

\begin{table}[H]
\centering\small
\setlength{\tabcolsep}{3pt}
\caption{Pre-registered criteria and outcomes (GH200). P2 lists plane/K--V sums for \texttt{d4h-all} and FP8.}
\label{tab:prereg}
\begin{tabular}{lp{3.4cm}p{1.9cm}l}
\toprule
 & Criterion & Result & Verdict \\
\midrule
P1 & KL/JVP in [0.8, 1.25] per arm (aggregate) & 7/21 arms & fail \\
P1 & Scale sweep $\mathrm{KL}(\epsilon)/\epsilon^2$ flat for linear arms & 1.02--1.03 & pass \\
P1d & log-log slope of KL on diagonal prediction in [0.8, 1.2] & slope 1.24 & fail \\
P2 & plane and K/V subsets sum to the whole within 20\% & 0.94/0.99, 0.99/1.00 & pass \\
P3 & group-size ratios predicted within $1.5\times$ & 1/4 ratios & fail \\
P4 & SR KL / $\tfrac12\sum F\,E[\Delta^2]$ in [0.5, 2]; SR vs RN direction & 0.85 & pass \\
P5 & numerical floors $\ge 10\times$ below smallest arm & $10^{6}\times$ & pass \\
E2 & amplification: R=4 leaves linear response at a finer format than R=1 & same (int4) & not supported \\
 
\bottomrule
\end{tabular}
\end{table}

\paragraph{Why the aggregate fails (post hoc, CPU).} We recomputed the sequence with the largest discrepancy for
\texttt{d4h-all} on CPU with the same weights; its measured KL matches the GH200 value to a ratio of
\DiagCpuOverGh. There the JVP predicts \DiagJvpOverKl{} times the measured KL. Figure~\ref{fig:sweep} shows that
the JVP is the correct small-perturbation limit: $\KL(\epsilon)/\epsilon^2$ rises from \DiagEpsQuarter{} to
\DiagEpsSixteenth{} to \DiagEpsSixtyFourth{} times $\KL(1)$ at $\epsilon=\tfrac14,\tfrac1{16},\tfrac1{64}$, and central
differences give \DiagFdSmall{} at step $\tfrac1{256}$ but \DiagFdLarge{} at step $\tfrac1{16}$. One of its
\DiagRows{} rows carries \DiagRowShare\% of the second-order term. In that row the model is confident (top
probability \DiagTopP) and the logit tangent reaches about \DiagMaxTangent{} for tail tokens. Linearizing the network
but keeping the output softmax exact makes the prediction worse, by \DiagTangentLo--\DiagTangentHi{} times on that
sequence for the four rotated or FP8 formats, so the saturation is inside the network: the root-mean-square logit
change per unit perturbation falls from \DiagTangentRmsSmall{} at step $\tfrac1{256}$ to \DiagTangentRmsLarge{} at
step $\tfrac1{16}$. Section~\ref{sec:theory} (T7) localizes it on five such sequences and confirms it, re-registered, on
\TSevenNewSeqs{} new ones: in all \TSevenScored{} scored cases loops 1 and 2 hold at least \TSevenLoopsMin\% of the excess. The single dominant row and the agreement of differences with the JVP were
shown on one sequence, and the tangent-logit comparison on \DiagTangentSeqs.

\paragraph{Rounding and group size.} Stochastic rounding on the INT4 grid measures \SrOverExpected{} of the
prediction built from the expected squared error, and is \SrOverRnMeas{} times worse than round-to-nearest against a
predicted \SrOverRnPred{} (P4). Group-size effects (P3) are not predicted: the comparisons pass through the same
outlier sequences, which change the ranking.

\section{Theory tests in full}\label{app:theory}

\paragraph{Design.} After the first batch we wrote down seven predictions. Each names a source theory, a numerical
prediction and a falsification condition, and all were frozen before the second set of GPU jobs. They are tested on
data not used to form them: \ConfSeqs{} held-out GSM8K chains for Ouro (\ConfRows{} scored rows, \ConfArms{}
formats including three new group-size arms), \MathSeqs{} MATH-500 chains (\MathRows{} rows) and Huginn-0125.
Table~\ref{tab:theory} lists predictions, measurements and verdicts. Its last column is post hoc and never changes a
verdict. A third pre-registration, at the end of this section, re-tests the open Huginn points and T7 on new data
(Table~\ref{tab:third}).

\paragraph{Huginn.} Huginn-0125 runs a prelude of \HugPrelude{} layers once, iterates a core block of \HugCore{}
layers $R=\HugR$ times from a random initial state, and finishes with a coda of \HugCoda{} layers; attention has
\HugHeads{} heads of dimension \HugHeadDim{} \cite{huginn}. We quantize the $R\times\HugCore$ core planes and keep
prelude and coda KV exact. The random initial state is replaced by a deterministic draw seeded by the token ids, so
every pass over the same tokens sees the same state. E1 uses Huginn's own greedy chains from the third batch:
\OwnGen{} GSM8K prompts generated with a causal state (position $t$ seeded by tokens $0..t$), capped at \OwnCap{}
tokens (\OwnStop{} stopped, \OwnLength{} hit the cap, median \OwnGenMedian{} tokens). Teacher forcing reproduced
every generated token (\OwnMismatch{} mismatches). E1 runs on \OwnSeqs{} of them (\OwnRows{} rows,
\OwnArms{} formats). The second batch first tested T3 and T4 on \HugEOneSeqs{} Ouro chains decoded to text and
re-encoded for Huginn (\HugEOneRows{} rows), with a state seeded by the whole padded row; we report those verdicts
as registered. E2 uses \HugETwoSeqs{} re-encoded chains (\HugETwoRows{} rows) in both batches, so that $\kappa$ is
measured on identical inputs at $R\in\{4,8,16,32\}$.

\paragraph{T1: the aggregate is not self-averaging.} Let the per-sequence excess $r_s=Q_s/\KL_s$ of second order
over measured KL have a power-law tail, $P(r>x)\sim x^{-\alpha}$. For $\alpha<1$ the mean does not exist and
$\sum_s Q_s$ is dominated by its largest term; for $1<\alpha<2$ the mean exists but the variance does not, and the
aggregate converges slower than $N^{-1/2}$. On the held-out chains the light-tailed formats aggregate
inside $[0.8,1.25]$ (\texttt{d4-all} \ConfAggrDfourall, FP8 \ConfAggrFeightall, BF16 \ConfAggrBfsixteen) and the
heavy-tailed ones outside it (\texttt{r4h-g128} \ConfAggrRfourhgonetwoeight, \texttt{r2h} \ConfAggrRtwoh), while
all medians fall inside. Two numbers miss: \texttt{r4h-g128} deviates by \ConfDevRfourhgonetwoeight{} instead of at
least \PrDev, and its Hill index \cite{hill} is \ConfHillRfourhgonetwoeight{} instead of below 1. Its largest sequence holds
\ConfTopShareRfourhgonetwoeight\% of the second-order sum, against \ExplTopShareRfourhgonetwoeight\% in the first
batch. The Hill indices keep their order (\texttt{r2h} \ConfHillRtwoh, \texttt{r4h-g128}
\ConfHillRfourhgonetwoeight, \texttt{d4h-all} \ConfHillDfourhall, FP8 \ConfHillFeightall, \texttt{d4-all}
\ConfHillDfourall). The heavy tail is real, but on the larger sample it sits at $1<\alpha<2$.

\paragraph{T2: group size as a screening length; rotation as Gaussianization.} Per-group scaling lets every group of
$g$ elements set its own range, much as a screening length sets the range of an interaction. On a uniform grid the
error is $\Delta^2\propto\mathrm{range}_g^2$. The range of $g$ samples grows like $\sqrt{2\ln g}$ for a Gaussian tail
and like $g^{1/\alpha}$ for a power-law tail, so the two tails predict different group-size laws. A Hadamard rotation
mixes coordinates and makes each close to Gaussian \cite{quarot}. The predictions came from exact CPU quantization of
planes from \ScrChains{} further chains. Unrotated keys have kurtosis \KurtPlainK{} (values \KurtPlainV), rotated
ones \KurtRotK{} (\KurtRotV). The unrotated key error grows by \ScrPlainKRatio{} from $g=16$ to $g=128$, which
corresponds to $\alpha\approx\TailAlphaK$. Table~\ref{tab:groups} compares median measured KL ratios with those
error ratios. All five fall in their bands on both distributions, and the unrotated ratio stays above the Gaussian
law's \GaussLaw. At these finite group sizes the rotated ratios, like the exact error ratios they track, lie below
that asymptotic law. What T2 confirms is that measured KL follows exact error ratios in both branches and that the
unrotated branch is not Gaussian; that rotation makes coordinates Gaussian rests on the kurtosis. Unlike P3 of the first batch, which compared aggregate ratios through the diagonal Fisher, T2 uses
medians, which the outlier sequences do not move.

\paragraph{T3: quenched and annealed rounding.} For a fixed input, round-to-nearest (RN) gives a fixed, quenched
error; stochastic rounding (SR) gives an annealed one \cite{stochround}. With a uniform fractional part $f$, RN has
mean square error $E[\min(f,1-f)^2]=\tfrac1{12}$ of a step squared and SR has $E[f(1-f)]=\tfrac16$, a factor of 2. SR
measures \SrRnConf{} times RN on held-out GSM8K, \SrRnMath{} on MATH-500 and \SrRnHug{} on Huginn. The SR
measurement divided by the prediction from the expected squared error is \SrExpConf, \SrExpMath{} and \SrExpHug.
On Huginn's own chains the two ratios replicate at \OwnSrRn{} and \OwnSrExp.

\paragraph{T6: distribution independence.} On MATH-500, values carry \MathVShare\% of the diagonal Fisher mass and
the final plane \MathFinalShare\%, against \ConfVShare\% and \ConfFinalShare\% on held-out GSM8K. The set of formats
whose median lies in the band differs from the first batch's by \MathSetDiff{} format.

\paragraph{T7: where the saturation sits.} On the \LocSeqs{} most discrepant sequences and two formats, we restricted
$\Delta$ to one layer (all loops) or to one (loop, layer) cell, and recorded the JVP and the KL at $\epsilon=1$ and
$\tfrac1{16}$. Layers and loops are numbered from 1. The pre-registered metric, the top three layers' share of the
whole-sequence JVP, is ill-posed: restricted JVPs omit cross-layer terms and do not add, and the share ranges from
\LocShareThreeLo{} to \LocShareThreeHi. It fails, and the falsification condition does not occur (at most
\LocMaxSpread{} layers carry 5\% of the excess; the threshold was \LocFalsify{}). Post hoc, we use each layer's share of the
excess $\mathrm{JVP}-\KL$ instead. In \LocTopHalf{} of \LocCases{} cases one layer holds at least half of it. For
\texttt{r4h-g128} on the sequence where the JVP exceeds the KL \LocARatio-fold, layer \LocALayer{} holds
\LocAShare\%. Layer \LocALayer{} is the top layer in \LocLayerSixCount{} cases, and the largest cell lies in loop 1
or 2 in all \LocEarlyCells. The third batch re-registered T7 with this excess share and tested it on
\TSevenNewSeqs{} new sequences selected by a fixed rule from the held-out chains, restricting $\Delta$ to one layer or to
one loop. Of \TSevenCases{} cases, \TSevenScored{} have JVP/KL of at least 1.5 and are scored. The median share of the
top three layers is \TSevenTop{} (range \TSevenTopLo--\TSevenTopHi), loops 1 and 2 hold at least \TSevenLoopsMin\% of
the excess in every scored case, and at most \TSevenMaxLayers{} layers carry 5\% of it. All three criteria pass. The
top layer varies: layer \LocALayer{} leads in only \TSevenLayerSix{} of \TSevenScored{} cases. The saturation sits in a
few layers of the first two loops, but not in one fixed layer.

\section{Huginn: separation, the switching sequence and spectra}\label{app:huginn}

\paragraph{T4: on Huginn's fixed point, coherent rounding errors add only weakly.} Huginn is trained so that its
latent state approaches a fixed point over the loops, as in equilibrium models \cite{deq}. Adjacent late planes then
nearly coincide, so RN should give them nearly equal errors. The cross-plane Fisher terms would then add, and the
diagonal prediction would underestimate RN but not SR. The planes do converge: on Huginn's own chains the relative
change of keys between loops falls monotonically from \OwnConvKFirst{} to \OwnConvKLast, and is \OwnConvKSixteen{}
between loops 16 and 17. The predicted underestimate is small. On the re-encoded chains of the second batch the
median measured/diagonal ratio is \HugDiagDfourall{} for RN and \HugDiagDfourallsr{} for SR, a ratio of \TfourA, below
the falsification threshold \PrTfourAFals. On the own chains it is \OwnDiagDfourall{} against \OwnDiagDfourallsr,
a ratio of \OwnTfourA.

The third batch measured the two candidate reasons directly, on \SplitSeqs{} own chains with one sequence at a
time. In the Fisher metric $\langle a,b\rangle=\sum_t\mathrm{Cov}_{p_t}(a,b)$, with $J_\ell$ the logit tangent of plane
$\ell$ and $\Delta_\ell$ its storage error, it compares three normalized quantities for adjacent planes. The first is the
correlation of the two errors seen through one Jacobian, $\rho^J=\langle J_{\ell+1}\Delta_\ell,J_{\ell+1}\Delta_{\ell+1}\rangle$.
The second is the alignment of the two Jacobians on one error, $A=\langle J_\ell\Delta_\ell,J_{\ell+1}\Delta_\ell\rangle$.
The third is the realized coherence $X=\langle J_\ell\Delta_\ell,J_{\ell+1}\Delta_{\ell+1}\rangle$. Over the last eight
planes the RN errors are correlated ($\rho^J=\SplitRhoJ$, element cosine \SplitRho), while SR, the negative control, is
not (\SplitRhoJSr). The Jacobians are nearly orthogonal ($A=\SplitAlign$), so the realized coherence is only
$X=\SplitX$. Cross-plane terms make up \SplitCross\% of the RN second-order term and \SplitCrossSr\% of the SR term;
adjacent pairs account for \SplitAdjShare\% of the RN cross term. This is the pre-registered class (Table~\ref{tab:third}).
Post hoc, the Jacobians alternate with parity in the earlier loops. For loops 9--24 the alignment is \SplitMidAOne{}
at distance 1 and \SplitMidATwo{} at distance 2 (per pair down to \SplitAOneMin{} and up to \SplitATwoMax), and the
realized coherence there is negative (\SplitMidX). One reading is a linearized loop map with an eigenvalue near $-1$
in the middle loops, which fades in the late ones. The fourth batch measured dominant Ritz values of the loop map
(Table~\ref{tab:fourth}, C): for three ordinary sequences they are mostly real and negative in the middle loops
(\SpecOtherNegMid{} cells, \SpecOtherNegLo{} to \SpecOtherNegHi), but so are the late ones, so the mode exists and is not
confined to the middle loops. With \ConvKrylov{} steps every cell converges (\ConvCells{} of \ConvOf) and the picture is
unchanged: negative dominant values in \ConvMidNeg{} middle cells and \ConvLateNeg{} late ones (Table~\ref{tab:fifth}, C).

The final plane holds \OwnFinalShare\% of the Fisher mass on the own chains (\HugFinalShare\% on the re-encoded
ones), more than the predicted ceiling of \PrTfourC\% but far below Ouro's \OuroFinalShareExpl\%. The last five
planes hold \OwnLastFive\%, and planes 1--20 together \OwnFirstTwenty\%. Values carry \OwnVShare\% of the mass, against
\FisherV\% on Ouro.

Outside the predictions, typical linear response is good on Huginn. The median measured/JVP ratio is
\OwnMedLo--\OwnMedHi{} for \OwnInBand{} of \OwnArms{} formats and \OwnMedDtwoh{} for \texttt{d2h}. The aggregate behaves
differently on the two chain sets. On the re-encoded chains it falls below the JVP (\texttt{d4-all}
\BorrowAggrDfourall, \texttt{r2h} \BorrowAggrRtwoh), as on Ouro. On the own chains it rises above it (\OwnAggrDfourall{}
and \OwnAggrRtwoh), and one sequence of \TopTokens{} tokens accounts for that. There, KL is \TopRatioDfourall{} and
\TopRatioRtwoh{} times the JVP and carries \TopShareDfourall\% and \TopShareRtwoh\% of the two formats' KL. Its scale
sweep $\KL(\epsilon)/(\epsilon^2\mathrm{JVP})$ reads \TopScaleQuarter, \TopScaleHalf, \TopRatioDfourall{} and \TopScaleTwo{} at
$\epsilon=\tfrac14,\tfrac12,1,2$: doubling the perturbation from $\tfrac12$ to 1 multiplies the KL by \TopJump. A response
switches on near the quantization scale, the opposite of the saturation seen on Ouro. The fourth batch located it
(Table~\ref{tab:fourth}, A). One of the sequence's \SwitchRows{} rows carries \SwitchRowShare\% of the excess (KL
\SwitchRowKl{} of \SwitchWholeKl; the other rows sum to \SwitchRestKl, against a JVP of \SwitchWholeJvp{} for the whole
sequence), and its argmax flips between $\epsilon=\tfrac12$ and $\tfrac34$. The row is not a near tie: its reference
margin is \SwitchMargin{} logits, the \SwitchMarginPct{}th percentile of the sequence (median \SwitchMarginMedian). Restricting
the error to one layer puts all of it in the last core layer (KL \SwitchLayerKl{} against a JVP of \SwitchLayerJvp; the
other three layers are linear within \SwitchOtherLayersMax); restricting it to one loop puts it in loop \SwitchLoop{} (KL
\SwitchLoopKl, JVP \SwitchLoopJvp), while the single-loop errors of loops 25--32, where the Fisher mass sits, stay linear
(KL/JVP \SwitchLateLinLo--\SwitchLateLinHi). Residual INT2 flips the same row at $\epsilon=\SwitchRtwohEps$ with an error
\SwitchRtwohRatio{} times smaller. The sequence's loop map is the only one of \SpecSeqs{} measured whose dominant Ritz
value in loops 10--22 is real, positive and above one (\SpecSwitchLo--\SpecSwitchHi, \SpecSwitchConv{} of
\SpecSwitchMidCells{} cells converged; \SpecSwitchLate{} at loops 30--32): an error injected in loop \SwitchLoop{} grows
about \SpecAmplify-fold before the map contracts again, which is how a middle-loop error in one layer flips a row with a
margin of \SwitchMargin{} logits. On all \StatSeqs{} own chains (Table~\ref{tab:fifth}, A; \StatKrylov{} Arnoldi steps,
\StatConvCells{} of \StatCellsTotal{} middle-loop cells converged), \StatExpanding{} chains (\StatExpandingPct\%) have such an
expanding middle-loop mode, and the switching sequence ranks \StatRank{} by its largest real part (\StatSwitchRe; the
largest is \StatTopRe). Expansion is necessary for the switch but does not produce one: the expanding chains' median
KL/JVP is \StatMedExp{} against \StatMedNon{} for the others, and the Spearman correlation between the largest real part
and $\log$ KL/JVP is \StatRho{} (\StatRhoAbs{} for the modulus). Without the three largest
excesses the aggregates are \TrimAggrDfourall{} and \TrimAggrRtwoh, and the median sweep over all sequences stays within
\OtherScaleMedLo--\OtherScaleMedHi. Residual coding gains much more than on Ouro: per row, residual INT2 gives
\OwnKlRowRtwoh{} and direct INT2 \OwnKlRowDtwoh, a factor of \OwnResidRatio{} against \OuroResidRatio{} on Ouro, and
residual INT4 (\OwnKlRowRfourh) comes close to BF16 (\OwnKlRowBfsixteen). Converged planes leave small residuals.

\section{Weight sharing on Huginn}\label{app:kappa}

\paragraph{T5: $\kappa(R)$ and the correlation length of loop gradients.} Write $C_{\ell\ell'}=\sum_i
E[g_{\ell,i}g_{\ell',i}]\Delta W_i^2$, so that $\kappa=\mathbf 1^{\!\top}C\mathbf 1/\mathrm{tr}\,C$. If the
correlation $\rho$ of loop gradients decays with loop distance over a correlation length $\xi$, then
$\kappa(R)\approx1+2\sum_{k\ge1}(1-k/R)\,\rho(k)$ grows with $R$ and saturates once $R\gg\xi$. If converged late
loops had nearly identical gradients, $\rho\to1$ and $\kappa$ would grow linearly in $R$. With \DrawsOld{} Fisher draws
per batch, the second batch found $\kappa$ growing \KappaGrowth-fold from $R=4$ to $R=32$, formats agreeing within
\KappaSpreadMax\%, and the Fisher-free ratio agreeing with $\kappa$ where it was measured. The predicted structure did
not hold: the correlation rose in \MonoViol{} of \MonoSteps{} steps along the diagonal, and pairs of late loops
(25--32) correlated at \LatePairs, against \EarlyPairs{} for early ones (1--8).

The third batch repeated $R=4$ and $R=32$ on the same batches with \DrawsNew{} draws. The KL of every batch and format
matched the second batch bitwise (\KappaGateChecked{} values), so only the sampled labels of the Fisher differ.
Table~\ref{tab:kappa} merges the two batches. With \DrawsNew{} draws, $\kappa(32)=\KappaThirtyTwo\pm\KappaSE$ (jackknife over
draws), not \KappaThirtyTwoOld. Four disjoint subsets of \DrawsOld{} draws give \KappaSubAll: \DrawsOld-draw estimates
scatter widely, and the second batch's value lies above all four, so sampling is the likely cause but not a proven one.
The growth remains, at \KappaGrowthNew-fold, with formats within \KappaSpreadNew\%.

The size of $\kappa(32)$ also depends on the estimator. $\kappa$ is a ratio of diagonal Fisher terms, and at $R=32$ the
diagonal shared prediction is only \FisherJvpAggr{} of the exact second order (per batch \FisherJvpLo--\FisherJvpHi;
\FisherJvpFour{} at $R=4$). Replacing the numerator by the exact second order,
$\mathrm{JVP}_\text{all}/\tfrac12\mathrm{tr}\,C$, gives \KappaMixedFour, \KappaMixedEight, \KappaMixedSixteen{} and
\KappaMixedThirtyTwo{} at $R=4,8,16,32$. Both readings grow; their values at $R=32$ differ by almost a factor of two.
INT8 is still linear at $R=32$ (measured/JVP \LinThirtyTwoInteight) and gives $\kappa=\KappaThirtyTwoInteight$, so the
growth is a property of the gradients, not of saturation. INT4 and FP8 leave the linear regime there
(\LinThirtyTwoIntfour{} and \LinThirtyTwoFpeight): the INT4 scale sweep $\KL(\epsilon)/(\epsilon^2\,\mathrm{JVP})$ reads
\HugScaleThirtyTwoQuarter, \HugScaleThirtyTwoHalf{} and \HugScaleThirtyTwoTwo{} at $\epsilon=\tfrac14,\tfrac12,2$,
against \HugScaleShallowLo--\HugScaleShallowHi{} at every $\epsilon$ for $R\le16$. Per batch, $\kappa(32)$ ranges from
\KappaBatchLo{} to \KappaBatchHi, and all \KappaBatchAbove{} batches exceed $\kappa(16)$, itself a \DrawsOld-draw estimate. The count-sketch estimate of $C$
gives \KappaGram{} against the exact \KappaThirtyTwo.

The fourth batch measured the exact second order at $R=32$ on the same batches (Table~\ref{tab:fourth}, B; the KL of
every batch and format equals the third batch's bitwise, \ExactGate{} values).
$\kappa_2=\mathrm{JVP}_\text{all}/\sum_\ell\mathrm{JVP}_\ell$ is \ExactKappaIntfour{} for INT4 and
\ExactKappaInteight{} for INT8, where the diagonal $\kappa$ is \ExactDiagIntfour{} and \ExactDiagInteight. The registered
expectation that $\kappa$ is a property of the gradients alone, and so the same for both formats within 25\%, fails: the
local second-order coherence depends on the direction of the weight error, which the diagonal Fisher averages out. INT8 is in the
linear regime ($\KL_\text{all}/\mathrm{JVP}_\text{all}=\ExactLinAllInteight$, single loops \ExactLinLoopInteight), and its
measured coherence $\KL_\text{all}/\sum_\ell\KL_\ell=\ExactCohInteight$ agrees with \ExactKappaInteight; for it the diagonal
Fisher overstates the shared term (\ExactDiagSharedInteight{} of the exact) and the single-loop terms (\ExactRInteight).
For INT4 the whole error saturates (\ExactLinAllIntfour) while its single-loop errors in loops 6--16 are super-quadratic
(KL/JVP \ExactMidLo--\ExactMidHi{} per loop, against \ExactLateLo--\ExactLateHi{} in loops 25--32 and
\ExactInteightLo--\ExactInteightHi{} for INT8 everywhere), the same middle-loop amplification as the switch above; its
diagonal shared term is \ExactDiagSharedIntfour{} of the exact one, so the \FisherJvpAggr{} above is an INT4 figure. Per
batch, $\kappa_2$ ranges from \ExactBatchLo{} to \ExactBatchHi{} (INT4) and \ExactBatchLoInteight{} to
\ExactBatchHiInteight{} (INT8). The measured coherence at $R=32$ is \ExactCohIntfour{} for INT4 and \ExactCohInteight{} for
INT8, close to each other and well below the INT4 $\kappa_2$; Table~\ref{tab:kappa} lists it. On
\OwnKappaSeqs{} of Huginn's own chains with the causal state (Table~\ref{tab:fourth}, D), the diagonal $\kappa(32)$ is
$\OwnKappaThirtyTwo\pm\OwnKappaSE$ (\DrawsNew{} draws, jackknife \OwnKappaRelSE\%; per batch
\OwnKappaBatchLo--\OwnKappaBatchHi), \KappaSigmaGap$\sigma$ from the re-encoded chains' value, with
$\kappa(4)=\OwnKappaFour$ and formats within \OwnKappaSpread\%.

The fifth batch asked which property of the direction matters (Table~\ref{tab:fifth}, B; INT4 KL and JVP bitwise equal
to the fourth batch's). The registered hypothesis was that coherence is carried by the systematic part of the error, the
part that moves the fixed point the same way in every loop: zero-mean stochastic rounding on the INT4 grid should then
match INT8, pure shrinkage ($\Delta=-10^{-3}W$) should exceed INT4, and the group-mean bias of the INT4 error should carry
most of it. All three directions give $\kappa_2$ between \DirLo{} and \DirHi{} (SR \DirSr, bias \DirBias,
shrinkage \DirShrink), between INT8's \ExactKappaInteight{} and INT4's \ExactKappaIntfour, and the INT4 error is no more
aligned with $W$ than INT8's (cosine \DirCosIntfour{} against \DirCosInteight). The direction dependence is real but not
along this axis. Per batch, $\kappa_2$ spans \DirSrBatchLo--\DirSrBatchHi{} for SR and
\DirShrinkBatchLo--\DirShrinkBatchHi{} for shrinkage, more than the spread between directions: on eight chains a
direction's $\kappa_2$ is set by a few batches, and the INT4/INT8 difference is a property of these chains, not
a constant.

The sixth batch tested the obvious reading of that chain dependence and was falsified (Table~\ref{tab:sixth}). Per chain,
the four INT4-sized directions of the fourth and fifth batches rank the same eight re-encoded chains alike (Spearman
0.76--0.93), so we registered that $\kappa_2$ rises with the largest real part of a chain's middle-loop
dominant Ritz value: an expanding map should align every loop's error along one direction. On \SixSeqs{} own chains, the
eight with the most expanding middle-loop modes (\SixExpReLo--\SixExpReHi) against the eight with the most negative
(\SixConReHi{} to \SixConReLo, period-two modes of modulus just above one), the correlation is \SixRhoIntfour{} for INT4
and \SixRhoShrink{} for shrinkage, and the period-two chains are the more coherent ones: median $\kappa_2$
\SixMedConIntfour{} against \SixMedExpIntfour{} for INT4 (\SixConRatioIntfour$\times$) and \SixMedConShrink{} against
\SixMedExpShrink{} for shrinkage (\SixConRatioShrink$\times$; shrinkage is linear on every chain, measured/JVP
\SixShrinkLinLo--\SixShrinkLinHi). The chain dependence itself replicated out of sample: INT4 and shrinkage rank the
\SixSeqs{} new chains alike (Spearman \SixRepl). A post hoc reading that fits both this ranking and the parity
alternation of Figure~\ref{fig:lag}: along a period-two mode the state alternates, so the weight error's effect on it
alternates too, and the two alternations cancel in propagation, so the alternating component of the error accumulates
across loops; a positive expanding mode has no such phase lock. The seventh pre-registration tested its signature, the
lag structure of pairwise loop cross terms $c_{\ell m}=\tfrac12(\mathrm{JVP}_{\ell m}-\mathrm{JVP}_\ell-\mathrm{JVP}_m)$
in the middle loops, on four chains of each group (Table~\ref{tab:seventh}). Even lags exceed the neighbouring odd ones in
\SevAltPtwo{} of the four period-two chains and in \SevAltExp{} of the four expanding ones, and the even-lag means differ by
\SevEvenRatio$\times$, not the registered $2\times$; the purest alternation (odd lags down to \SevStrongOdd, even lags at
least \SevStrongEven) belongs to an expanding chain whose $\kappa_2$ is \SevStrongKappa, below one. The measured
pairs carry only \SevCovLo--\SevCovHi\% of the total cross term $\mathrm{JVP}_\text{all}-\sum_\ell\mathrm{JVP}_\ell$,
because \SevLateLo--\SevLateHi\% of the single-loop second order sits in loops 25--32. Writing
$\kappa=1+\rho_\text{eff}(N_\text{eff}-1)$ with $N_\text{eff}=(\sum_\ell\mathrm{JVP}_\ell^{1/2})^2/\sum_\ell\mathrm{JVP}_\ell$,
the two groups overlap in $N_\text{eff}$ and differ in $\rho_\text{eff}$ (\SevRhoPtwoLo--\SevRhoPtwoHi{} against
\SevRhoExpLo--\SevRhoExpHi{} for shrinkage), far below the middle-loop short-lag correlations: the difference lies in late
and long-range pairs, which this batch did not measure.

The eighth batch measured all of it (Table~\ref{tab:eighth}). Because the JVP is a quadratic form in the coefficients
$a_\ell$ of the tangent $\sum_\ell a_\ell\Delta_\ell$, four block JVPs (loops 1--8, 9--24, 25--28, 29--32), the six
unions of two blocks, and one JVP with signs $(-1)^\ell$ split the cross term exactly: into within- and between-block
parts, and into the sums $E$ and $O$ over even and odd lags, with $\mathrm{JVP}_\text{all}-\sum_\ell\mathrm{JVP}_\ell=2(E+O)$.
On the sixth batch's \SixSeqs{} chains every whole JVP matched bitwise and the decomposition closed to $\EightClose$ of
$\mathrm{JVP}_\text{all}$. For shrinkage the registered predictions held: late-involving pairs carry \EightLateCov{} of
the chain-to-chain covariance of $\kappa-1$ and a median \EightLateMed{} of the cross term, and $\kappa$ rises with the
odd-lag balance $O/(|E|+|O|)$ (Spearman \EightOdd). Two post hoc checks qualify both. Under equal correlation of all
pairs the late share would be \EightLateGeom, so late pairs are under-correlated for their mass (\EightLateEnrich$\times$)
and the large share only follows the single-loop mass; pairs within loops 9--24 hold \EightMidMeas{} of the cross term
against \EightMidGeom{} under equal correlation, \EightMidEnrich$\times$ enriched, as the third batch's middle-block share
suggested. The odd-lag correlation is partly algebraic, since $\kappa<1$ with $E>0$ forces the balance below $-\tfrac12$;
on the \EightOddSubN{} chains with $\kappa>1.05$ it is \EightOddSub. Since $\kappa-1=2E/\sum_\ell\mathrm{JVP}_\ell+
2O/\sum_\ell\mathrm{JVP}_\ell$, the variance of $\kappa-1$ splits exactly into covariance shares of the two lag classes,
free of that coupling: odd lags carry \EightOddShare{} and even lags \EightEvenShare{} (INT4: odd \EightIntOddShare;
post hoc). In each of the \EightBelowOne{} chains with $\kappa<1$ a negative
odd-lag sum outweighs the even-lag sum, by a narrow margin where both are large (\EightEvenNine{} and \EightOddNine{} of
$\sum_\ell\mathrm{JVP}_\ell$ for the chain of the seventh batch's purest alternation): parity alternation can be strong
across all loop pairs, and it cancels coherence rather than building it. This is a property of the cross terms, not of
the loop map's spectrum: the sixth batch's period-two chains, with negative dominant Ritz values, have positive odd-lag
sums (median \EightOddPtwo, against \EightOddExp{} for the expanding chains; Spearman between the largest middle-loop real
part and $O/\sum_\ell\mathrm{JVP}_\ell$ is \EightReOdd). For INT4, recorded but not scored, the pattern is reversed: pairs within loops 9--24
carry \EightIntMidCov{} of the covariance and \EightIntMidMeas{} of the cross term (\EightIntMidEnrich$\times$ their
equal-correlation share), late-involving pairs \EightIntLateCov. Which chains are coherent is shared across error
directions; where the coherence sits is not.

With \DrawsNew{} draws the off-diagonal part $\kappa-1=\KappaOffTotal$ splits by loop blocks as follows: middle loops (9--24)
among themselves \BlkMidMid{} (\BlkMidShare\%), early with middle \BlkEarlyMid, middle with late \BlkMidLate, late among
themselves \BlkLateLate, early among themselves \BlkEarlyEarly, early with late \BlkEarlyLate. The late loops hold
\LateFisherShare\% of the diagonal Fisher, but adjacent late loops correlate at only \LateAdjLo--\LateAdjHi, and late pairs
correlate at \LatePairsNew{} against \EarlyPairsNew{} for early ones. The coherence comes from the middle loops, and that
is the same finding as T4: near the fixed point, loops neither add KV rounding errors coherently nor align their weight
gradients. Post hoc, the correlation alternates with the parity of the loop distance (Figure~\ref{fig:lag}). Mean
correlations at distances 1 to 6 are \LagOne, \LagTwo, \LagThree, \LagFour, \LagFive, \LagSix; in the middle loops
distance 1 averages \MidLagOne{} and distance 2 \MidLagTwo, and $\rho(\ell,\ell+2)$ reaches \LagTwoMax{} where
$\rho(\ell,\ell+1)$ peaks at \LagOneMax. The KV Jacobians of T4 show the same parity, measured independently with exact
JVPs.

\section{Registrations three to nine}\label{app:registrations}

\paragraph{Third pre-registration.} The third batch fixed \ThirdCriteria{} criteria before its two jobs (Table~\ref{tab:third}). They
replicate the Huginn E1 findings on \OwnSeqs{} of Huginn's own chains (H), separate the two T4 explanations (S),
re-measure $\kappa$ with \DrawsNew{} draws (K) and re-register T7 with the excess share (T7). Thresholds that use earlier
results say so in the registration; these criteria confirm observed effects rather than predict new ones. No
falsification condition occurred. Three criteria failed: H-c, by \OwnTfourAMiss; the secondary S5, which tested parity only in
the late loops where it is absent; and K-d, because $\kappa(32)$ fell below the band built around the \DrawsOld-draw value.

\begin{table}[H]
\centering\footnotesize\setlength{\tabcolsep}{4pt}
\caption{Third pre-registration (GH200). H: Huginn E1 on \OwnSeqs{} own greedy chains; S: separation of the T4
explanations on \SplitSeqs{} chains (late = both planes among the last eight); K: $\kappa$ with \DrawsNew{} Fisher draws on the
second batch's E2 batches; T7: localization on \TSevenNewSeqs{} new Ouro sequences, \TSevenScored{} scored cases.}
\label{tab:third}
\begin{tabular}{llll}
\toprule
 & Pre-registered criterion & Measured & Verdict \\
\midrule
H-a & median KL/JVP in [0.8, 1.25] for $\ge$7/8 formats & 7/8 & pass \\
H-b & SR/RN in [1.5, 2.5]; SR / expected in [0.6, 1.4] & 1.68; 0.96 & pass \\
H-c & RN/SR measured/diagonal $<$1.2 & 1.21 & fail (not falsified) \\
H-d & final plane 25--50\%, V 40--66\%, last five $\ge$60\% & 40.8\%, 58\%, 89\% & pass \\
H-e & \texttt{d2h}/\texttt{r2h} KL $\ge$100 & 346 & pass \\
H-f & plane change monotone, $<$0.1 after loop 16 & 0.54$\to$0.014 & pass \\
S1 & RN late error cosine $\ge$0.7; SR $\le$0.1 & 0.76; 0.00 & pass \\
S2 & RN late $\rho^J\ge$0.7; SR $|\rho^J|\le$0.1 & 0.87; 0.02 & pass \\
S3 & class J: $\rho^J\ge0.5$, alignment $A<0.5$ & $\rho^J$ 0.87, $A$ 0.13: J & as predicted \\
S4 & cross fraction RN $-$ SR $<$0.2 & 13.9\% $-$ 2.5\% & pass \\
S5 & (secondary) late $A_2>A_1$ & 0.075 vs 0.13 & fail \\
K0 & KL identical to the second batch, per (R, batch, format) & 36/36 & pass \\
K-a & $\kappa(32)/\kappa(4)\ge$1.5 & 2.53 & pass \\
K-b & format spread $<$10\% & 1.6\% & pass \\
K-c & jackknife s.e.\ $\le$10\% & 0.21 (8.0\%) & pass \\
K-d & $\kappa(32)$ in [2.75, 4.58] & 2.68 & fail \\
K-e & late pairs below early pairs & 0.061 vs 0.19 & pass \\
K-f & middle block $\ge$40\% of $\kappa-1$ & 51\% & pass \\
T7a & median top-3 excess share $\ge$0.6 & 0.76 & pass \\
T7b & median loops 1--2 excess share $\ge$0.7 & 1.00 & pass \\
T7c & $\ge$2/3 of cases with fewer than 8 layers at $\ge$5\% & 19/19 & pass \\
 
\bottomrule
\end{tabular}
\end{table}

\paragraph{Fourth pre-registration.} The fourth batch fixed \FourthCriteria{} criteria before one job (Table~\ref{tab:fourth}):
the localization of the switching sequence (A), the exact second order behind $\kappa(32)$ (B), the spectrum of the loop
map (C) and $\kappa$ on Huginn's own chains (D). No falsification condition occurred; \FourthFailed{} criteria failed. A2
and A4 predicted a near-tie row and a late-plane origin, and the switch is neither; A5 compared the controls' top-row
share of an excess that is zero for them, an ill-posed criterion. B1 and B2 predicted a format-independent exact $\kappa$
above two, and the exact value depends on the format. C1 and C3 predicted the $-1$ mode to be confined to the middle
loops: the dominant values are negative in \SpecMidNeg\% of the middle cells and \SpecLateNeg\% of the late ones, with
only \SpecConverged{} of \SpecCells{} dominant values converged after \SpecKrylov{} Arnoldi steps, so the spectrum is
indicative only; the late loops contract (\SpecLateContract{} cells with $|\lambda|<1$, C2).

\begin{table}[H]
\centering\footnotesize\setlength{\tabcolsep}{4pt}
\caption{Fourth pre-registration (GH200). A: the switching Huginn sequence under direct INT4 with \SwitchControls{} control
sequences; B: exact second order at $R=32$ on the third batch's E2 batches, INT4 and INT8; C: dominant Ritz values of the
loop map at loops 10--32 on \SpecSeqs{} own chains, \SpecKrylov{} Arnoldi steps per cell; D: $\kappa$ on \OwnKappaSeqs{} own
chains with \DrawsNew{} draws.}
\label{tab:fourth}
\begin{tabular}{llll}
\toprule
 & Pre-registered criterion & Measured & Verdict \\
\midrule
A1 & top-3 rows hold $\ge$80\% of the excess & 99.0\% (one row) & pass \\
A2 & that row's margin in the lowest 20\% of the sequence & 2.55 logits, 38th pct. & fail \\
A3 & top-3 layers hold $\ge$60\% & 100\% (one layer) & pass \\
A4 & loops 25--32 hold $\ge$60\% & 46\%; loop 8 dominates & fail \\
A5 & controls: top-3 rows $<$50\% & excess $\approx0$, share undefined & fail (ill-posed) \\
B0 & KL of every batch and format equals the third batch's & 12/12 bitwise & pass \\
B1 & $\kappa_2(32)\ge$2.0 for INT4 and INT8 & 4.09; 1.65 & fail \\
B2 & INT4/INT8 ratio in [0.8, 1.25] & 2.48 & fail \\
B3 & INT8 $\KL$ coherence within 30\% of $\kappa_2$ & 1.74 vs 1.65 & pass \\
B4 & diagonal/exact single-loop ratio in [0.4, 1.0] & 0.85 & pass \\
C1 & middle loops: negative dominant Ritz value in $\ge$2/3 of cells & 50\% & fail \\
C2 & late loops: $|\lambda|<1$ in $\ge$2/3 & 11/12 & pass \\
C3 & negative fraction middle $>$ late & 50\% vs 58\% & fail \\
D1 & $\kappa(32)/\kappa(4)\ge$1.5 & 2.25 & pass \\
D2 & $\kappa(32)$ in [1.74, 3.62] & 2.43 $\pm$ 0.14 & pass \\
D3 & format spread $<$10\% & 1.1\% & pass \\
D4 & jackknife s.e.\ $\le$12\% & 5.6\% & pass \\
 
\bottomrule
\end{tabular}
\end{table}

\paragraph{Fifth pre-registration.} The fifth batch fixed \FifthCriteria{} criteria (Table~\ref{tab:fifth}): expanding
modes across all own chains (A), three synthetic error directions for the exact $\kappa$ (B) and the spectrum with
\ConvKrylov{} Arnoldi steps (C). No falsification condition occurred; \FifthFailed{} criteria failed, among them the three
main predictions: an expanding middle-loop mode does not predict a sequence's excess (A2, A4), the systematic part of a
weight error does not carry the loop coherence (B1--B3), and the $-1$ mode is not specific to the middle loops once every
cell converges (C1, C3).

\begin{table}[H]
\centering\footnotesize\setlength{\tabcolsep}{4pt}
\caption{Fifth pre-registration (GH200). A: dominant Ritz values at loops 13, 16 and 19 of all \StatSeqs{} own chains
(\StatKrylov{} Arnoldi steps) joined to the third batch's E1; B: exact second order at $R=32$ on the fourth batch's E2
batches for stochastic-rounding INT4, the group-mean part of the INT4 error and a pure shrinkage; C: the fourth batch's
four sequences with \ConvKrylov{} steps, converged cells only.}
\label{tab:fifth}
\begin{tabular}{llll}
\toprule
 & Pre-registered criterion & Measured & Verdict \\
\midrule
A1 & the switching sequence expands again; its max.\ middle-loop Re ranks $\le$3 of 64 & +1.23, rank 2 & pass \\
A2 & Spearman(max.\ Re, $\log$ KL/JVP) $\ge$0.3 & $0.21$ & fail \\
A3 & chains with an expanding middle-loop mode $\le$25\% & 9/64 = 14\% & pass \\
A4 & $\ge$3 of the 5 highest-excess chains expand & 2/5 & fail \\
B0 & INT4 KL and JVP equal the fourth batch's & 12/12 bitwise & pass \\
B1 & $\kappa_2$(SR INT4) within 25\% of INT8's 1.65 & 2.44 & fail \\
B2 & $\kappa_2$(shrinkage) $\ge$ INT4's 4.09 & 2.21 & fail \\
B3 & $\kappa_2$(group-mean bias) $\ge$3.0 & 2.37 & fail \\
B4 & $|\cos(\Delta,W)|$ INT4 $>$ INT8 & $0.0026$ vs $-0.0002$ (both $\approx0$) & pass (trivial) \\
C0 & converged cells $\ge$24/32 with 72 steps & 32/32 & pass \\
C1 & middle loops: negative dominant value in $\ge$2/3 of converged cells & 10/20 & fail \\
C2 & late loops: $|\lambda|<1$ in $\ge$2/3 & 10/12 & pass \\
C3 & negative fraction middle $>$ late & 10/20 vs 8/12 & fail \\
 
\bottomrule
\end{tabular}
\end{table}

\paragraph{Sixth pre-registration.} The sixth batch fixed \SixthCriteria{} criteria and one falsification condition
(Table~\ref{tab:sixth}) for the hypothesis that a chain's local second-order coherence rises with the expansion of its middle-loop map.
The falsification condition occurred: the correlation is negative and the expanding chains are the less coherent ones.
H3, the out-of-sample replication of the chain dependence across error directions, passed.

\begin{table}[H]
\centering\footnotesize\setlength{\tabcolsep}{4pt}
\caption{Sixth pre-registration (GH200): local second-order $\kappa_2$ at $R=32$ for INT4 and shrinkage on \SixSeqs{} own chains, the
eight most expanding and the eight most negative middle-loop dominant modes of the fifth batch.}
\label{tab:sixth}
\begin{tabular}{llll}
\toprule
 & Pre-registered criterion & Measured & Verdict \\
\midrule
G & all 16 chains measured, including the switching one & 16/16 & pass \\
H1 & Spearman(max.\ middle-loop Re, $\log\kappa_2$ INT4) $\ge$0.5 & $-0.18$ & fail \\
H2 & expanding / contracting median $\kappa_2$ (INT4) $\ge$1.5 & 0.61 (2.64 vs 4.34) & fail \\
H3 & per-chain Spearman(INT4, shrinkage) $\ge$0.6 & 0.84 & pass \\
H4 & Spearman(max.\ Re, $\log\kappa_2$ shrinkage) $\ge$0.5 & $-0.36$ & fail \\
F & falsification: H1 $<0$ and expanding median $<$ contracting & both hold & falsified \\
 
\bottomrule
\end{tabular}
\end{table}

\paragraph{Seventh pre-registration.} The seventh batch fixed \SeventhCriteria{} criteria (Table~\ref{tab:seventh}) for
the phase-lock reading of the sixth batch. No falsification condition occurred, but both discriminating predictions failed
(P1, P2). P3 passed only as a determinism check: the same chains, host, build and code reproduce the sixth batch's arrays
bitwise (\SevDeterm{} of \SevDetermOf), so it is not a replication. P4 holds for both groups.

\begin{table}[H]
\centering\footnotesize\setlength{\tabcolsep}{4pt}
\caption{Seventh pre-registration (GH200): normalized cross terms $\hat c_k=c_{\ell,\ell+k}/(\mathrm{JVP}_\ell\mathrm{JVP}_{\ell+k})^{1/2}$
for shrinkage, averaged over anchors $\ell=9,11,\dots,21$, on the four most coherent period-two chains and four expanding
chains of the sixth batch.}
\label{tab:seventh}
\begin{tabular}{llll}
\toprule
 & Pre-registered criterion & Measured & Verdict \\
\midrule
G & all 8 chains, both directions & 8/8 & pass \\
P1 & $\hat c_2>\hat c_1,\hat c_3$ in $\ge$3/4 period-two chains and $\le$1/4 expanding ones & 3/4 vs 3/4 & fail \\
P2 & median even-lag $\hat c$ period-two $\ge2\times$ expanding & 0.54 vs 0.44 & fail \\
P3 & median $\kappa_2$ period-two $\ge1.5\times$ expanding & 3.40 vs 1.12 & pass (determinism check) \\
P4 & period-two median $\hat c_4/\hat c_2\ge0.5$ & 0.90 & pass (not discriminating) \\
 
\bottomrule
\end{tabular}
\end{table}

\paragraph{Eighth pre-registration.} The eighth batch fixed \EighthCriteria{} criteria, a gate and one falsification
condition (Table~\ref{tab:eighth}) for a closed-form second-order decomposition of the cross term. The gate and all three criteria passed
and the falsification condition did not occur. The equal-correlation baseline, the restriction of H3 to $\kappa>1$ and the
INT4 decomposition are post hoc; the INT4 single-loop JVPs were taken from the sixth batch under the bitwise gate on the
whole JVP.

\begin{table}[H]
\centering\footnotesize\setlength{\tabcolsep}{4pt}
\caption{Eighth pre-registration (GH200): closed-form second-order decomposition of the cross term for shrinkage on the sixth batch's
\SixSeqs{} chains, from block, block-union and alternating-sign JVPs; H2 and H3 use the \EightLateN{} chains with
$|\kappa-1|\ge0.05$.}
\label{tab:eighth}
\begin{tabular}{llll}
\toprule
 & Pre-registered criterion & Measured & Verdict \\
\midrule
G & whole JVP bitwise equal to the sixth batch; closure $<10^{-4}$ & 16/16; $2\times10^{-6}$ & pass \\
H1 & late-involving share of the chain-to-chain covariance of $\kappa-1$ $\ge$0.6 & 0.63 & pass \\
H2 & median late-involving share of the cross term $\ge$0.5 & 0.59 (15 chains) & pass \\
H3 & Spearman($\kappa$, $O/(|E|+|O|)$) $\ge$0.5 & 0.88 & pass \\
F & falsification: middle-block covariance share $\ge$0.5 & 0.29 & not met \\
 
\bottomrule
\end{tabular}
\end{table}

\paragraph{Ninth registration.} Two runs tested the eighth batch's post hoc readings as predictions on chains never
used for a closed-form second-order decomposition: \NineChains{} own chains drawn with a fixed seed from the \(64-16\) not used by the sixth
batch, and \NineChainsB{} more drawn with a second seed from the remaining ones, registered and frozen before any output
of the first run was read; both directions measured their own single-loop JVPs (Table~\ref{tab:ninth}). The pooled
\(24\)-chain analysis was registered as secondary. The first sample contains no chain with \(\kappa<1\) (Var\((\kappa-1)\)
\NineVarA), the second \NineBelowOneB{} (\NineVarB). No falsification condition occurred. The middle-block enrichment
and the direction dependence replicate in both runs; the odd-lag share replicates at about one half, not at the original
\EightOddShare, and the spectrum link does not replicate. The late-involving share of the covariance of \(\kappa-1\) is
\NineLateCov{} in the pooled sample, again a mass effect (enrichment \NineLateEnr). The second run used a different GH200
host and a different build of the same PyTorch minor version; its quantities are ratios within one run.

\begin{table}[H]
\centering\footnotesize\setlength{\tabcolsep}{4pt}
\caption{Ninth registration (GH200): two independent samples of unused own chains; thresholds come from the eighth
batch's post hoc values (in parentheses). Pooled: registered secondary analysis over both samples.}
\label{tab:ninth}
\begin{tabular}{lP{6.0cm}rrrll}
\toprule
 & Registered criterion & run 1 & run 2 & pooled & run 1 & run 2 \\
\midrule
R1 & middle-block enrichment over equal correlation $\ge$1.5 (2.2) & 2.63 & 3.07 & 2.82 & pass & pass \\
R2 & odd-lag share of the covariance of $\kappa-1$ $\ge$0.5 (0.87) & 0.43 & 0.52 & 0.48 & fail & pass \\
R3 & Spearman($\kappa$, odd-lag balance) on $\kappa>1.05$ $\ge$0.4 (0.64) & 0.33 & 0.79 & 0.60 & fail & pass \\
R4 & INT4 middle-block covariance share $>$ shrinkage's (0.53 vs 0.29) & 0.55 vs 0.25 & 0.45 vs 0.25 & 0.50 vs 0.25 & pass & pass \\
R5 & Spearman(largest middle-loop Re, $O/\sum_\ell\mathrm{JVP}_\ell$) $<0$ ($-0.33$) & $-0.10$ & $0.03$ & $0.01$ & pass & fail \\
 
\bottomrule
\end{tabular}
\end{table}

\paragraph{Tenth registration.} Sensitivity-guided formats on Ouro (Table~\ref{tab:tenth}; main paper,
Table~\ref{tab:alloc}). The registration froze the rule, not the widths: costs $D_c(b)=\tfrac12\sum_iF_{ii}\Delta_i(b)^2$
and $E_c(b)=\sum_i\Delta_i(b)^2$ of every (plane, K$|$V, layer) cell were computed in the job on the first batch's 64
chains, a greedy allocation (start at 2 bits, take the largest cost reduction per stored bit that fits) wrote the widths
and the predicted median ratios to checksummed files, and the scoring stages read those files. Budgets are the bytes of
uniform INT2, uniform INT4 and \texttt{d4h}. Allocated formats store at most their baseline's bytes; all of them matched
exactly. \TenPassed{} of \TenCriteria{} criteria passed and no falsification condition occurred.

\begin{table}[H]
\centering\footnotesize\setlength{\tabcolsep}{4pt}
\caption{Tenth registration (GH200): sensitivity-guided KV formats on Ouro, \TenGsm{} held-out GSM8K and \TenMath{}
MATH-500 chains. Ratios are medians of per-sequence KL ratios. The 19\% budget equals the all-2-bit start of the greedy
rule, so its allocations equal the uniform format by construction.}
\label{tab:tenth}
\begin{tabular}{lP{6.4cm}ll}
\toprule
 & Registered criterion & Measured & Verdict \\
\midrule
G & allocation file checksums match; allocated bytes $\le$ baseline & match; 12, 20, 31 & pass \\
F1a/b/c & held-out GSM8K: median KL ratio to the uniform format $<1$ at 19\,/\,31\,/\,48\% & 1.00 / 0.60 / 0.15 & fail / pass / pass \\
F2a/b/c & measured / calibration-predicted median ratio in [0.5, 2] & 1.00 / 1.08 / 1.00 & pass / pass / pass \\
F3a/b & Fisher allocation beats error-only allocation (31, 19\%) & 0.60 / 1.00 & pass / fail \\
F4 & beats the one-line hand rule at 31\% & 0.48 & pass \\
F5 & one width per (plane, K$|$V) within $1.25\times$ of per-layer & 1.66 & fail \\
F6a/b/c & MATH-500: median ratio $<1$ at 19\,/\,31\,/\,48\% & 1.00 / 1.08 / 0.16 & fail / fail / pass \\
F7 & 19\% allocation stays linear (KL/JVP in [0.8, 1.25]) & 1.38 & fail \\
 
\bottomrule
\end{tabular}
\end{table}

\paragraph{Eleventh registration.} Bit allocation by measured single-cell costs on Ouro (Table~\ref{tab:eleventh};
main paper, Table~\ref{tab:alloc}). The registration was written after the tenth registration's tail had been seen, so
every primary criterion is scored on \ElevenMath{} MATH-500 chains that no registration had used; the GSM8K chains, all
used by the tenth registration, are a seen-set replication (S1/S2). On the tenth registration's \ElevenCalChains{}
calibration chains each of the 192 cells was stored alone at 2, 4, 8 and 16 bits with every other cell exact, and the
one-step KL of every sequence is its cost $C_c(b)$; the same greedy rule allocated the 25, 31 and 48\% budgets (the bytes
of K4V2, INT4 and \texttt{d4h}). The tenth registration's diagonal-Fisher formats were reused verbatim, a diagonal-Fisher
allocation at 25\% and one with a 4-bit floor for values were frozen before the job, and K4V2 and K2V4 are the uniform
formats at 25\%. Thresholds come from the tenth registration (sequences worse $\le 11\%$; twice its 48\% median and
aggregate). \ElevenPassed{} of \ElevenCriteria{} criteria passed and the falsification condition (T3) did not occur. The
Fisher formats' KL values on the seen GSM8K chains agree bitwise with the tenth registration. Unscored: on the tenth
registration's five worst sequences, formats that keep one group of the Fisher allocation's 2-bit cells and store
every other cell exactly split its KL by cell group.

\begin{table}[H]
\centering\footnotesize\setlength{\tabcolsep}{4pt}
\caption{Eleventh registration (GH200): bit allocation by measured single-cell KL on Ouro, scored on \ElevenMath{}
unseen MATH-500 chains (T, P), the \ElevenCalChains{} calibration chains (A) and \TenGsm{} seen GSM8K chains (S). Ratios
are of per-sequence KL; aggregate = ratio of sums.}
\label{tab:eleventh}
\begin{tabular}{lP{6.4cm}ll}
\toprule
 & Registered criterion & Measured & Verdict \\
\midrule
G & checksums of costs and formats; frozen formats verbatim; bytes $\le$ budget; all calibration batches & match; 16, 20, 31; 10/10 & pass \\
T1 & 31\%: measured-cost allocation, aggregate KL ratio to INT4 $<1$ & 0.69 & pass \\
T2a/b & median ratio $<1$; sequences worse $\le 11\%$ & 0.66 / 6\% & pass / pass \\
T3 & aggregate ratio to the diagonal-Fisher allocation $<1$ (falsification if not) & 0.051 & pass \\
T4a/b/c & 48\%, diagonal-Fisher allocation replicated: median, aggregate $\le 0.30$, worse $\le 5\%$ & 0.17 / 0.3032 / 0.6\% & pass / fail / pass \\
T5a/b & 48\%, measured-cost allocation: median and aggregate ratio $<1$ & 0.14 / 0.14 & pass / pass \\
T6a/b & 25\%: aggregate and median ratio $<1$ against both K4V2 and K2V4 & 0.84, 0.074 / 0.76, 0.078 & pass / pass \\
T7 & 31\%, Fisher allocation without 2-bit values: aggregate ratio $<1$ & 14.5 & fail \\
P1 & measured / predicted median ratio in [0.5, 2] & 1.25 & pass \\
A1/A2 & calibration chains: whole KL / sum of single-cell costs in [0.5, 2] (allocation, INT4) & 0.85 / 1.01 & pass / pass \\
S1/S2 & seen GSM8K chains: aggregate ratio to INT4 and to the diagonal-Fisher allocation $<1$ & 0.53 / 0.10 & pass / pass \\
 
\bottomrule
\end{tabular}
\end{table}

\paragraph{Twelfth registration.} Why the first loop's first-layer keys fail at 2 bits (Table~\ref{tab:twelfth}).
On the eleventh registration's \ElevenCalChains{} calibration chains the cell was stored at 2 bits on one subset of its
elements at a time, every other element and cell exact: each of the KV heads, and five classes of key positions that
partition every sequence (the first token, tokens 1--3, the rest of the few-shot prefix shared by all prompts, the
record's own prompt, the generated tokens). Every chain had been used before, so the registration tests a mechanism
on chains where the effect was seen, and one format that had never been measured: the second-order allocations with
only that cell at 4 bits. \TwelvePassed{} of \TwelveCriteria{} criteria passed and the falsification condition (the cell
explains neither the MATH tail nor does the guard remove it) did not occur. Unscored: on MATH the same head carries
\TwelveMathHeadShare\% of the cell's cost; the guarded allocation makes \TwelveGuardMathWorse\% of MATH sequences worse
than INT4.

\begin{table}[H]
\centering\footnotesize\setlength{\tabcolsep}{4pt}
\caption{Twelfth registration (GH200): the cost of the first loop's first-layer keys at 2 bits on the
\ElevenCalChains{} calibration chains, split by KV head and key position, and the diagonal-Fisher allocations with only
that cell at 4 bits on \ElevenMath{} MATH-500 and \TenGsm{} GSM8K chains. Shares are of the summed KL.}
\label{tab:twelfth}
\begin{tabular}{lP{6.4cm}ll}
\toprule
 & Registered criterion & Measured & Verdict \\
\midrule
G & the whole cell's cost reproduces the eleventh registration; format checksums & bitwise; match & pass \\
H1 & one KV head carries $\ge 50\%$ of the cell's 2-bit cost & 96\% (head 5) & pass \\
H2 & the first key position (attention sink) carries $\ge 50\%$ & $4\times10^{-9}$ & fail \\
H3 & MATH: the cell alone at 2 bits $\ge 50\%$ of the diagonal-Fisher allocation's KL & 80\% & pass \\
H4a/b & that allocation with the cell at 4 bits: aggregate ratio to INT4 $<1$ (MATH, GSM8K) & 0.85 / 0.59 & pass / pass \\
H4c & the same at 25\%: aggregate ratio to K4V2 $<1$ (MATH) & 0.90 & pass \\
 
\bottomrule
\end{tabular}
\end{table}

\paragraph{Thirteenth registration.} Free generation on the \ThirteenN{} GSM8K problems (Table~\ref{tab:thirteenth}):
3-shot chain-of-thought prompts, greedy, at most 1024 new tokens, the records' stop strings, FP32 compute, 24 cache
slots; prefill attends exact keys and values and every stored token then goes through the format. Answers are scored
by the gsm8k-cot strict-match rule, re-implemented without loading the dataset and checked beforehand against
lm-eval's verdicts on 2{,}200 stored generations (all agreed). A prompt's ending ``moved'' when the token ids differ
from exact storage's but the text up to the stop string is the same once trailing whitespace is removed. The threshold
for moved endings was half the teacher-forced switching rate on the same chains; it did not allow for answers that
diverge before the ending row. \ThirteenPassed{} of \ThirteenCriteria{} criteria passed and the falsification condition
occurred.

\begin{table}[H]
\centering\footnotesize\setlength{\tabcolsep}{4pt}
\caption{Thirteenth registration (GH200): free generation on \ThirteenN{} GSM8K problems with simulated KV formats,
compared per problem with exact storage (accuracy \ThirteenAccExact\%). Accuracy differences in points.}
\label{tab:thirteenth}
\begin{tabular}{lP{6.4cm}ll}
\toprule
 & Registered criterion & Measured & Verdict \\
\midrule
G & format checksums; all eight formats generated every problem & match; 200 each & pass \\
F & the diagonal-Fisher allocation moves at least 5\% of answer endings (P1) & 4.5\% & falsified \\
P1 & same criterion, counted as a criterion & 4.5\% & fail \\
P2a/b/c & endings moved $\le 2\%$: guarded allocation, measured-cost allocation, INT4 & 0 / 0 / 0 & pass / pass / pass \\
P3a/b/c/d & accuracy within $\pm 3$ points of exact storage: INT4, Fisher, guarded, measured & $-4.0$ / $-2.5$ / $-5.5$ / $-2.5$ & fail / pass / fail / pass \\
 
\bottomrule
\end{tabular}
\end{table}

\paragraph{Fourteenth registration.} Free generation on the 200 MATH-500 problems (Table~\ref{tab:fourteenth}):
minerva 4-shot prompts, greedy, at most 512 new tokens, stop at the next ``Problem:'', 16 cache slots; the 200 GSM8K
problems were generated again with exact and BF16 storage. MATH answers are the last boxed expression (else the
``Final Answer'' sentence) after minerva's normalization, copied from lm-eval. String accuracy compares the normalized
answer with the normalized reference; symbolic dependencies were unavailable in that run. The later audit in
Appendix~\ref{app:math-scoring} adds symbolic scoring and corrects the earlier description of string accuracy as a lower
bound on the official lm-eval metric: the extractors differ.
Answer-level criteria use the \FourteenParse{} problems where exact storage gives an answer. Exact storage reproduced
both sets of chains token for token. \FourteenPassed{} of \FourteenCriteria{} criteria passed and the falsification
condition occurred.

\begin{table}[H]
\centering\footnotesize\setlength{\tabcolsep}{4pt}
\caption{Fourteenth registration (GH200): free generation on 200 MATH-500 problems with the diagonal-Fisher allocation
with and without its one 2-bit key cell guarded, and BF16 storage as the smallest perturbation. Answer rates on the
\FourteenParse{} problems exact storage answers; points.}
\label{tab:fourteenth}
\begin{tabular}{lP{6.4cm}ll}
\toprule
 & Registered criterion & Measured & Verdict \\
\midrule
G & format checksums; every format generated every problem & match; 200 each & pass \\
F & the guarded allocation reproduces exact storage's MATH answer on no more problems than the unguarded one & 93 vs 95 & falsified \\
Q1a/b & guarded reproduces $\ge 5$ points more answers; one-sided sign test $p<0.05$ & $-$1.8 / 0.79 & fail / fail \\
Q2 & guarded string accuracy $\ge 2$ points higher & $-$1.0 & fail \\
Q3 & answer-change rate of the unguarded allocation $\ge 10$ points above BF16 storage's & 15.2 vs 0.9 & pass \\
Q4 & GSM8K: BF16 storage rewrites $\le 20\%$ of generations & 3\% & pass \\
 
\bottomrule
\end{tabular}
\end{table}

\paragraph{Fifteenth registration.} Free generation at 48\% of BF16 bytes (Table~\ref{tab:fifteenth}): \texttt{d4h} and
the two allocations of the eleventh registration at its bytes, on the same 200 GSM8K and 200 MATH-500 problems. An
answer changed when its extracted answer differs from exact storage's (GSM8K: every problem; MATH: the
\FourteenParse{} problems exact storage answers, against the fourteenth registration's exact arm, which matches the
chains token for token); the two sets are pooled per problem, and the formats of the two previous registrations are
counted the same way. A one-sided sign test on discordant problems detects about a two-to-one split at the observed
counts (for example 20:10 of 30), which the registration stated. \FifteenPassed{} of \FifteenCriteria{} criteria passed.

\begin{table}[H]
\centering\footnotesize\setlength{\tabcolsep}{4pt}
\caption{Fifteenth registration (GH200): answer changes against exact storage on \FifteenPooled{} pooled GSM8K and MATH
problems at 48\% of BF16 bytes. ``A vs B; x:y'': changes of the two formats and their discordant problems.}
\label{tab:fifteenth}
\begin{tabular}{lP{5.4cm}P{2.4cm}l}
\toprule
 & Registered criterion & Measured & Verdict \\
\midrule
G & format checksums; every format generated every problem; earlier formats present & match; 200 each & pass \\
R1 & measured-cost allocation at 48\% changes fewer answers than \texttt{d4h} (one-sided sign test $p<0.05$) & 23 vs 32; 17:8, $p=0.054$ & fail \\
R2 & diagonal-Fisher allocation at 48\% changes fewer answers than \texttt{d4h} & 18 vs 32; 20:6, $p=0.005$ & pass \\
R3 & \texttt{d4h} (48\%) changes fewer answers than uniform INT4 (31\%) & 32 vs 43; 19:8, $p=0.026$ & pass \\
R4 & eight formats: Spearman $\rho$(one-step aggregate KL, answer changes) $\ge 0.6$ & 0.83 & pass \\
 
\bottomrule
\end{tabular}
\end{table}

\paragraph{Sixteenth registration.} Free generation on the \SixteenN{} GSM8K test problems that no earlier registration
used (Table~\ref{tab:sixteenth}). The full test set was a local copy, index-identical to the documents behind the
first 200 problems; rebuilding their 3-shot prompts from it reproduced all 200 texts and token ids. Settings as in the
thirteenth registration; formats: exact, uniform INT4, \texttt{d4h}, BF16, the two allocations at 48\%, and the
measured-cost allocation at INT4's bytes (secondary). \SixteenPassed{} of \SixteenCriteria{} criteria passed.

\begin{table}[H]
\centering\footnotesize\setlength{\tabcolsep}{4pt}
\caption{Sixteenth registration (GH200): free generation on \SixteenN{} unused GSM8K problems. Strict-match accuracy: discordant
problems (first format right only : second right only); answer changes against exact storage.}
\label{tab:sixteenth}
\begin{tabular}{lP{5.6cm}P{2.2cm}l}
\toprule
 & Registered criterion & Measured & Verdict \\
\midrule
G & format checksums; every primary format generated all 1119 problems & match; complete & pass \\
A1 & uniform INT4 lowers accuracy against exact storage & 49:37, $p=0.12$ & fail \\
A2 & \texttt{d4h} (48\%) scores higher than uniform INT4 & 47:35, $p=0.11$ & fail \\
C1 & diagonal-Fisher allocation at 48\% changes fewer answers than \texttt{d4h} & 53 vs 104; $p<10^{-6}$ & pass \\
C2 & measured-cost allocation at 48\% changes fewer answers than \texttt{d4h} & 49 vs 104; $p<10^{-6}$ & pass \\
C3 & \texttt{d4h} changes fewer answers than uniform INT4 & 104 vs 143; $p=0.0002$ & pass \\
C4 & BF16 storage changes $\le 25\%$ as many answers as uniform INT4 & 9 of 143 & pass \\
 
\bottomrule
\end{tabular}
\end{table}

\paragraph{Seventeenth registration.} The four formats whose accuracy losses on the first 200 problems were the
largest (uniform K4V2 and the measured-cost allocation at 25\%, the diagonal-Fisher allocation with and without its
one 2-bit key cell guarded at 31\%) on the \SixteenN{} problems of the sixteenth registration, against its exact
samples (Table~\ref{tab:seventeenth}). This job's exact storage reproduced the first 200 of them token for token.
The problems had been used for other formats only. \SeventeenPassed{} of \SeventeenCriteria{} criteria passed.
Answer changes: K4V2 \SeventeenKfourChanges, measured-cost 25\% \SeventeenKlTwentyfiveChanges, diagonal Fisher
\SeventeenFisherChanges{} (fewer than uniform INT4 despite the largest one-step aggregate KL), guarded
\SeventeenGuardChanges.

\begin{table}[H]
\centering\footnotesize\setlength{\tabcolsep}{4pt}
\caption{Seventeenth registration (GH200): strict-match accuracy on \SixteenN{} GSM8K problems against exact storage (discordant
problems, exact right only : format right only) and the rank test over ten formats.}
\label{tab:seventeenth}
\begin{tabular}{lP{5.6cm}P{2.2cm}l}
\toprule
 & Registered criterion & Measured & Verdict \\
\midrule
G & format checksums; all four formats on all 1119 problems; this job's exact storage reproduces the reference on 200 & match; complete; 200/200 & pass \\
B1 & 25\% measured-cost allocation lowers accuracy against exact storage & 101:52, $p=5\times10^{-5}$ & pass \\
B2 & guarded diagonal-Fisher allocation (31\%) lowers accuracy & 44:45, $p=0.58$ & fail \\
B3 & uniform K4V2 (25\%) lowers accuracy & 90:78, $p=0.20$ & fail \\
B4 & ten formats: Spearman $\rho$(one-step aggregate KL, answer changes) $\ge 0.6$ & 0.82 & pass \\
 
\bottomrule
\end{tabular}
\end{table}

\paragraph{Eighteenth registration.} The seventeenth registration's post hoc comparison of the 25\% measured-cost
allocation with uniform K4V2, registered and tested on 1119 GSM8K \emph{training} problems that no registration used
(ranked by a hash of their index; same 3-shot prompt as the test problems; the model may have seen them, so only the
paired differences are read; Table~\ref{tab:eighteenth}). Exact storage scored \EighteenAccExact\%, K4V2
\EighteenAccKfour\% and the measured-cost allocation \EighteenAccKl\%. \EighteenPassed{} of \EighteenCriteria{} criteria
passed.

\begin{table}[H]
\centering\footnotesize\setlength{\tabcolsep}{4pt}
\caption{Eighteenth registration (GH200): strict-match accuracy at 25\% of BF16 bytes on 1119 unused GSM8K training problems
(discordant problems, first format right only : second right only).}
\label{tab:eighteenth}
\begin{tabular}{lP{5.6cm}P{2.2cm}l}
\toprule
 & Registered criterion & Measured & Verdict \\
\midrule
G & format checksums; all three formats on all 1119 problems & match; complete & pass \\
D1 & 25\% measured-cost allocation scores below uniform K4V2 (falsification if not) & 88:60, $p=0.013$ & pass \\
D2 & 25\% measured-cost allocation scores below exact storage & 89:54, $p=0.002$ & pass \\
 
\bottomrule
\end{tabular}
\end{table}

\paragraph{Nineteenth registration.} The 25\% measured-cost allocation with only its 2-bit key cells raised to 4
bits (0.875 more stored bits per element over the planes; a diagnostic, not an equal-bytes format), on both sets of
1119 problems of the sixteenth to eighteenth registrations, against their stored references; this job's exact storage
reproduced the first 200 problems of each set token for token (Table~\ref{tab:nineteenth}). \NineteenPassed{} of
\NineteenCriteria{} criteria passed.

\begin{table}[H]
\centering\footnotesize\setlength{\tabcolsep}{4pt}
\caption{Nineteenth registration (GH200): strict-match accuracy of the 25\% measured-cost allocation with its 2-bit key cells raised
to 4 bits, pooled over 1119 test and 1119 training problems (discordant problems, first right only : second right only).}
\label{tab:nineteenth}
\begin{tabular}{lP{5.6cm}P{2.2cm}l}
\toprule
 & Registered criterion & Measured & Verdict \\
\midrule
G & format checksums; both sets of 1119 problems complete; exact storage reproduces both references on 200 & match; complete; 200/200 & pass \\
E1 & raising only the 25\% allocation's 2-bit key cells to 4 bits raises its accuracy (pooled; falsification if not) & 150:117, $p=0.025$ & pass \\
E2 & the raised format and uniform K4V2 score the same within noise (two-sided $p\ge 0.05$) & 106:138, $p=0.047$ & fail \\
 
\bottomrule
\end{tabular}
\end{table}

\paragraph{Twentieth registration.} Measured costs against the diagonal Fisher on Huginn-0125 ($R=\HugR$, causal
initial state, FP32), run on \TwentyHW. A block cell is four adjacent loop planes of one core layer's keys or values (64
cells). Budgets count the core planes; prelude and coda KV remain exact and are excluded. Each block cell was stored
alone at 2, 4, 8 and 16 bits on \TwentyCalChains{} calibration chains (\TwentyCalBatches{}
batches), restarting the forward pass at the block's first loop, and the diagonal Fisher was taken per plane on the same
chains. The greedy rule of the tenth registration allocated one width per block from measured costs or from summed Fisher
costs (and, unscored, one width per plane from Fisher costs) at 25, 31, 33 and 48\% of BF16 bytes; all were scored on the
third registration's \TwentySeqs{} chains, where uniform INT4 reproduced that registration's GH200 per-chain KL (aggregate
ratio \TwentyReproAgg, $\rho=\TwentyReproRho$; Table~\ref{tab:twentieth}). \TwentyPassed{} of \TwentyCriteria{} criteria
passed. Measured costs beat the uniform formats at every budget, but at INT4's bytes they tie with Fisher costs
(\TwentyKlFbBetter{} sequences better, \TwentyKlFbWorseN{} worse, $p=\TwentyKlFbP$). The largest block-cell ratio of
measured to Fisher cost at 2 bits, \TwentyOutCell, is \TwentyOutRatio$\times$ the median ratio. Post hoc, the two costs
rank the 64 cells alike (Spearman \TwentyRankTwo{} at 2 bits, \TwentyRankFour{} at 4 bits), and one sequence (the fourth
registration's switching sequence) holds \TwentyTopShareLo--\TwentyTopShareHi\% of both allocations' aggregates at INT4's
bytes and of the Fisher allocation's at 33\%, so these aggregate ratios mostly compare that sequence under two formats; at 33\% of the bytes the Fisher allocation's aggregate is
\TwentyFbUniThirtythree{} of the uniform format's because of it, while the measured-cost allocation keeps that sequence's
tail out (\TwentyKlFbLo{} of the Fisher allocation's aggregate). This is a qualitatively similar concentration of KL;
it does not establish that the same cell-level mechanism operates in both models.

\begin{table}[H]
\centering\footnotesize\setlength{\tabcolsep}{4pt}
\caption{Twentieth registration (\TwentyHWName): bit allocation on Huginn-0125 by measured block costs and by
diagonal-Fisher block costs (one width per four loops, K$|$V and core layer) on \TwentySeqs{} chains; ratios of
per-sequence KL. R1 is reported, not scored.}
\label{tab:twentieth}
\begin{tabular}{lP{5.6cm}P{2.6cm}l}
\toprule
 & Registered criterion & Measured & Verdict \\
\midrule
G & weights and package checksums; identity replay of 80 chains; block restart bitwise and $\le 10^{-6}$ of the full pass; evaluation chains = the third registration's 64; at least 6 calibration batches; floors $<10^{-6}$; budgets met & match; 0 hard; 0; match; 13/13; $9.7\times10^{-9}$; met & pass \\
R1 & uniform INT4 reproduces the third registration per chain (aggregate in [0.95, 1.05], $\rho\ge 0.95$; reported) & 1.0002, $\rho=0.9998$ & pass \\
H1a & measured-cost block allocation below the diagonal-Fisher block allocation at INT4's bytes (aggregate) & 1.065 & fail \\
H1b & the same, median & 0.975 & pass \\
H1c & the same aggregate at 25, 33 and 48\% of BF16 bytes & 0.94, 0.13, 1.29 & fail \\
F1 & falsified if the INT4-bytes aggregate exceeds 1.25 & 1.065 & not falsified \\
H2 & measured-cost allocation below the uniform format of equal bytes at every budget (aggregate) & 0.18, 0.58, 0.14, 0.061 & pass \\
H2b & the same at INT4's bytes, median & 0.034 & pass \\
A1 & block costs add up: measured-cost allocation's KL / sum of its block costs on the calibration chains in [0.5, 2] & 0.84 & pass \\
A2 & the same for uniform INT4 & 1.03 & pass \\
P1 & held-out median ratio to uniform INT4 / the calibration chains' predicted ratio in [0.5, 2] & 0.88 & pass \\
O1 & question: does a block cell cost $>30\times$ the median measured/Fisher ratio at 2 bits? & 7.2$\times$ & no \\
 
\bottomrule
\end{tabular}
\end{table}

\paragraph{Twenty-first registration.}\label{app:checkpoint-v22} The Ouro-1.4B readings on a second checkpoint of the same family, Ouro-2.6B
(48 core layers, $R=4$, 16 KV heads of width 128; its tokenizer is byte-identical to Ouro-1.4B's), run on one GH200.
All thresholds came from the Ouro-1.4B results and were fixed before any Ouro-2.6B measurement. The chains are
Ouro-1.4B's token sequences, teacher-forced; they are not the larger model's own generations. The pipeline repeats
the first, tenth and eleventh registrations without code changes: a sensitivity map with four formats on the first
registration's \TwentyoneSeqs{} GSM8K chains, diagonal-Fisher and measured single-cell costs of the 384 cells
(4 planes, K$|$V, 48 layers) at 2, 4, 8 and 16 bits on the same chains, the greedy allocation at 25, 31 and 48\% of
BF16 bytes, and scoring on \TwentyoneMath{} of the eleventh registration's 176 fresh MATH chains. The measured
calibration stopped at its registered deadline after \TwentyoneCalBatches{} of \TwentyoneCalBatchesTotal{} batches
(\TwentyoneCalChains{} of 64 chains); the Fisher costs use all 64. \TwentyonePassed{} of \TwentyoneCriteria{} criteria
passed (Table~\ref{tab:twentyfirst}).

Where the error lives replicates. Values carry \TwentyoneFisherV\% of the diagonal Fisher mass, keys carry
\TwentyoneMeasK\% of the measured INT4 KL, and the median per-sequence KL/JVP of INT4 is \TwentyoneLinMed. The
measured-cost allocations beat uniform INT4 (\TwentyoneKlIntAgg{} of its aggregate), \texttt{d4h} at 48\% and K4V2 at
25\%. The allocation result against the diagonal Fisher does not replicate. At INT4's bytes the measured-cost
allocation's aggregate KL on MATH is \TwentyoneKlFisherAgg{} of the Fisher allocation's (median \TwentyoneKlFisherMed;
\TwentyoneKlFisherWorse\% of chains worse), so the registered falsification criterion T3 failed. The calibration chains
(\TwentyoneKlFisherIns) and GSM8K chains 65--200 (\TwentyoneKlFisherGsm) point the same way, as does the unscored
comparison at 48\% (\TwentyoneKlFisherFortyeight). The Fisher allocation is \TwentyoneFisherIntAgg{} of uniform INT4's
aggregate, against 13.6 on Ouro-1.4B (F1 failed). The aggregate KL/JVP of INT4 is \TwentyoneLinAgg, so no saturating
tail pulls it below 0.90 (L2 failed). On the registered question O1, one cell stands out: \TwentyoneOutCell{} has
\TwentyoneOutRatio$\times$ the median measured/Fisher cost ratio at 2 bits (median \TwentyoneMedRatio; rank
\TwentyoneOutRankC{} of 384 by measured cost and \TwentyoneOutRankD{} by Fisher cost). It is not Ouro-1.4B's
first-loop, first-layer keys.

Post hoc, both allocator families at 25, 31 and 48\% of the bytes store that cell at 4 or 8 bits.
The Ouro-1.4B failure, a Fisher allocation that leaves the outlier key cell at 2 bits, therefore does not
arise here. With no such tail to remove, the measured-cost allocation loses to the Fisher allocation by a small margin
on most chains; its costs also come from fewer chains. This reading is post hoc and rests on one checkpoint and one
calibration. The registered result is that, on Ouro-2.6B, measured cell costs did not beat diagonal-Fisher costs.

\begin{table}[H]
\centering\footnotesize\setlength{\tabcolsep}{4pt}
\caption{Twenty-first registration (GH200): the Ouro-1.4B readings on Ouro-2.6B. Sensitivity map on \TwentyoneSeqs{}
GSM8K chains; allocations scored on \TwentyoneMath{} MATH chains; ratios of per-sequence KL. Thresholds and the
quoted Ouro-1.4B values come from earlier registrations. O1 is a registered question, not a criterion.}
\label{tab:twentyfirst}
\begin{tabular}{lP{5.6cm}P{2.6cm}l}
\toprule
 & Registered criterion & Measured & Verdict \\
\midrule
G & weights and package checksums; tests; frozen arms verbatim; budgets met; at least 4 calibration batches; at least 100 fresh MATH chains; at least 48 E1 chains & match; 8 batches; 175 MATH; 64 E1 & pass \\
R1 & values carry at least 70\% of the diagonal Fisher mass (Ouro-1.4B: 94\%) & 96\% & pass \\
R2 & keys carry at least 60\% of measured \texttt{d4h-all} KL (1.4B: 89\%) & 88\% & pass \\
L1 & median per-sequence KL/JVP of \texttt{d4h-all} in [0.8, 1.25] (1.4B: 1.01) & 0.96 & pass \\
L2 & aggregate KL/JVP of \texttt{d4h-all} below 0.90 (1.4B: 0.79) & 0.95 & fail \\
T1 & MATH: measured-cost allocation below uniform INT4 at equal bytes (aggregate) & 0.60 & pass \\
T2 & the same, median & 0.56 & pass \\
T3 & MATH: measured-cost allocation below the diagonal-Fisher allocation at INT4's bytes (aggregate; falsification if not) & 1.134 & falsified \\
T5 & MATH: measured-cost allocation below \texttt{d4h} at 48\% (median and aggregate) & 0.14, 0.14 & pass \\
T6 & MATH: measured-cost allocation below K4V2 at 25\% (median and aggregate) & 0.47, 0.54 & pass \\
F1 & MATH: the diagonal-Fisher allocation above uniform INT4 at equal bytes (aggregate; 1.4B: 13.6) & 0.53 & fail \\
A1 & single-cell costs add up: in-sample KL / sum of costs in [0.5, 2] & 0.71 & pass \\
O1 & question: does one cell's measured/Fisher ratio at 2 bits exceed 30$\times$ the median? & 125$\times$ (keys of core layer 18 in loop 3) & yes \\
 
\bottomrule
\end{tabular}
\end{table}

\paragraph{Admission and truncation.} The initial launch ended E1 with zero sequences because its deadline offsets
assumed a longer allocation. An amendment changed only those offsets before criterion data were produced; the
aborted archive is preserved. The scored run passed its frozen-file checks and 48 CPU tests. E1 completed 64/64
chains, Fisher calibration 64/64, in-sample evaluation 64/64 and secondary GSM8K 136/136. Measured calibration reached
\CheckpointCalBatches/22 batches (\CheckpointCalN/64 chains, \CheckpointCalMin{} minutes of recorded batch time),
and MATH reached \CheckpointMathN/176 chains before their deadlines. The omitted MATH ID is 301. The registered
minimum gates require four calibration batches, 100 MATH chains and 48 E1 chains, and pass together with the frozen
allocation, provenance and byte-budget gates. Passing those minima does not complete the planned protocol.
The measured-cost and Fisher allocations therefore use different calibration coverage; their ranking is conditional
on this time-bounded run. Length-ordered execution also makes the truncated calibration subset unrepresentative of
the full 64-chain sample. Nsight Compute lacked performance-counter permission; its failure is retained and contributes
no performance evidence.

\paragraph{Independent audit.} Independent CPU execution of the frozen analyzer exactly reproduces the original analysis, including all failures;
a separate standard-library reader recomputes all 11 criterion values, byte budgets and gates from released numerical
records. Neither check replaces missing GPU cases or establishes independently timestamped registration.

\section{Deviations from the registrations}\label{app:deviations}

\paragraph{Deviations from the pre-registration.} The batch table listed \MathPlanned{} MATH-500 chains; the frozen
driver ran \MathSeqs. The Huginn stages first failed on code, not criteria, and were rerun after three fixes: a
Hadamard rotation of order \HugHeadDim, block-wise FP8 quantization for widths not divisible by 128, and per-loop
recomputation of activations after E2 ran out of GPU memory at $R=16$ (peak \HugETwoOomGB\,GB; rerun peak
\HugETwoPeakGB\,GB). E2's results at $R=4$ and $8$ were seen before the $R=16$ and $32$ rerun. The only change to the
frozen analysis code was reading results from two directories; no criterion or threshold changed. The frozen code
tests only the late-versus-early half of T5c. The monotonicity half and its strict pairwise definition were computed
for this paper after the $R=32$ results were seen; the verdict does not depend on them, because the frozen half fails
by itself. The defect of the T7 metric
was recognized after its results. The third batch ran as registered; the analyses of the reversed Huginn aggregate,
of the Jacobian parity by plane range, of disjoint draw subsets and of the mixed $\kappa$ reading are post hoc. The
fourth batch ran as registered. Its registration said the spectrum criteria would use converged Ritz values; the frozen
code scores every cell, and we report the code's verdict, since filtering to converged cells leaves too few to score.
The defect of A5 was recognized after its results. The per-loop reading of the INT4 single-loop errors is post hoc. The
fifth batch's direction stage ran out of GPU memory at its start in the first job (five error directions resident at
once) and was rerun in a second job after a change that keeps the inactive directions on the host; its A and C stages
come from the first job, and no measurement or criterion changed. The sixth batch ran on a different GH200 host with a
different build of the same PyTorch minor version; its quantities are ratios of JVPs and do not depend on bitwise
agreement with the earlier batches. Its mechanism paragraph is post hoc. The seventh batch ran on the same host and build,
and the coverage and $\rho_\text{eff}$ decomposition of its results are post hoc. The eighth batch ran on the same host
and build with the same chain order, so its whole JVPs could be checked bitwise against the sixth batch. In the tenth registration the smallest budget equalled the start of the greedy rule (all cells at the 2-bit floor), so its three criteria could not pass; this was missed when the registration was frozen. The coarse allocation (one width per plane and K$|$V) could not move bytes at the INT4 budget either, so its criterion compares uniform INT4 with the per-layer allocation. The eleventh registration ran as registered and its calibration was not truncated. Its cell-group split of the
tail, the cost ratio and upgrade rank of the first loop's first-layer keys, and the reading that every format storing that
cell at 2 bits has a heavy tail are post hoc; they correct the tenth registration's post hoc attribution of its tail to
2-bit value cells. The twelfth registration ran as registered on chains used before; the reading of its switching rows
(the row after an answer's two newlines, ``Q'' flipping to a newline) is post hoc. The thirteenth registration ran as
registered; the link between its moved endings and the teacher-forced switching sequences is post hoc. The fourteenth
registration ran as registered, including its two secondary formats. The fifteenth registration ran as registered; its
job name was changed after the registration's checksum was taken (the batch script's checksum is in the package). The sixteenth registration ran as registered; the ordering of its answer changes by one-step KL is post hoc. The seventeenth registration ran as registered; its comparison of the 25\% measured-cost allocation with K4V2
and the key-cell reading are post hoc; the eighteenth registered that comparison and ran as registered. The nineteenth ran as registered. The twentieth was frozen for a GH200; a first job there passed its identity gate
and was then cancelled on a stale status reading, and a second attempt on a 48\,GB GPU ran out of memory in its third
calibration batch. Neither reached a criterion and neither is scored. An amendment frozen before the third run moved it
to \TwentyHW{} with the same PyTorch version built for x86 and a shorter time limit, and made R1 a cross-hardware check; no criterion or
threshold changed. Its Nsight Compute pass failed for lack of counter permissions and is not reported. The twenty-first was frozen for one GH200 in a booked window. Its first launch
used stage deadlines sized for a longer window, so the sensitivity stage stopped before its first batch and nothing
was scored; that output is kept unscored. An amendment frozen before the second launch shortened only the stage
deadlines and moved the output directory; no criterion, threshold, input or code changed. Triton came from the
Python package index instead of the PyTorch index (same version). The measured calibration stopped at its
deadline after \TwentyoneCalBatches{} of \TwentyoneCalBatchesTotal{} batches, which the registration allows (at
least four). Its Nsight Compute pass and a second, unregistered attempt after the archive both failed for lack of
counter permissions.

\section{Negative results}\label{app:negative}


\begin{itemize}
\item \textbf{Mature-value drafting does not transfer to Ouro (H100).} Reading final-loop values in a depth-2
  self-speculative draft lowers draft--target agreement from \DasIdentity{} to \DasMatureV{} on GSM8K. Values
  converge across loops more slowly than keys (mean cosine to the final loop at loop~1: \CosKLoopOne{} for keys,
  \CosVLoopOne{} for values) \cite{das}.
\item \textbf{Residual KV reads more bytes than BF16} (Table~\ref{tab:bytes}).
\item \textbf{FP8 KV without a split-KV kernel is not faster} at equal concurrency (\S\ref{sec:repr}).
\item \textbf{The diagonal Fisher is not a calibrated aggregate predictor} (P1d, slope \POneDSlope), nor are
  group-size ratios (P3).
\item \textbf{No amplification by weight sharing} between $R=1$ and $R=4$ on Ouro (\S\ref{sec:weights}). On
  Huginn, INT4 and FP8 weight errors leave the linear regime at $R=32$ while INT8 does not; that observation is post
  hoc and is not the pre-registered test.
\item \textbf{A fixed point does not make errors add coherently} (T4a, T5c). Huginn's converged late planes have
  strongly correlated rounding errors, but nearly orthogonal output Jacobians, and its late loops do not align their
  weight gradients.
\item \textbf{The second batch overstated $\kappa(32)$} (\KappaThirtyTwoOld{} with \DrawsOld{} Fisher draws, against
  $\KappaThirtyTwo\pm\KappaSE$ with \DrawsNew; K-d).
\item \textbf{The local second-order coherence depends on the format} (B1, B2): \ExactKappaIntfour{} for INT4 and \ExactKappaInteight{}
  for INT8 at $R=32$, where the diagonal $\kappa$ gives \ExactDiagIntfour{} for both. \textbf{The switching sequence is
  neither a near tie nor a late-plane effect} (A2, A4).
\item \textbf{Neither an expanding loop-map mode nor the systematic part of a weight error explains the exceptions}
  (fifth batch): \StatExpandingPct\% of the own chains have an expanding middle-loop mode and one of them switches;
  stochastic rounding, group-mean bias and pure shrinkage all give $\kappa_2$ of \DirLo--\DirHi.
\item \textbf{Coherence does not rise with an expanding loop map} (sixth batch, falsified): chains with period-two
  middle-loop modes are \SixConRatioIntfour--\SixConRatioShrink$\times$ more coherent than chains with expanding ones.
\item \textbf{Parity alternation of loop cross terms does not carry the coherence} (seventh batch): it appears in both
  groups, and its purest case cancels to $\kappa_2=\SevStrongKappa$.
\item \textbf{The aggregate's tail is not mean-divergent} (T1). The heavy-tailed formats keep a tail index between
  1 and 2 on held-out chains, not below 1.
\item \textbf{The pre-registered localization metric is ill-posed} (T7): restricted JVPs omit cross terms.
\item \textbf{Measured block costs do not beat diagonal-Fisher block costs on Huginn} (twentieth registration, H1a and
  H1c): \TwentyKlFbAgg{} of the Fisher allocation's aggregate at INT4's bytes and \TwentyKlFbLo--\TwentyKlFbHi{} at the
  other budgets; no block cell's measured/Fisher cost ratio exceeds 30 times the median (registered question O1, Table~\ref{tab:twentieth}).
\item \textbf{Measured cell costs do not beat diagonal-Fisher cell costs on a second Ouro checkpoint} (twenty-first
  registration, T3 falsified): on Ouro-2.6B the measured-cost allocation's aggregate KL at INT4's bytes is
  \TwentyoneKlFisherAgg{} of the Fisher allocation's, and the Fisher allocation does not fail in the tail (F1,
  \TwentyoneFisherIntAgg{} of uniform INT4's). An outlier cell exists (O1, \TwentyoneOutRatio$\times$ the median
  ratio), but neither allocator family stores it at 2 bits (Table~\ref{tab:twentyfirst}).
\end{itemize}

\section{Paired accuracy audit}\label{app:paired-accuracy}

The eighteenth registration (historical run label v19) compares exact FP32 storage, uniform K4V2 and the measured-cost
allocation at 25\% of BF16 bytes on the same \PairedN{} GSM8K training problems. The problems were unused by earlier
experiment batches; this does not establish that they were absent from model training. Strict-match rescoring of all
\PairedGenerations{} saved generations reproduces the recorded verdicts: exact storage solves \PairedExactCorrect{}
problems, K4V2 solves \PairedKfourCorrect{}, and the measured-cost allocation solves \PairedKlCorrect{}. Input, format
and generation-file checksums match their saved metadata, and the
three output ID sets match the input set with no duplicates.

\begin{table}[H]
\centering\small
\caption{Paired strict-match accuracy differences on 1119 training problems (candidate minus reference, percentage points).
Intervals are post hoc, approximate 95\% intervals without multiplicity adjustment. Discordant counts are
reference-only correct : candidate-only correct. The first two one-sided exact sign tests were registered; the
K4V2--exact comparison is secondary.}
\label{tab:paired-accuracy}
\begin{tabular}{lrrrr}
\toprule
Candidate vs. reference & $\Delta$ (pp) & 95\% interval & Discordant & $p$ \\
\midrule
Measured cost vs. K4V2 & -2.50 & [-4.63, -0.38] & 88:60 & 0.013 \\
Measured cost vs. Exact & -3.13 & [-5.22, -1.04] & 89:54 & 0.002 \\
K4V2 vs. Exact & -0.63 & [-2.75, +1.50] & 77:70 & 0.310 \\
 
\bottomrule
\end{tabular}
\end{table}

For paired correctness differences $d_i\in\{-1,0,1\}$, with mean $\bar d$ and $m$ discordant problems out of $n$,
the standard error is $\sqrt{(m/n-\bar d^2)/(n-1)}$. The intervals in Table~\ref{tab:paired-accuracy} are
$100(\bar d\pm1.96\,\mathrm{SE})$. These descriptive intervals were added after the original experiments and do not
alter the registered criteria, tests or ledger. In particular, the K4V2--exact interval permits a loss of about
2.75 points: failure to reject a difference is not evidence of equivalence. The comparison between the measured-cost
allocation and K4V2 supports a strict-score loss in this paired sample. Appendix~\ref{app:scoring-robustness} shows that
this difference is sensitive to answer extraction. It does not establish a general accuracy ranking or estimate
population accuracy beyond the tested checkpoint, prompts and split.

\section{Answer-extraction sensitivity audit}\label{app:scoring-robustness}

\paragraph{Scope and provenance.} This post hoc audit asks whether the 25\%-budget comparison survives a change of
answer extraction. It uses two filters already present in the frozen GSM8K-CoT task configuration:
\texttt{strict-match}, which requires the phrase ``The answer is'' followed by a numeric match, and
\texttt{flexible-extract}, which selects the last numeric match. Both then apply the same saved exact-match
normalization against the reference answer. The strict results remain primary. The analysis plan was saved before
computing flexible scores, after strict parser failures were known; it is not a prospective registration and adds no
criterion to the ledger.

The test sample contains 1119 problems: exact outputs are from historical v17 and the two quantized arms from v18,
with matching input and tokenizer hashes. All training-sample outputs are from v19, the eighteenth registration.
We rechecked raw-file hashes, completeness, unique IDs and input/output ID equality, and reproduced every saved strict
answer, correctness and parse-failure flag across 6714 generations. No capped, unparseable or incorrect output was
discarded; no model was rerun. An independent CPU rescore of the original files on Arrhenius reproduced the complete
indicator-table checksum. The upstream lm-evaluation-harness v0.4.9 \texttt{RegexFilter} implementation also matched
all 13,428 extractions across the two rules. This validates extraction on the saved text, not the full inference harness.

\begin{table}[H]
\centering\small\setlength{\tabcolsep}{4pt}
\caption{Extraction sensitivity at 25\% of BF16 bytes ($n=1119$ separately for each split). Accuracies are percentages;
$\Delta$ is measured-cost allocation minus K4V2 in percentage points. Approximate paired 95\% intervals and one-sided
exact sign tests are descriptive, with no multiplicity adjustment. Discordant counts: K4V2-only : allocation-only correct.
All flexible scores and this combined analysis are post hoc; the original registered tests are unchanged.}
\label{tab:scoring-robustness}
\begin{tabular}{llrrrlrr}
\toprule
Split & Rule & Exact & K4V2 & Alloc. & $\Delta$ [95\% interval] & Discordant & $p$ \\
\midrule
Test & Strict & 60.41 & 59.34 & 56.03 & -3.31 [-5.56, -1.05] & 102:65 & 0.003 \\
Test & Flexible & 78.19 & 73.10 & 72.92 & -0.18 [-2.37, +2.01] & 79:77 & 0.468 \\
Train & Strict & 68.10 & 67.47 & 64.97 & -2.50 [-4.63, -0.38] & 88:60 & 0.013 \\
Train & Flexible & 82.31 & 78.91 & 78.91 & +0.00 [-1.92, +1.92] & 60:60 & 0.536 \\
 
\bottomrule
\end{tabular}
\end{table}

\paragraph{The relative gap depends on extraction.} Under flexible extraction, the test difference is only two
correct answers (\ScoreTestFlexibleAllocCorrect{} versus \ScoreTestFlexibleUniformCorrect{}), and the training counts
are equal (\ScoreTrainFlexibleAllocCorrect{} each). The intervals still allow losses of about 2 points and gains of
similar size, so these results do not establish equivalence. Both formats remain below exact under flexible extraction:
on test, allocation-minus-exact is $\ScoreTestFlexibleAllocExactDelta$ points
($[\ScoreTestFlexibleAllocExactLo,\ScoreTestFlexibleAllocExactHi]$) and K4V2-minus-exact is
$\ScoreTestFlexibleUniformExactDelta$ ($[\ScoreTestFlexibleUniformExactLo,\ScoreTestFlexibleUniformExactHi]$);
on training, both differences are $\ScoreTrainFlexibleAllocExactDelta$ points, with intervals
$[\ScoreTrainFlexibleAllocExactLo,\ScoreTrainFlexibleAllocExactHi]$ for the allocation and
$[\ScoreTrainFlexibleUniformExactLo,\ScoreTrainFlexibleUniformExactHi]$ for K4V2.

\paragraph{Parseability and limitations.} Strict extraction fails on
\ScoreTestStrictExactUnparsed/\ScoreTestStrictUniformUnparsed/\ScoreTestStrictAllocUnparsed{} test outputs and
\ScoreTrainStrictExactUnparsed/\ScoreTrainStrictUniformUnparsed/\ScoreTrainStrictAllocUnparsed{} training outputs
(exact/K4V2/allocation); flexible extraction fails on
\ScoreTestFlexibleExactUnparsed/\ScoreTestFlexibleUniformUnparsed/\ScoreTestFlexibleAllocUnparsed{} and
\ScoreTrainFlexibleExactUnparsed/\ScoreTrainFlexibleUniformUnparsed/\ScoreTrainFlexibleAllocUnparsed{}, respectively.
The released indicators retain each output's parseability, correctness under both rules and stopping reason.
They show that many strict failures yield the reference number under flexible extraction, but a last-number heuristic
can also select an intermediate or unrelated number and can overturn a correct strict answer. Thus this audit does not
prove that formatting causes the entire gap, or adjudicate the reasoning. The training sample is unused by earlier
experimental batches, not guaranteed absent from model training. We do not pool the splits. The supported conclusion
at 25\% is that the strict-score advantage of K4V2 over the measured-cost allocation is extraction-sensitive.

\paragraph{Extension to all complete formats (post hoc).} After observing the 25\% result, we saved a separate
analysis plan and applied the same two rules to every full 1119-problem arm from historical v17--v20. The extension
contains 12 test arms and four training arms, \AllRows{} saved generations in total (including the earlier audit).
It excludes the 200-problem pilot, partial exact reproducibility controls, MATH and ongoing runs. Every strict
verdict is reproduced; input, tokenizer, format and raw-output hashes are verified. The complete exported indicator
table matches an independent Arrhenius rescore, and the upstream extractor matches both rules on every output.
Table~\ref{tab:scoring-all-arms} reports all arms, including both guarded variants; no arm or capped output is dropped.

\begin{table}[H]
\centering\small\setlength{\tabcolsep}{5pt}
\caption{All complete 1119-problem arms from historical v17--v20, rescored post hoc. Accuracies are percentages;
the last column is flexible accuracy minus exact, in percentage points with an approximate paired 95\% interval.
Nominal bytes include scales, relative to BF16; the actual allocation simulation stores dense FP32 planes.
``Measured'' means measured single-cell costs. Guarded variants use additional bytes, so guard comparisons are not
equal-budget comparisons. All intervals are unadjusted for multiplicity.}
\label{tab:scoring-all-arms}
\begin{tabular}{llrrrl}
\toprule
Split & Format & Bytes (\%) & Strict & Flexible & Flexible $-$ exact [95\% interval] \\
\midrule
Test & Exact & 200.00 & 60.41 & 78.19 & -- \\
Test & BF16 & 100.00 & 60.59 & 78.28 & +0.09 [-0.09, +0.26] \\
Test & Uniform INT4 & 31.25 & 59.34 & 76.50 & -1.70 [-3.23, -0.16] \\
Test & d4h & 48.44 & 60.41 & 77.84 & -0.36 [-1.76, +1.04] \\
Test & K4V2 & 25.00 & 59.34 & 73.10 & -5.09 [-7.32, -2.86] \\
Test & Fisher-31 & 31.25 & 61.22 & 78.73 & +0.54 [-0.97, +2.04] \\
Test & Fisher-31 guarded & 31.32 & 60.50 & 77.84 & -0.36 [-1.96, +1.25] \\
Test & Measured-31 & 31.25 & 60.32 & 78.19 & +0.00 [-1.42, +1.42] \\
Test & Fisher-48 & 48.44 & 60.59 & 78.55 & +0.36 [-0.54, +1.25] \\
Test & Measured-48 & 48.44 & 60.14 & 77.84 & -0.36 [-1.22, +0.50] \\
Test & Measured-25 & 25.00 & 56.03 & 72.92 & -5.27 [-7.50, -3.04] \\
Test & Measured-25 guarded & 26.37 & 58.18 & 75.51 & -2.68 [-4.58, -0.78] \\
Train & Exact & 200.00 & 68.10 & 82.31 & -- \\
Train & K4V2 & 25.00 & 67.47 & 78.91 & -3.40 [-5.27, -1.52] \\
Train & Measured-25 & 25.00 & 64.97 & 78.91 & -3.40 [-5.32, -1.47] \\
Train & Measured-25 guarded & 26.37 & 65.77 & 80.07 & -2.23 [-4.08, -0.39] \\
 
\bottomrule
\end{tabular}
\end{table}

\paragraph{Compare methods directly, and retain within-problem dependence.} Comparing each method separately
against exact does not test the difference between methods. Table~\ref{tab:scoring-all-comparisons} therefore reports
all 11 format pairs fixed in the extension plan, under both rules. It also reports the change in each paired gap:
for correctness indicators $c_i$ and $r_i$ of candidate and reference,
\[
u_i=(c_i^{\rm flex}-r_i^{\rm flex})-(c_i^{\rm strict}-r_i^{\rm strict}),\qquad
\mathrm{SE}(\bar u)=\sqrt{\frac{\sum_i(u_i-\bar u)^2}{n(n-1)}}.
\]
The interaction interval is $100(\bar u\pm1.96\,\mathrm{SE})$. This retains the correlation among all four scores on a
problem. A significant contrast under one rule and a non-significant contrast under another does not itself test this
interaction. All intervals here are descriptive, post hoc and unadjusted; none changes a registered verdict.

\begin{table}[H]
\centering\footnotesize\setlength{\tabcolsep}{3.5pt}
\caption{Direct format contrasts and extraction interactions, all in percentage points with approximate paired
95\% intervals. Interaction: flexible difference minus strict difference. Each split has 1119 paired problems;
the splits are kept separate. All 11 planned contrasts are reported without multiplicity adjustment.}
\label{tab:scoring-all-comparisons}
\begin{tabular}{lllll}
\toprule
Split & Candidate $-$ reference & Strict [95\% interval] & Flexible [95\% interval] & Interaction [95\% interval] \\
\midrule
Test & Measured-25 $-$ K4V2 & -3.31 [-5.56, -1.05] & -0.18 [-2.37, +2.01] & +3.13 [+1.56, +4.69] \\
Test & Measured-31 $-$ Uniform INT4 & +0.98 [-0.71, +2.67] & +1.70 [+0.09, +3.31] & +0.71 [-0.47, +1.90] \\
Test & Measured-48 $-$ d4h & -0.27 [-1.72, +1.19] & +0.00 [-1.44, +1.44] & +0.27 [-0.61, +1.14] \\
Test & Fisher-31 $-$ Uniform INT4 & +1.88 [+0.23, +3.53] & +2.23 [+0.55, +3.92] & +0.36 [-0.86, +1.57] \\
Test & Fisher-48 $-$ d4h & +0.18 [-1.24, +1.60] & +0.71 [-0.66, +2.09] & +0.54 [-0.32, +1.39] \\
Test & Fisher-31 guarded $-$ Fisher-31 & -0.71 [-2.05, +0.62] & -0.89 [-2.20, +0.42] & -0.18 [-1.11, +0.75] \\
Test & Measured-25 guarded $-$ Measured-25 & +2.14 [+0.08, +4.21] & +2.59 [+0.49, +4.70] & +0.45 [-0.97, +1.86] \\
Test & Measured-25 guarded $-$ K4V2 & -1.16 [-3.18, +0.86] & +2.41 [+0.46, +4.37] & +3.57 [+2.10, +5.05] \\
Train & Measured-25 $-$ K4V2 & -2.50 [-4.63, -0.38] & +0.00 [-1.92, +1.92] & +2.50 [+1.00, +4.00] \\
Train & Measured-25 guarded $-$ Measured-25 & +0.80 [-1.17, +2.78] & +1.16 [-0.68, +3.01] & +0.36 [-0.95, +1.67] \\
Train & Measured-25 guarded $-$ K4V2 & -1.70 [-3.54, +0.15] & +1.16 [-0.58, +2.90] & +2.86 [+1.38, +4.34] \\
 
\bottomrule
\end{tabular}
\end{table}

The measured-25 versus K4V2 gap changes by $\AllTestMeasuredTwentyfiveInteractionDelta$ points on test
($[\AllTestMeasuredTwentyfiveInteractionLo,\AllTestMeasuredTwentyfiveInteractionHi]$) and
$\AllTrainMeasuredTwentyfiveInteractionDelta$ on training
($[\AllTrainMeasuredTwentyfiveInteractionLo,\AllTrainMeasuredTwentyfiveInteractionHi]$).
At 31\%, measured cost and Fisher both have positive point differences against uniform INT4 under both rules;
the measured-cost flexible interval and both Fisher intervals exclude zero before multiplicity adjustment. These
are exploratory gains on one test sample, not an independently confirmed winner. The measured-31 interaction
interval includes zero, so the difference in its two rules' interval verdicts is not evidence of a rule effect.
At 48\%, both allocation-versus-uniform intervals include zero under both rules.

Guarding all 2-bit keys in measured-25 uses \AllGuardBytes\% of BF16 bytes. Under flexible extraction it gains
$\AllTestMeasuredGuardFlexibleDelta$ points over unguarded measured-25 on test
($[\AllTestMeasuredGuardFlexibleLo,\AllTestMeasuredGuardFlexibleHi]$), but the training gain of
$\AllTrainMeasuredGuardFlexibleDelta$ has interval
$[\AllTrainMeasuredGuardFlexibleLo,\AllTrainMeasuredGuardFlexibleHi]$. Its comparison with K4V2 also changes with
extraction, and it spends more bytes; neither comparison isolates a universal key-cell failure mechanism.
The single-cell Fisher guard uses \AllFisherGuardBytes\% of BF16 bytes and does not resolve a task improvement
over its unguarded reference under either rule. The released results include all parse failures, rule transitions,
capped counts and discordant pairs. This audit tests the saved outputs, not new model inference or reasoning quality,
and does not establish transfer to another checkpoint or MATH.

\section{MATH scoring audit}\label{app:math-scoring}

\paragraph{Protocol and provenance.} This post hoc audit independently rescores all \MathScoreOutputs{} saved
generations from the fourteenth and fifteenth registrations: nine formats on the same \MathScoreN{} problems.
It keeps every capped, incorrect and unparseable output, rather than conditioning on exact storage producing an answer.
Input IDs, raw-file checksums, completeness and all stored extraction, correctness and parse-failure flags are verified.
The analysis plan was saved after the historical score totals were known and before computing these paired contrasts;
it adds no criterion to the registration ledger. Table~\ref{tab:math-scoring} reports all formats and scoring rules.

\textit{Boxed symbolic} accepts historical normalized-string equality or symbolic equivalence under the pinned
lm-evaluation-harness v0.4.9.2 Minerva routine (SymPy 1.14.0, ANTLR 4.11.0).
\textit{Answer-only math-verify} uses math-verify 0.8.0 with the saved reference answer, rather than the full reference
solution used upstream. \textit{Final-Answer symbolic} applies upstream Minerva's literal final-answer sentence extractor
and symbolic routine. It is a format-sensitivity diagnostic: most saved outputs do not match that sentence pattern.
Thus string equality is a lower bound only on our explicitly augmented boxed-symbolic rule, not on the official
Minerva metric. None of these heuristics adjudicates reasoning or establishes that a changed answer is correct.

\begin{table}[H]
\centering\small\setlength{\tabcolsep}{5pt}
\caption{Post hoc MATH scores on all \MathScoreN{} problems per format. Scores and capped-output rates are percentages;
bytes are relative to BF16. The last two scoring columns diagnose alternative extraction semantics and are not
official full-harness task scores. Exact uses dense FP32 storage. The Fisher guard spends additional bytes.}
\label{tab:math-scoring}
\begin{tabular}{lrrrrrr}
\toprule
Format & Bytes & String & Boxed symbolic & Answer-only & Final-Answer & Capped \\
 & (\%) & & & math-verify & symbolic & \\
\midrule
Exact & 200.00 & 43.0 & 50.0 & 56.0 & 3.5 & 46.0 \\
BF16 & 100.00 & 43.5 & 50.5 & 56.0 & 4.0 & 44.5 \\
Rotated INT4 & 31.25 & 43.5 & 49.0 & 56.0 & 6.0 & 41.5 \\
Fisher 31 & 31.25 & 44.5 & 52.0 & 57.5 & 5.5 & 40.5 \\
Fisher 31 + guard & 31.32 & 43.5 & 49.5 & 53.5 & 5.0 & 41.5 \\
Measured 31 & 31.25 & 43.5 & 49.0 & 53.5 & 7.5 & 43.0 \\
Direct INT4 & 48.44 & 42.5 & 48.5 & 55.5 & 3.5 & 46.5 \\
Fisher 48 & 48.44 & 43.5 & 50.0 & 56.0 & 3.0 & 42.5 \\
Measured 48 & 48.44 & 43.0 & 49.5 & 57.0 & 3.5 & 44.0 \\
 
\bottomrule
\end{tabular}
\end{table}

\paragraph{Result and uncertainty.} Exact accuracy rises from \MathScoreExactString\% under normalized-string
equality to \MathScoreExactSymbolic\% under boxed-symbolic scoring. All direct allocator-versus-uniform contrasts in
Table~\ref{tab:math-scoring-paired} have paired intervals containing zero. In particular, measured-cost allocation at
31\% has the same boxed-symbolic accuracy as rotated INT4; neither this result nor the guard contrast establishes
equivalence. At \MathScoreMinCap--\MathScoreMaxCap\% capped outputs, the 512-token protocol itself is a substantial
restriction on interpreting task quality. These are new analyses of existing generations, not an independent task
replication. Intervals condition on the frozen allocations, use a paired normal approximation and have no multiplicity
adjustment. The released indicators also reproduce every non-exact format's comparison with exact under all four rules.

\begin{table}[H]
\centering\footnotesize\setlength{\tabcolsep}{4pt}
\caption{Candidate minus reference on the same \MathScoreN{} MATH problems. Differences and paired 95\% intervals
are in percentage points; discordant counts are reference-only : candidate-only correct. All contrasts are post hoc.}
\label{tab:math-scoring-paired}
\begin{tabular}{lllll}
\toprule
Candidate vs. reference & Boxed symbolic & Discordant & Answer-only math-verify & Discordant \\
\midrule
Measured 31 vs. Rotated INT4 & +0.00 [-3.93, +3.93] & 8:8 & -2.50 [-7.20, +2.20] & 14:9 \\
Fisher 31 vs. Rotated INT4 & +3.00 [-2.18, +8.18] & 11:17 & +1.50 [-4.14, +7.14] & 15:18 \\
Measured 31 vs. Fisher 31 & -3.00 [-7.59, +1.59] & 14:8 & -4.00 [-9.35, +1.35] & 19:11 \\
Fisher 31 + guard vs. Fisher 31 & -2.50 [-7.20, +2.20] & 14:9 & -4.00 [-9.35, +1.35] & 19:11 \\
Measured 48 vs. Direct INT4 & +1.00 [-2.10, +4.10] & 4:6 & +1.50 [-2.78, +5.78] & 8:11 \\
Fisher 48 vs. Direct INT4 & +1.50 [-2.30, +5.30] & 6:9 & +0.50 [-3.55, +4.55] & 8:9 \\
Measured 48 vs. Fisher 48 & -0.50 [-4.04, +3.04] & 7:6 & +1.00 [-3.17, +5.17] & 8:10 \\
 
\bottomrule
\end{tabular}
\end{table}

\paragraph{Historical rescore discrepancy.} The independent audit reproduces every historical string and
boxed-symbolic verdict. Answer-only math-verify differs on one percentage-answer problem in all nine formats:
the saved reference is $10$, whereas the output includes $10\%$, which the pinned parser interprets as $0.1$.
The earlier rescore's runtime was not fully recorded, so we retain both results and all nine discrepancies rather
than infer its exact implementation. This illustrates unit sensitivity of answer-only scoring; it is not evidence
of nine new model failures. The public table uses the pinned audit, preserves the historical file's checksum and
does not change the registered string-score or answer-agreement verdicts.

\section{Earlier kernel measurements}\label{app:historical-kernels}

An earlier H100 implementation study used the engine's paged KV cache rather than the dense simulated allocation
path. With the original decode kernel, FP8 KV ran at \NonsplitLo--\NonsplitHi{} of BF16 throughput at equal concurrency;
with a split-KV kernel that reads per-vector scales, the ratio was \FpeightLo--\FpeightHi{} over \FpeightPoints{}
points, and capacity at a fixed pool rose \CapRatio$\times$. The unfused writer gave
\FpeightTorchWriteA--\FpeightTorchWriteC{}, showing that the writer also affects throughput. These historical
measurements concern that FP8 implementation. They do not establish a bandwidth or serving benefit for the mixed-bit
allocations evaluated in this paper.

\section{Interpretation limits}\label{app:interpretation}

\paragraph{Generation protocol and task measurements.} The Ouro mixed-bit generator attends exact KV throughout prefill and computes
the first output token before quantizing the stored prompt planes. Subsequent steps attend the stored, transformed KV
of previous tokens together with the current token's fresh KV. The free-generation results therefore test decode-time
storage perturbations, not a pipeline that also quantizes prefill attention. Teacher-forced one-step KL and free-generation
accuracy also follow different token trajectories. A high-KL delimiter row need not recur in free generation, and a
high-KL row within a MATH answer does not by itself show that the answer is wrong. Ratios of summed KL weight chains
by their reference KL, so a tail-dominated aggregate can differ from an unweighted per-chain comparison; neither
directly measures task correctness.

Both task evaluations use fixed few-shot templates, greedy decoding, stop strings and generation caps. The MATH pilot's
answer-agreement analysis conditions on the \FourteenParse{} of 200 problems for which exact storage yields a parsed
answer; it does not describe agreement on all problems. Matching exact is not an accuracy oracle: exact storage's
normalized-string accuracy is \FourteenAccExact\%. The full-sample symbolic audit in Appendix~\ref{app:math-scoring}
raises exact accuracy but does not resolve a direct allocator advantage on this pilot.
The guarded and unguarded allocations agree with exact on 93 and 95 of those 112 problems, respectively
(6:8 discordant pairs), so the failed improvement test does not establish equivalence or identical answers. The GSM8K
extraction audit rescores the same saved outputs under two rules, without adjudicating reasoning or supplying an
independent generation sample. Unused training-split questions need not be absent from model pretraining. These
experiments do not measure robustness to prompt changes, alternative decoding policies or other task domains.

\paragraph{Allocation uncertainty and transfer.} Isolated-cell calibration measures each perturbation with the other
cells exact. The observed agreement with joint costs on the calibration sample does not guarantee additivity for other
allocations or prompts. The greedy search is not compared with a global optimum. Paired task intervals condition on
the checkpoint, prompts and selected format; they omit variability from resampling calibration chains and choosing a
new allocation, as well as multiplicity from the exploratory comparisons. Guarded formats can spend more bytes than
their references, so those contrasts do not isolate the benefit of a guard at fixed budget. Ouro allocates per loop,
layer and K/V cell, whereas Huginn groups four adjacent loops; changes in outlier strength or allocation gains can
reflect this granularity as well as model differences. The later second-Ouro test remains conditional on 1.4B-generated
chains and incomplete measured calibration. One Huginn checkpoint and the absence of a matched non-looped control
also limit architecture-level attribution.
Huginn's shared-weight decomposition covers two error directions and \SixSeqs{} plus \NineChains{}+\NineChainsB{}
chains, with no expanding chain in the new samples. Few Fisher draws overstate $\kappa(32)$
(\KappaThirtyTwoOld{} at \DrawsOld{} draws), and odd-lag terms explain about half, rather than most, of the
chain-to-chain variation out of sample. H100 and GH200 measurements are linked only by the reported reproduction checks.

\paragraph{Calibration and implementation cost.} Ouro's measured-cost calibration uses \ElevenCalChains{} chains and
\ElevenForwards{} forwards per batch, taking \ElevenCalMin{} minutes on one GH200. The study does not measure how many
requests would amortize that cost, how often a changed prompt distribution requires recalibration, or whether the
allocated widths remain stable across calibration samples. Huginn's budgets cover core KV only, with prelude and coda
KV left exact; they are not whole-cache compression fractions. Nominal storage includes group scales and offsets but does
not measure format-index storage, packing alignment or temporary workspace. Dense FP32 computation isolates storage
error without testing its interaction with lower-precision arithmetic. The earlier FP8 kernel measurements concern a
different implementation; they cannot supply packed mixed-bit throughput, latency, capacity or concurrency measurements.

\paragraph{Fair serving comparisons.} A shared attention reader can isolate cache-format effects but does not establish
competitiveness with an optimized BF16 serving implementation. A direct comparison requires the same checkpoint,
hardware, context, output length, batch and loop count, with compute precision and kernel differences stated explicitly.
Prefill and decode must be timed separately after warmup; compile time, retained cache storage and peak allocator memory
have different meanings. Methods that combine token sparsity and cache quantization also change the work performed;
switching off token selection need not restore the same dense kernel. Component ablations alone therefore cannot
attribute a speed difference to quantization. The separate L40S pilot below supplies single-request implementation
timings, but its reader fails the model-level numerical gate and it does not measure an accuracy--throughput tradeoff.
Concurrent serving and maximum feasible batch capacity remain unresolved.

\clearpage
\section{L40S frontend and component qualification}\label{app:l40s-qualification}

We separately tested a common BF16 Ouro-1.4B frontend and the public FlashLoop KIVI readers on one L40S
(SM89, CUDA 13.0, Torch 2.14) while other workloads used other GPUs on the same node. The checkpoint loader
compared all 269 state-dict tensors, containing 1,434,652,673 elements, with the pinned source. For fixed prompts of
lengths 1, 7, 65 and 129, the official and common frontends produced 16 outputs each using the same reference token
histories. A separate CPU readback finds identical float32 bit patterns for all $4\times16\times49{,}152$ logits and
no greedy-token disagreements. These short batch-one cases qualify that frontend comparison only; they do not measure
task accuracy, independently generated trajectories or long-context serving.

The original 20 synthetic full/selected K/V reader cases cover lengths 1, 7, 64, 65 and 129, including repeated and
invalid selected indices. Every output meets its numerical tolerance against an independent reconstruction of the
\emph{actual} packed words and FP16 metadata; the largest absolute difference is $1.91\times10^{-6}$. Nevertheless,
16/20 cases fail the separately specified bitwise packing comparison to a CPU FP32-division reference, so the complete
gate fails. A second test exercises selected cross-loop QK and PV for four loop depths and the same five lengths.
All 40 output arrays match the independently reconstructed stored payload in this sample, but 36/40 cases fail the
packing check. Loop-depth cases reuse the same packed inputs and are not independent samples. The two other fused
attention entry points are evaluated separately below.

A post hoc audit of the 40 unique packed streams finds 221 differing 4-bit codes among 376,832 values; all occur within
$9.54\times10^{-7}$ of a rounding midpoint under the FP32-division reference. Stored FP16 minima and scales match.
Replacing the reference scale calculation by multiplication with the FP32 reciprocal of 15 reproduces all observed
codes, and sampled K/V SM89 machine code uses that reciprocal. This post hoc reconstruction accounts for the observed
differences in these saved inputs, but does not reverse the original failed gate or prove equivalence on other inputs.
A further post hoc CPU counterfactual compares reader reconstructions from the actual payload and from the strict
FP32-division payload. The difference exceeds the original reader tolerance for 88 of 3,152 standalone output elements
and 32 of 5,360 cross-loop output elements, with respective maxima 0.06641 and 0.08429. These cases reuse inputs; the
counts are not independent observations. This is not a new CUDA or model-quality test.
The first cross-loop probe
used a shared Triton cache; the subsequent reader and frontend runs used private caches. These are component and
frontend observations for a different implementation, not validation of the simulated allocator or a performance
comparison. Synthetic raw tensors and the sampled disassembly are available; real-model logits remain local.

A separate L40S follow-up checks the fused dense-source and sparse-delta attention entry points against NumPy
reconstruction of their actual packed data. Each has 20 synthetic cases, using the same five base lengths, zero- and
two-token tails, and the loop depths supported by that entry point. All outputs and returned logits meet the fixed
$0.004+0.004|\mathrm{reference}|$ tolerance; maximum output errors are $1.53\times10^{-5}$ for dense and
$5.96\times10^{-8}$ for sparse. Sparse logit checking covers only the returned current-token logits. Inputs are reused
across loops. An initial attempt stopped after four dense cases because its sparse loop argument was outside the CUDA
ABI; that failure is retained. The corrected follow-up's separate strict packing check fails for all 40 records when
all K/V streams, including inactive streams, are checked. Thus this result qualifies the tested actual-payload
calculations only, not the packer or end-to-end model behavior.

An additional exploratory run completed eight fixed batch-one functional windows, each with a 129-token prompt and
128 outputs. The original readback failed because it required fused QK/PV calls from the all-token quantization-only
adapter; the upstream selected-reader condition instead routes this adapter through full GEMV. A separate post hoc
readback changes only that entry-point expectation and verifies all eight windows' finite logits, token positions and
loop receipts. The original readback failure and the failed packing gate remain. These windows do not establish
quantized-model correctness, task quality, speed or a matched serving comparison.
In a post hoc analysis against the BF16 Flash baseline on this one prompt, the first greedy-token divergence is at
output 100 for INT4 and KL31, output 53 for Fisher31, and output 4 for FlashLoop default and sparse BF16. The quant-only
adapter and dense BF16 produce the same 128 greedy tokens as the baseline on this prompt, although their logits differ.
Logit comparisons stop at the first divergence so their token histories agree. This single-prompt observation does not
measure task accuracy or repair the failed packing gate.

\clearpage
\section{Physical packed-cache pilot on L40S}\label{app:l40s-packed}

This separately registered pilot uses the same pinned Ouro-1.4B checkpoint and common BF16 frontend as the preceding
component checks, on one L40S while other GPUs on the host run other workloads. It evaluates the physical LoopKV
\texttt{PackedCell} payload and FP16 offset/scale metadata with a Triton packed-attention reader, rather than the
simulated dense-plane allocator. The original \PackedNativeTests{} native CUDA tests pass without skips. The first
Compute Sanitizer attempt wrote its report and captured test output to the same file, so its zero-error message is
not accepted; a separate-log rerun completes all \PackedNativeTests{} tests and reports zero memcheck errors.

\paragraph{Full-model numerical gate.} On one 129-token prompt, we compare INT4, KL31 and Fisher31 readers against
reconstruction from the same physical cache for 16 outputs, feeding each the reference-generated token history. All
192 prefilled K/V cells have matching payload and metadata bytes, and the prefill logits match exactly. Nevertheless,
all three cases fail the pre-specified model gate: relative logit L2 is \PackedRelativeRange\% versus a 0.5\% limit,
maximum per-step KL exceeds $10^{-4}$, and INT4 has \PackedIntMismatch{} greedy-token mismatch (the other two have none).
The gate also requires every logit error to fit $0.125+0.01|\mathrm{reference}|$. These are three configurations of
one prompt, not independent task-quality samples; they do not establish that either arithmetic path is the ground truth.

\paragraph{Exploratory same-input diagnosis.} Along the reference trajectory we compare both readers on the identical
query, diagonal and physical cache at each of 15 decode steps and all 96 loop/layer calls per step, for each arm.
All \PackedAttentionCalls{} attention outputs meet the original $3\times10^{-5}+3\times10^{-5}|\mathrm{reference}|$
kernel tolerance; the largest absolute error is $\PackedMaxAttentionError$ and the largest normalized error is
\PackedMaxAttentionScaled. Casting the two outputs to BF16 nevertheless changes \PackedIntBFChanges{},
\PackedKLBFChanges{} and \PackedFisherBFChanges{} values for INT4, KL31 and Fisher31, respectively, out of
$1440\times2048$ values per arm. Each shadow run reproduces the previous reference logits bitwise and its prefill
cache fingerprints. For the first BF16-change sample per arm, an independent float64 NumPy reconstruction puts both
reader outputs within $\PackedOracleMax$ absolute error. These observations are consistent with small arithmetic
differences crossing BF16 rounding boundaries and being amplified by later layers, but do not prove the complete
causal chain or reverse the failed full-model gate.

\begin{table}[htbp]
\centering
\caption{Diagnostic batch-one L40S timings and retained cache backing, medians of three fixed-output runs per arm and
context. Each window generates 128 tokens; prefill includes allocation, encoding and first argmax, while decode covers
127 calls and includes argmax and finite-logit checks. The model-level packed-reader gate fails, so these are not
quality-qualified speedups. Peak is PyTorch allocated memory for the whole process, including model and transients.}
\label{tab:l40s-packed-pilot}
\begin{tabular}{rlrrrr}
\toprule
Context & Reader & Prefill (s) & Decode (token/s) & Cache (MiB) & Peak (GiB) \\
\midrule
512 & BF16 + Flash & 0.172 & 17.49 & 480.00 & 3.39 \\
512 & LoopKV INT4 & 0.310 & 6.16 & 150.44 & 3.06 \\
512 & LoopKV KL31 & 0.298 & 3.97 & 150.44 & 3.06 \\
512 & LoopKV Fisher31 & 0.319 & 3.79 & 150.44 & 3.06 \\
\midrule
2048 & BF16 + Flash & 0.185 & 17.48 & 1632.00 & 5.21 \\
2048 & LoopKV INT4 & 0.313 & 6.14 & 510.44 & 4.12 \\
2048 & LoopKV KL31 & 0.309 & 6.14 & 510.44 & 4.13 \\
2048 & LoopKV Fisher31 & 0.316 & 6.15 & 510.44 & 4.13 \\
 
\bottomrule
\end{tabular}
\end{table}

\paragraph{Diagnostic timing and storage.} Table~\ref{tab:l40s-packed-pilot} reports 24 timed windows, after eight
warmups and two short baseline profiles, with the arm order rotated across repeats. The prompts repeat a fixed token
sequence and are not independent natural long documents. The baseline forces Flash SDPA, confirmed by operator profiles;
the packed reader uses FP64 scores and FP32 values before BF16 casting. Thus the measurements compare these actual
implementations, not cache bit width alone. Cache backing includes low-bit payload, scales, offsets and shared rotation
buffers; capacity is allocated for prompt length plus 128 even though the final stored length is prompt plus 127.
Peak allocator values include model weights and transient workspaces, while reserved memory also depends on previous
windows in the same process. Other host workloads and three repeats limit reproducibility. The packed implementation is
slower than the BF16 baseline in these median observations, and the numerical failure precludes a valid speed/quality
claim. No network, scheduler, continuous batching, multi-request throughput, maximum capacity or task accuracy is tested.

\clearpage
\section{Artifact scope}\label{app:artifact}

The available artifact contains the manuscript sources, generated number and table fragments, figure data, benchmark
and analysis code, and accompanying tests. Together these allow the analysis logic and manuscript values to be
inspected, and they are sufficient to rebuild both PDFs. A compact correctness table for the eighteenth registration,
its source-file checksums and a Python-standard-library analysis script also reproduce the counts, exact sign tests
and intervals in Table~\ref{tab:paired-accuracy}. A second table contains 6714 per-output indicators across the test and
training samples and reproduces Table~\ref{tab:scoring-robustness}, parseability counts and rule transitions. These
tables contain IDs and indicators, with no prompts or generated text. The expanded audit contains \AllRows{}
indicators and reproduces Tables~\ref{tab:scoring-all-arms}--\ref{tab:scoring-all-comparisons}, including direct
extraction-interaction intervals.
The MATH audit releases another \MathScoreOutputs{} text-free indicators with four scoring rules, stopping reasons,
raw-file and scorer hashes, pinned dependencies and all historical-score discrepancies. These reproduce
Tables~\ref{tab:math-scoring}--\ref{tab:math-scoring-paired} without symbolic dependencies; independent rescoring of
the private generated text additionally requires those dependencies and the pinned upstream source.

A historical evidence index maps the first 20 ledger entries to preserved controls, available registration documents,
package checksums, analysis summaries and run records. Executable-source files are exported by content hash from the
saved source archives, with each archive's recorded Git revision; they supplement the later revision in the main code
directory. Compressed numerical outputs retain their original bytes and checksums. A Python-standard-library script
recomputes the first batch's key-only KL shares and KL/JVP summaries, and all 35 numerical cells of the allocation
table, directly from these per-sequence files. It does not recompute the full verdict ledger or Fisher tensors.
The later Ouro-2.6B evidence bundle separately preserves the 11-criterion registration and amendment, all three archive
checksums (scored, aborted and profiler runs), numerical calibration and per-sequence records, and the independent
audit. Its converted sensitivity array retains a hash binding to the original tensor; the complete runtime-generated
model and freeze manifests were not independently recovered, although the archived preflight records their checks.
The L40S component bundle contains synthetic source inputs, packed tensors, outputs, independent CPU readers and
sampled disassembly. It does not redistribute the real-checkpoint frontend logits.
The physical packed-cache pilot additionally releases sanitized per-window timing and per-attention error summaries,
the clean memcheck logs and scripts that recompute their aggregate counts. Real-model logits, tokens, K/V samples and
profiler traces remain private; the public summary alone cannot reproduce the GPU execution.

The index distinguishes content integrity from historical timing and complete reproduction. The first batch has no
separately packaged registration document; one failed hardware trial lacks a saved archive-checksum receipt, and the
earliest Huginn repair revisions are not all separate source snapshots. Original runtime-generated freeze tables,
model weights, full prompt datasets, generated text, real-model binary tensors and profiler traces are not included. Thus these
materials expose the criteria and available measurement records without independently proving when they were fixed
or reproducing every GPU run end to end. The code path used for the reported format study
uses dense per-loop planes and simulated quantization; it is not a packed mixed-bit paged-KV implementation. We
therefore treat the single-cell result as specific to the reported checkpoints and prompts until further replication,
and we draw no serving-throughput or long-context conclusion from it.

\end{document}